# Towards a Generalized SA Model: Symbolic Regression-Based Correction for Separated Flows

Sun Xuxiang[1,2,3], Shan Xianglin[1,2,3], Liu Yilang[1,2,3], Zhang Weiwei[1,2,3,*]

1.School of Aeronautic, Northwestern Polytechnical University, Xi'an 710072, China;

2.International Joint Institute of Artificial Intelligence on Fluid Mechanics, Northwestern Polytechnical University, Xi'an, 710072, China;

3.National Key Laboratory of Aircraft Configuration Design, Xi'an 710072, China

*Corresponding author. E-mail: aeroelastic@nwpu.edu.cn

**Abstract**: This study focuses on the numerical simulation of high Reynolds number separated flows and proposes a data-driven approach to improve the predictive capability of the SA turbulence model. First, data assimilation was performed on two typical airfoils with high angle-of-attack separated flows to obtain a high-fidelity flow field dataset. Based on this dataset, a white-box model was developed using symbolic regression to modify the production term of the SA model. To validate the effectiveness of the modified model, multiple representative airfoils and wings, such as the SC1095 airfoil, DU91-W2-250 airfoil, and ONERA-M6 wing, were selected as test cases. A wide range of flow conditions was considered, including subsonic to transonic regimes, Reynolds numbers ranging from hundreds of thousands to tens of millions, and angles of attack varying from small to large. The results indicate that the modified model significantly improves the prediction accuracy of separated flows while maintaining the predictive capability for attached flows. It notably enhances the reproduction of separated vortex structures and flow separation locations, reducing the mean relative error in lift prediction at stall angles by 69.2% and improving computational accuracy by more than three times. Furthermore, validation using a zero-pressure-gradient flat plate case confirms the modified model's ability to accurately predict the turbulent boundary layer velocity profile and skin friction coefficient distribution. The findings of this study provide new insights and methodologies for the numerical simulation of high Reynolds number separated flows, contributing to more accurate modeling of complex flow phenomena in engineering applications.

# 1 Introduction

Turbulence represents a complex state of fluid flow characterized by a broad range of scales and swirling motion, often accompanied by intense fluctuations. It is observed across various systems, ranging from large-scale atmospheric currents to small-scale eddies within mixing tanks [1]. Many applications in engineering rely on the prediction of turbulent flows, for instance: aircraft design, atmospheric flows, gas turbine engines, reactive flows etc. Despite the increase in computational power over the last decades, the full solution of Navier–Stokes equations remains unfeasible for the majority of applications. Therefore, engineers have to rely on turbulence models to overcome this challenge.

Among the various approaches used to model turbulence, the Reynolds-Averaged Navier-Stokes (RANS) method stands as a widely adopted technique. RANS is based on the concept of decomposing the flow variables into time-averaged mean quantities and turbulent fluctuations. This separation allows the equations governing the mean flow to be solved, while the unresolved turbulent fluctuations are modeled using turbulence models. Its critical function in predicting fluid flow behavior is evidenced in a broad range of engineering applications, including aerodynamics[2], heat transfer[3], and combustion[4].

Turbulence modeling is one of key elements in both the RANS framework and Computational Fluid Dynamics (CFD)[5]. Significant endeavors have been directed towards formulating closure equations for the RANS equations, leading to the development of several renowned models. Examples include the one-equation model Spalart–Allmaras (SA)[6], and the two-equation models $k-\epsilon$[7] and $k-\omega$[8], alongside their numerous variations[5]. These models are characterized by sets of parameters calibrated using experimental data from canonical flows of simplified nature. However, owing to the tailored development of each model to address specific flow regimes, coupled with the apparent non-universality of their parameters, questions persist regarding their accuracy and associated uncertainties. On the other hand, despite advancements in computational capabilities, the use of high-fidelity simulations such as Direct Numerical Simulation (DNS) and Large Eddy Simulation (LES) remains limited for high Reynolds number flows commonly encountered in aerospace applications[9]. Therefore, improving the accuracy of Reynolds-Averaged Navier-Stokes (RANS) methods remains a worthwhile research area[10].

With the development of machine learning technologies, there has been increasing interest in using machine learning and data-driven approaches to enhance turbulence models[11, 12]. For example, high-fidelity data from LES/DNS can be used to train a data-driven model to replace or improve traditional turbulence models[13-18]. Ling and Templeton et al.[19, 20] constructed model invariants in the form of Galilean invariants and employed machine learning to predict unphysical eddy viscosity regions from traditional model calculations. By classifying these regions using methods like Support Vector Machines (SVMs), they significantly improved the ability to detect errors in RANS results. Zhu et al.[17] developed a purely data-driven turbulent black-box algebraic model using a single-layer neural network and achieved coupled computations with solvers. Their results demonstrated that the constructed model achieved comparable accuracy to the SA model, along with higher computational efficiency. It also exhibited generalization capabilities for different flow conditions and geometries, validating the feasibility of such replacement models.

In cases where high-fidelity data is difficult to obtain (e.g., for high Reynolds number flows), inverse flowfield modeling using data assimilation has emerged as an effective approach[21-24]. Duraisamy et al.[25], Singh et al.[26-28], and Parish et al.[29] established a framework combining flowfield inversion with machine learning (FIML). In this framework, spatially distributed correction coefficients are added to turbulence models, optimized using adjoint methods, and then learned through neural networks. Building on the FIML framework, prior physical constraints for non-equilibrium turbulent effects can be introduced during flowfield inversion[30, 31], while corrections for turbulence anisotropy can be incorporated into nonlinear eddy viscosity models by introducing correction coefficients [32]. Wu et al. used a curved backward-facing step case as their study object to perform flowfield inversion corrections to the destruction term in the ω equation of the SST model[22]. Based on the high-fidelity flowfield data obtained from inversion, they developed an explicit algebraic model mapping local flow features to correction fields using Symbolic Regression (SR). This model demonstrated superior performance compared to the traditional SST model across different test cases and exhibited good generalization capabilities. Additionally, other works by this research team[33, 34] further highlighted the broad potential of combining flowfield inversion with symbolic regression in data-driven turbulence modeling.

Unlike the adjoint method for flowfield inversion, Yang et al. [35]employed an Ensemble Kalman Filter (EnKF) to invert correction coefficients in the $k - \omega - \gamma - A_r$ transition model and trained neural network and random forest models based on the inverted data. Some studies have also used data assimilation methods to correct coefficients of classical turbulence models[36, 37]. For instance, Li et al.[37] determined the coefficients of a $k - \omega$ model in order to minimize the velocity error with respect to high-fidelity simulations. The optimization problem can then be treated with, for example, gradient-based techniques or ensemble-based method. Recently, as an integration of modeling and data assimilation approaches, there have been efforts to directly construct neural network models based on observed data using flowfield inversion techniques[38-41].

In the data-driven modeling paradigm, the generalization ability of a model is a crucial performance metric. Generalization refers to a model's capability to make accurate predictions on unseen datasets and serves as a key indicator of model performance. Due to the inherent complexity of turbulence problems, current data-driven turbulence models exhibit limited generalization capabilities, typically performing well only on geometries and flow conditions similar to the training data. This limitation is closely related to factors such as model selection, training data size, feature selection, and the incorporation of physical principles. Wu et al.[34] proposed four levels of model generalization capability within the FIML framework:

(1) Generalization to similar geometries.

(2) Generalization to simple attached flows.

(3) Generalization to completely different geometries and flow conditions with the same physical mechanisms. For example, a model trained on separation flows dominated by an adverse pressure gradient should perform well on other adverse pressure gradient-driven separation flows.

(4) Generalization across a range of test cases with entirely different separation characteristics, geometries, and inflow conditions. For instance, a model should accurately predict both separation caused by an adverse pressure gradient and separation induced by bluff-body geometries. They noted that most current data-driven turbulence models remain at the first level of generalization, with only a few models[22, 42] achieving the third level. They also emphasized the importance of the second level for data-driven turbulence models. McConkey et al.[43] supported this perspective. They evaluated the generalization ability of random forests, neural networks, and XGBoost[44] using a dataset consisting of periodic hills, bumps, curved backward-facing steps, square ducts, and contraction-expansion pipes. Their findings led them to recommend XGBoost for turbulence modeling due to its strong generalization ability and relatively low tuning cost. They also observed that the constructed models generalized well to flows similar to the training data but lacked generalization to unseen flow types. As a result, they concluded that machine learning methods are best suited for developing specialized models for specific flow types, a common requirement in industrial applications. Li et al.[45] conducted a comparative study on the generalization capabilities of three frameworks: TBNN, FIML, and PIML. Their results indicated that FIML exhibited superior generalization performance.

The findings from the 2022 NASA Turbulence Modeling Workshop further confirmed that most current machine learning models can only generalize to flow conditions similar to those in the training dataset. In his invited talk at the workshop, Spalart, the creator of the SA model, pointed out several critical issues with existing machine-learning-based turbulence models, particularly regarding generalization and interpretability. He emphasized that no "universal" turbulence model has emerged from current research, with many models failing even to provide reasonable boundary layer predictions for turbulent flat plates. Despite these shortcomings, Spalart did not entirely dismiss the potential paradigm shift that machine learning could bring but stressed that overcoming deep-seated challenges is necessary. He suggested that, at this stage, machine learning should play more of a "supporting role" rather than acting as an "architect" in independently designing complete turbulence models. Overall, the generalization ability of data-driven turbulence models remains a major challenge faced by researchers, requiring further in-depth analysis and exploration.。

The aforementioned studies demonstrate the potential of building or improving turbulence models through data-driven approaches; however, most of these works focus on low Reynolds number separated flows with simple geometries or high Reynolds number attached flows. Limited attention has been given to high Reynolds number complex separated flows commonly encountered in industrial applications. This paper specifically addresses this aspect and aims to develop a reasonable data-driven model to enhance the simulation accuracy of the classical SA turbulence model for high Reynolds number separated flows.

This work is organized as follows: Section 2 introduces the main numerical computation methods. Section 3 explains the key aspects of symbolic regression modeling, including sample point selection and the construction of input and output features. Section 4 presents the computational results of the corrected model on multiple test cases, comparing them with experimental data and the results of the SA model. Finally, Section 5 provides a summary and outlook.

## 2 Numerical Methods

The RANS equations are defined as follows in the conservation form:

$$\frac{\partial \rho}{\partial t} + \frac{\partial (\rho u_i)}{\partial x_i} = 0,$$

$$\frac{\partial (\rho u_i)}{\partial t} + \frac{\partial (\rho u_i u_j)}{\partial x_j} = -\frac{\partial p}{\partial x_i} + \frac{\partial}{\partial x_j}(\tau_{ij}^l + \tau_{ij}^t),$$

$$\frac{\partial (\rho E)}{\partial t} + \frac{\partial (\rho E u_j)}{\partial x_j} = -\frac{\partial (p u_j)}{\partial x_i} + \frac{\partial}{\partial x_j}\left[\left(\tau_{ij}^l + \tau_{ij}^t\right)u_i\right] + \frac{\partial}{\partial x_j}\left(q_j - c_p \rho \overline{T'u_j'}\right)$$

The quantity $\tau_{ij}^t$ is known as the Reynolds-stress tensor. Here we consider the classical Boussinesq approximation for the Reynolds stress tensor:

$$\tau_{ij}^t = \mu_t \left(\frac{\partial \overline{u}_i}{\partial x_j} + \frac{\partial \overline{u}_j}{\partial x_i}\right) - \frac{2}{3}\rho k \delta_{ij}$$

$$k = \frac{1}{2}\overline{u'_k u'_k} = \frac{1}{2}\left(\overline{u'^2} + \overline{v'^2} + \overline{w'^2}\right)$$

where $\mu_t$ is the eddy viscosity and $k$ is the turbulent kinetic energy, usually accounted in the pressure term.

We use the particular Spalart–Allmaras (SA) models in this work[46]. This model takes into account the convection, diffusion, production, and destruction of the eddy viscosity in a single transport equation as follows:

$$\mu_t = \rho\hat{v}f_{v1}$$

$$\frac{D\hat{v}}{Dt} = C_{b1}(1-f_{t2})\hat{S}\hat{v} + \frac{1}{\sigma}\left[\frac{\partial}{\partial x_j}\left((v+\hat{v})\frac{\partial \hat{v}}{\partial x_j}\right) + C_{b2}\frac{\partial \hat{v}}{\partial x_i}\frac{\partial \hat{v}}{\partial x_i}\right] - \left(C_{w1}f_w - \frac{C_{b1}}{\kappa^2}f_{t2}\right)\left(\frac{\hat{v}}{d}\right)^2$$

The production, destruction, and transportation terms in the SA turbulence model read

$$P = C_{b1}(1-f_{t2})\hat{S}\hat{v},$$

$$D = \left(C_{w1}f_w - \frac{C_{b1}}{\kappa^2}f_{t2}\right)\left(\frac{\hat{v}}{d}\right)^2$$

$$T = \frac{1}{\sigma}\left[\frac{\partial}{\partial x_j}\left((v+\hat{v})\frac{\partial \hat{v}}{\partial x_j}\right) + C_{b2}\frac{\partial \hat{v}}{\partial x_i}\frac{\partial \hat{v}}{\partial x_i}\right]$$

where $d$ is the nearest distance to the wall. Functions used in the model are defined by:

$$f_{v1} = \frac{\chi^3}{\chi^3 + c_{v1}^3}, \chi = \frac{\hat{v}}{v}, \hat{S} = \Omega + \frac{\hat{v}}{\kappa^2 d^2}f_{v2}, \Omega = \sqrt{2W_{ij}W_{ij}}$$

$$f_{v2} = 1 - \frac{\chi}{1+\chi f_{v1}}, f_w = g\left[\frac{1+c_{w3}^6}{g^6 + c_{w3}^6}\right]^{1/6}, g = r + c_{w2}(r^6 - r),$$

$$r = \min\left[\frac{\hat{v}}{\hat{S}\kappa^2 d^2}, 10\right], f_{t2} = c_{t3}\exp(-c_{t4}\chi^2), W_{ij} = \frac{1}{2}\left(\frac{\partial u_i}{\partial x_j} - \frac{\partial u_j}{\partial x_i}\right),$$

$$c_{b1} = 0.1355, \sigma = 2/3, c_{b2} = 0.622, \kappa = 0.41,$$

$$c_{w2} = 0.3, c_{w3} = 2, c_{v1} = 7.1, c_{t3} = 1.2, c_{t4} = 0.5, c_{w1} = \frac{c_{b1}}{\kappa^2} + \frac{1+c_{b2}}{\sigma}$$

The SA model provides high accuracy for equilibrium flows, but fails to accurately capture the separation and recovery in non-equilibrium wall-bounded separating flows. Our study aims to address this drawback by using a data-driven-based correction of the production term of SA model:

$$\frac{D\hat{v}}{Dt} = \beta(\mathbf{x})P(\hat{v},\mathbf{U}) - D(\hat{v},\mathbf{U}) + T(\hat{v},\mathbf{U})$$

## 3 Symbolic Regression Modeling

By performing data assimilation on two typical separated flows, high-fidelity flowfield data and spatial distributions of the correction coefficients for the SA production term were obtained. These two representative flows are:

（1）The separation flow over an SC1095 airfoil under the inflow conditions of Ma=0.6, angle of attack (AoA)=9.17°, Re=4.9×10$^6$. In this type of flow, the standard SA model predicts the flow separation point too early, resulting in a separated vortex occupying approximately 80% of the upper surface under the influence of the shock wave and adverse pressure gradient. This leads to a rapid decrease in lift, with the numerically simulated lift coefficient being smaller than the experimental value. During the data assimilation process, adjustments were made to the spatial distribution of the correction coefficients for the SA model's production term, which delayed the flow separation point, suppressed separation to some extent, and reduced the size of the separated vortex. This resulted in better agreement between numerical simulation results and experimental data. A comparison of the velocity fields before and after assimilation is shown in Figure 1.

（2）The separation flow over a DU91-W2-250 airfoil under the inflow conditions of Ma=0.15, AoA=15.0º, Re=3.0×10$^6$. This case represents a typical trailing-edge separation for a thick airfoil. In this type of flow, the standard SA model predicts the flow separation point too far downstream, resulting in a smaller separated vortex. Consequently, the lift coefficient obtained through numerical simulation is larger than the experimental value within the airfoil's stall angle range. The data assimilation process enhanced the separation by reducing $\beta$ at the leading edge of the upper surface, which moved the separation point upstream and increased the size of the separated vortex. This resulted in better agreement between computed results and experimental data. The velocity field and streamline comparisons before and after assimilation for this case are shown in Figure 2. These two types of flows are highly representative of airfoil separation phenomena, so the data obtained from these cases were used as the sample dataset for constructing the data-driven model.

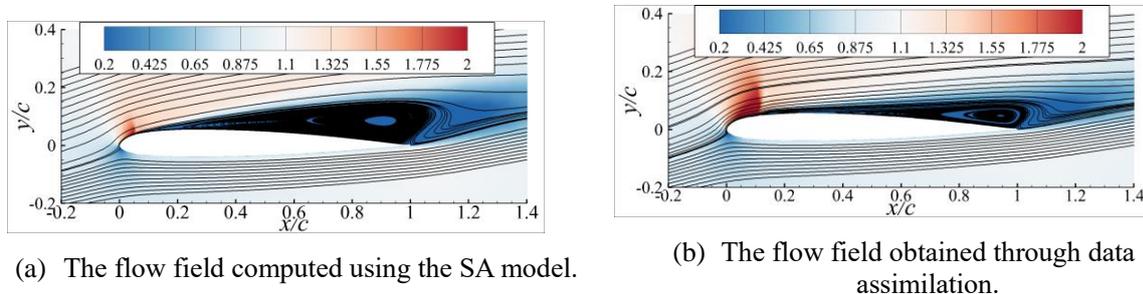

(a) The flow field computed using the SA model.  (b) The flow field obtained through data assimilation.

Figure 1 Comparison of the flow field for the SC1095 airfoil under conditions Ma=0.6, AoA=9.17°, and Re=4.9×10$^6$. The left figure shows the computed flow field and streamline distribution using the standard SA model, while the right figure shows the optimal flow field and streamline distribution obtained by assimilating the generation term coefficient distributions of the SA model.

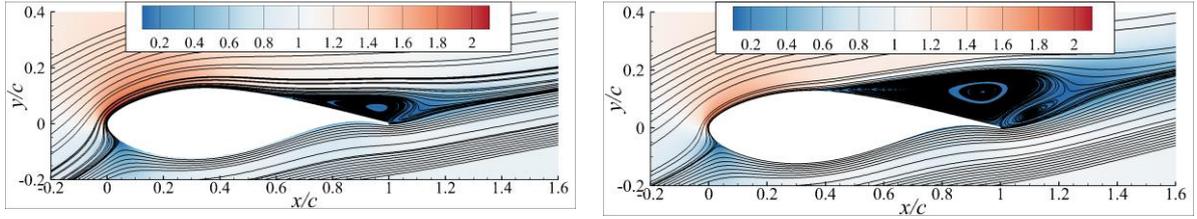

(a) The flow field computed using the SA model.

(b) The flow field obtained through data assimilation.

Figure 2 The flow field comparison for the DU91-W2-250 airfoil under the incoming flow conditions of Ma=0.15, AoA=15.0º, Re=3.0×10⁶. The left figure shows the computed flow field and streamline distribution using the standard SA model, while the right figure illustrates the optimal flow field and streamline distribution obtained by assimilating the generation term coefficient distributions of the SA model.

After obtaining the assimilated dataset, we employed symbolic regression to establish a model linking local flowfield characteristics to the correction coefficient $β$. Compared to neural network-based models, methods based on symbolic regression require relatively smaller amounts of data[22, 33], necessitating downsampling and simplification of samples. Representative sample points are selected for symbolic regression modeling. In the relationship expression constructed in this study, the model input consists of time-averaged flow features, while the output corresponds to the correction coefficient of the source term at each grid cell, whose spatial distribution is illustrated in Figure 3. From this spatial distribution, it can be observed that in most regions of the flowfield, the $β$ value is close to 1, indicating that no correction is applied to the source term. Significant variations in $β$ occur only near the wall, which is why the training sample data points were primarily selected from this region. Specifically, 2,000 random points near the wall were chosen, resulting in a total of 4,000 data points used for training the symbolic regression model. Table 1 lists the features used for modeling along with their corresponding expressions. During the modeling process, numerical scaling and transformation of the features were necessary to ensure consistency in magnitude, which facilitates model training. Referring to previous studies[13, 33], certain feature transformations were applied, with the transformation formulas provided in the last column of Table 1.

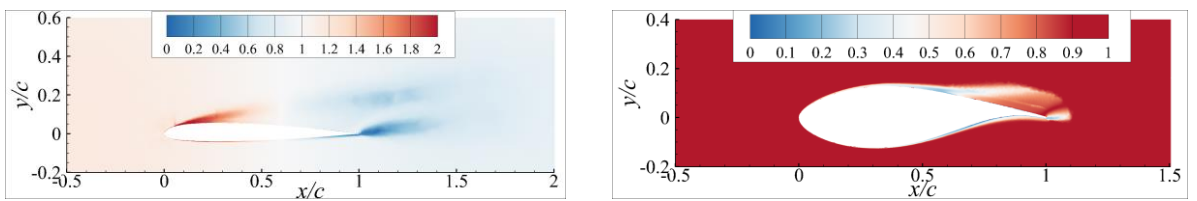

(a) The spatial distribution of the $β$ field for the SC1095 airfoil.

(b) The spatial distribution of the $β$ field for the DU91-W2-250 airfoil.

Figure 3 The spatial distribution of the $β$ field obtained through data assimilation for different airfoil geometries in the training dataset.

Among the features listed in Table 1, $x_0$ and $x_1$ represent the magnitudes of the rotation rate tensor and the strain rate tensor, respectively. The feature $x_2$ characterizes the entropy of the flowfield [17, 21] while $x_3$ is the dimensionless Q-criterion. The terms $x_4$ and $x_5$ are invariants defined based on the rotation rate tensor and the strain rate tensor, respectively [1, 14, 22]. The feature $x_6$ denotes the minimum distance from the grid cell center to the wall. The term $x_7$ represents the magnitude of the pressure gradient, where $\varepsilon$ in the transformation formula is a small constant introduced to ensure numerical stability, set to $10^{-9}$ n actual computations. The feature $x_8$ is used for boundary layer identification[47]. This expression is employed in the DDES method as a shielding function to ensure that the RANS approach is still applied within the boundary layer.

Table 1 Input feature expressions and transformation formulas for symbolic regression modeling.

| Feature | Expression | Transformation |
|---|---|---|
| $x_0$ | $\sqrt{2W_{ij}W_{ij}}, W_{ij} = \frac{1}{2}\left(\frac{\partial U_i}{\partial x_j} - \frac{\partial U_j}{\partial x_i}\right)$ | $\log(x+1)$ |
| $x_1$ | $\sqrt{2S_{ij}S_{ij}}, S_{ij} = \frac{1}{2}\left(\frac{\partial U_i}{\partial x_j} + \frac{\partial U_j}{\partial x_i}\right)$ | $\log(x+1)$ |
| $x_2$ | $P\rho^{-\gamma} - 1$ | — |
| $x_3$ | $\dfrac{x_0^2 - x_1^2}{x_0^2 + x_1^2}$ | — |
| $x_4$ | $S_{ij}S_{ji}$ | — |
| $x_5$ | $W_{ij}W_{ji}$ | — |
| $x_6$ | $d$ | — |
| $x_7$ | $\sqrt{\dfrac{\partial P}{\partial x_i}\dfrac{\partial P}{\partial x_i}}$ | $\log(x+\varepsilon)$ |
| $x_8$ | $1 - \tanh(8r_d)^3 \quad r_d = \dfrac{v_l + v_t}{\kappa^2 d^2 \sqrt{U_{ij}U_{ij}}}$ | — |

All fundamental flow variables used for feature calculations are dimensionless, and their nondimensionalization follows the convention below:

$$x_i = \frac{\overline{x_i}}{L}, \rho = \frac{\overline{\rho}}{\rho_\infty}, U_i = \frac{\overline{U_i}}{V_\infty}, P = \frac{\overline{P}}{\rho_\infty a_\infty^2}, \mu = \frac{\overline{\mu}}{\mu_\infty}$$

where the subscript "∞"denotes the freestream value, and the superscript "-" indicates a dimensional quantity.In summary, the algebraic relationship to be constructed in this study is formulated as follows:

$$\beta = f(x_0,...,x_8)+1$$

When $\beta$=1, it indicates that the original SA model is used without any modifications. Since $\beta$ remains close to 1 in most regions of the flowfield, a prior assumption of 1 is incorporated into the algebraic relationship being constructed, while the deviation near the wall is modeled using symbolic regression.

In this study, the symbolic regression model is built using the PySR library. The predefined operator set is listed in Table 2, and the loss function is defined as the sum of squared errors between the predicted and target values, specifically given by:

$$loss = \sum_i (Y_{i,pred} - Y_{i,target})^2$$

Table 2 Operators used in symbolic regression.

| Operator Type | Operators |
| --- | --- |
| Unary | $\exp(x_i)$, $\tanh(x_i)$ |
| Binary | $x_i + x_j, x_i - x_j, x_i * x_j, x_i / x_j$ |

Although the symbolic regression model offers several advantages, it also has certain limitations, such as insufficient model expressiveness, strict requirements for feature selection, and difficulties in effective modeling on large-scale datasets. In this study, the dataset used for model construction consists of 4,000 data points, primarily obtained from adverse pressure gradient flowfields, with a lack of data from favorable pressure gradient and zero-pressure gradient scenarios. This limitation affects the model's generalizability and correction performance, particularly leading to incorrect activation of corrections in certain regions of the flowfield or under attached flow conditions, which may result in deviations or even errors in the final computation. Attempts to expand the dataset and incorporate a broader range of flowfield data for retraining revealed significant challenges in the modeling process. To prevent incorrect activation of the correction formula, it is necessary to apply post-hoc modifications to the derived equations based on relevant physical priors. This process of posterior correction integrates physical knowledge with data-driven modeling, a widely adopted approach in recent studies [22, 33, 34, 48]. After the corrections, the model is embedded into the CFD solver for bidirectional coupled flow simulations. In the following test cases, results obtained from the model-corrected computations are denoted as "SR".

## 4 Validation of the SR model

This section presents the computational results of the SA urbulence model with SR corrections under different geometries and inflow conditions, along with comparisons to the standard SA model and experimental data. The test cases include the SC1095 airfoil, ONERA-M6 wing, DU91-W2-250 airfoil, S809 airfoil, S814 airfoil, S805 airfoil, curved backward-facing step, hump, and zero-pressure-gradient flat plate. Schematic diagrams of the different airfoil shapes are shown in Figure 4.

The SC1095 airfoil and ONERA-M6 wing cases primarily examine the performance of the corrected model in transonic shock-induced separation flows. The DU91-W2-250 airfoil and S-series wind turbine airfoils are used to assess the model's ability to correct separated flows at high angles of attack. In these two categories of test cases, the first type is characterized by large-scale separation induced by shock waves, where the correction should appropriately suppress the size of separation vortices and shift the separation point downstream. Conversely, in the second type, the opposite correction is required—advancing the flow separation point to improve agreement between the computational and experimental results. The curved backward-facing step and hump cases are included to evaluate the corrected model's performance in different separation mechanisms. Finally, the zero-pressure-gradient flat plate case is used to assess the corrected model's behavior in simple attached flow over a smooth surface, where it should yield results consistent with the standard SA model.

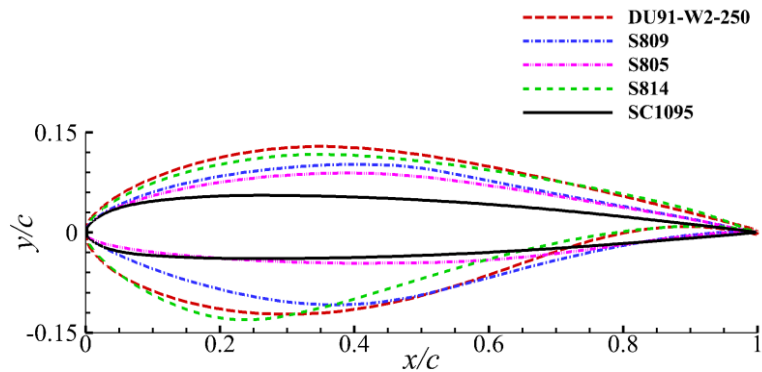

Figure 4 Schematic diagram of the airfoil shape.

### 4.1 SC1095 Airfoil

This section presents the coupled computational results of the corrected model for the SC1095 airfoil. Since the training data for the corrected model includes assimilated data from the SC1095 airfoil at Ma=0.6, AoA=9.17º, Re=4.9×10$^6$ the primary objective here is to evaluate the model's generalization capability under different inflow conditions.

Figure 5 compares the lift and drag coefficients under the inflow conditions of Ma=0.6

and Re=4.9×10$^6$. In the figure, the green solid line and black arrows indicate the training data conditions, the black solid line represents the experimental results, the red solid line corresponds to the results obtained using the SA model with SR corrections, and the blue dashed line represents the results from the original SA model. Unless otherwise specified, subsequent figures in this chapter follow this same notation. The comparison in Figure 5 shows that the corrected model effectively improves the computational results in the stall angle of attack region, yielding lift and drag coefficients that better align with experimental values. Under low angle-of-attack conditions, where the flow remains attached, the corrected model's performance is nearly identical to that of the standard SA model. This indicates that the model is only activated in separated flow conditions while maintaining consistency with the standard SA model in attached flow regions.

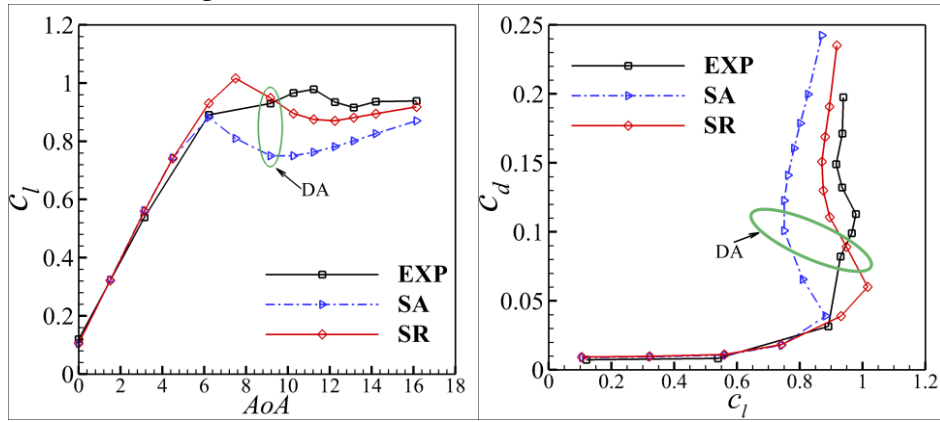

Figure 5 Comparison of lift and drag coefficients for the SC1095 airfoil at Ma=0.6, Re=4.9×10$^6$

Figure 6 presents the comparison of surface pressure coefficient distributions at angles of attack AoA=6.22º and 9.17º. The results indicate that the corrected model provides a more accurate prediction of the surface pressure coefficient distribution. Figure 7 shows the velocity fields and streamlines obtained from both models for the SC1095 airfoil at Ma=0.6, AoA=9.17º, Re=4.9×10$^6$. It can be observed that the shock wave position predicted by the corrected model is further downstream, and the model effectively suppresses flow separation, reducing the size of the separation vortex. As a result, the computed flowfield aligns more closely with the experimental observations. Additionally, the comparison of the surface friction coefficient in Figure 8 further confirms that the corrected model suppresses flow separation, shifting the separation point downstream. These findings collectively demonstrate that the corrected model effectively improves the prediction of shock-induced separation flows.

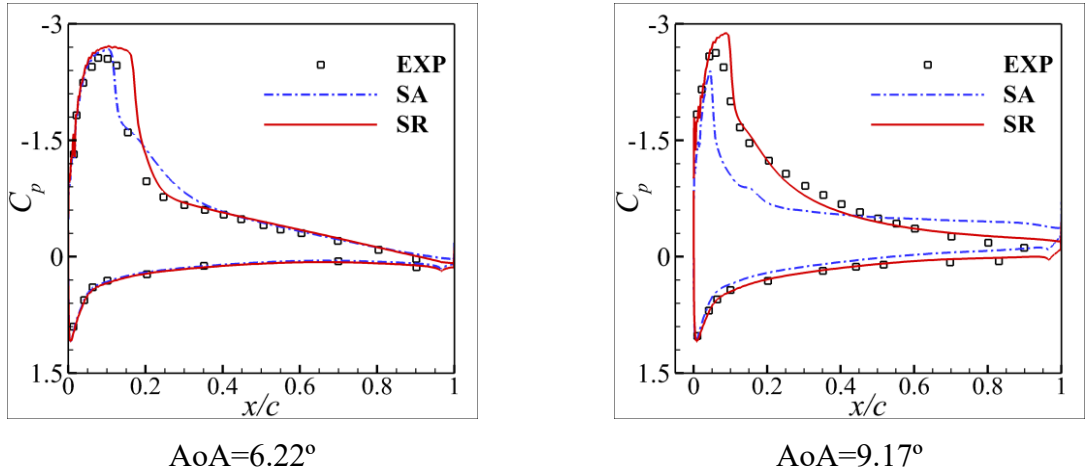

AoA=6.22º           AoA=9.17º

Figure 6 Comparison of wall pressure coefficient results for different angles of attack on the SC1095 airfoil at Ma=0.6, Re=4.9×10$^6$

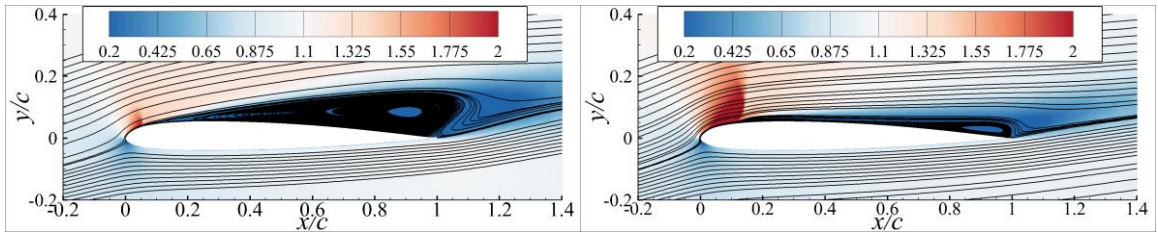

Figure 7 Flow field of the SC1095 airfoil at Ma=0.6, AoA=9.17º, Re=4.9×10$^6$ using different models: Left figure shows results from the SA model; right figure shows results from the SA model with SR correction.

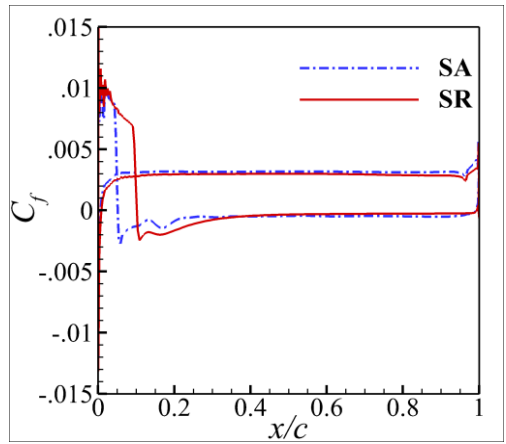

Figure 8 Comparison of skin friction coefficient distributions for the SC1095 airfoil at Ma=0.6, AoA=9.17º, Re=4.9×10$^6$ using different models.

Figure 9 presents the relative error of the lift coefficient computed by the standard SA model and the corrected model compared to experimental data at stall angles of attack. The results show that, in contrast to the SA model, the corrected model significantly reduces simulation errors across different angles of attack. For lift predictions throughout the stall region, the relative error of the SA model is 15.94%, whereas the SR model reduces this error substantially to 5.32%, representing a 66.62% reduction in computational error and a threefold improvement in accuracy. This demonstrates that the corrected model exhibits strong applicability in separated flow regions and significantly enhances the accuracy of RANS-based simulations for separated flows.

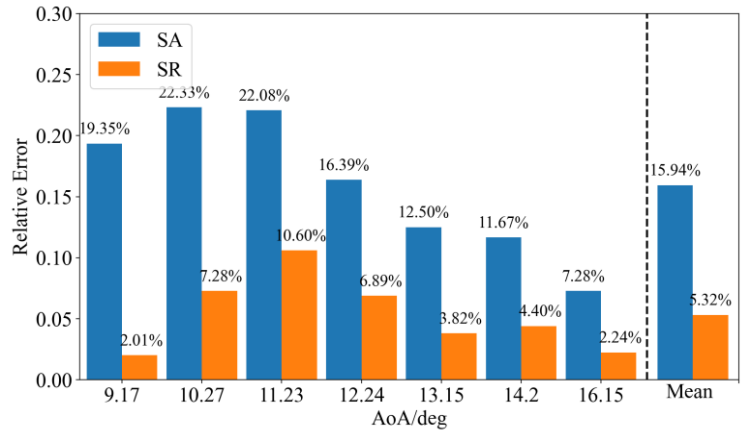

Figure 9 Comparison of relative errors in lift coefficients at stall angle of attack for the SC1095 airfoil at Ma=0.6, Re=4.9×10$^6$
Figure 10 Comparison of lift and drag coefficients for the SC1095 airfoil at Ma=0.5, Re=4.34×10$^6$

Figure 10 presents a comparison of the lift and drag coefficients computed using the two models under the inflow conditions of Ma=0.5 and Re=4.34×10$^6$. The results show that in the stall angle of attack region, the corrected model's calculations are in excellent agreement with experimental data, indicating that the model has good generalization capability across different flow states. Figure 11 further shows the relative error of the lift coefficient at the stall angle of attack for this inflow condition. It can be observed that, compared to the SA model, the corrected model significantly reduces prediction errors across different angles of attack. For the entire stall region, the relative error of the lift coefficient for the SR model decreases substantially from 12.28% (for the SA model) to 1.56%, reducing the computational error by 87.30% and improving the accuracy by a factor of 7.87.

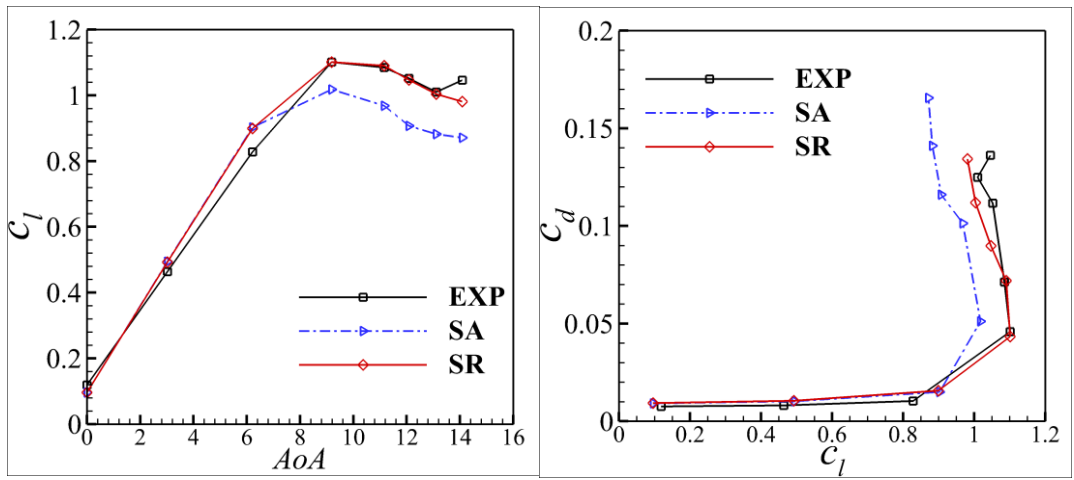

Figure 10 Comparison of lift and drag coefficients for the SC1095 airfoil at Ma=0.5, Re=4.34×10$^6$

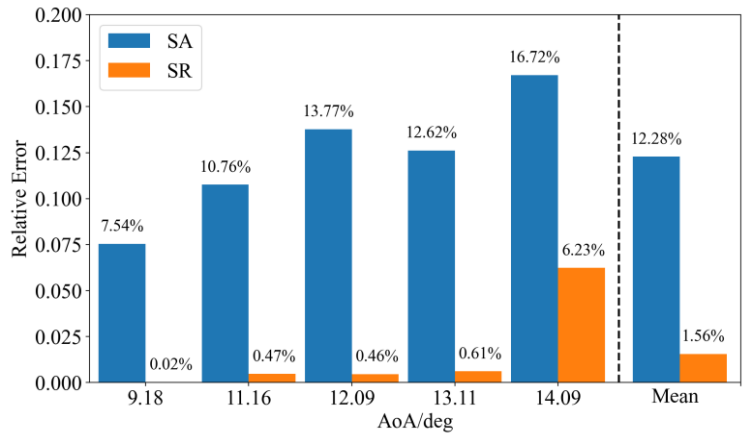

Figure 11 Comparison of relative errors in lift coefficients at stall angle of attack for the SC1095 airfoil at Ma=0.5, Re=4.34×10$^6$

## 4.2 ONERA-M6

This section presents a comparison of the model's results in three typical computational states for the ONERA-M6 wing case. The freestream Mach number is 0.84, with inflow AoA of 3.06º、5.06º、6.06º, and the Reynolds number based on the average aerodynamic chord length is 11.72×10$^6$. Among the three angles of attack, the first corresponds to attached flow, while the other two represent shock-induced separated flow. This case mainly investigates the performance of the corrected model in three-dimensional attached flow and separated flow. Figure 12 to Figure 14 present the comparison of surface pressure coefficient distributions at different wing sections under various inflow angles of attack. As shown in Figure 12, under the attached flow condition at 3.06º , the results of both models are very similar, indicating that the modified model can maintain consistency with the SA model in three-dimensional attached flow cases.  In the two separated flow cases at higher angles of attack (Figure 13 and Figure 14) , the corrected model significantly outperforms the original SA model, indicating that the correction has been activated and plays a positive role in these typical separated flow scenarios. By examining the comparison of surface streamlines from different models in separated flow (Figure 15 and Figure 16), it can be seen that for the two sections near the wing root（$\eta$=0.2/0.44）the flow remains attached, and thus, the pressure coefficient results from both models remain consistent. However, in the sections closer to the wingtip, shock-induced separated flows are observed. In these sections, the corrected model effectively corrects the flow, providing more accurate shock position predictions compared to the SA model. As a result, the corrected model gives a more precise indication of the pressure variation behind the shock, leading to computational results that align more closely with experimental data.

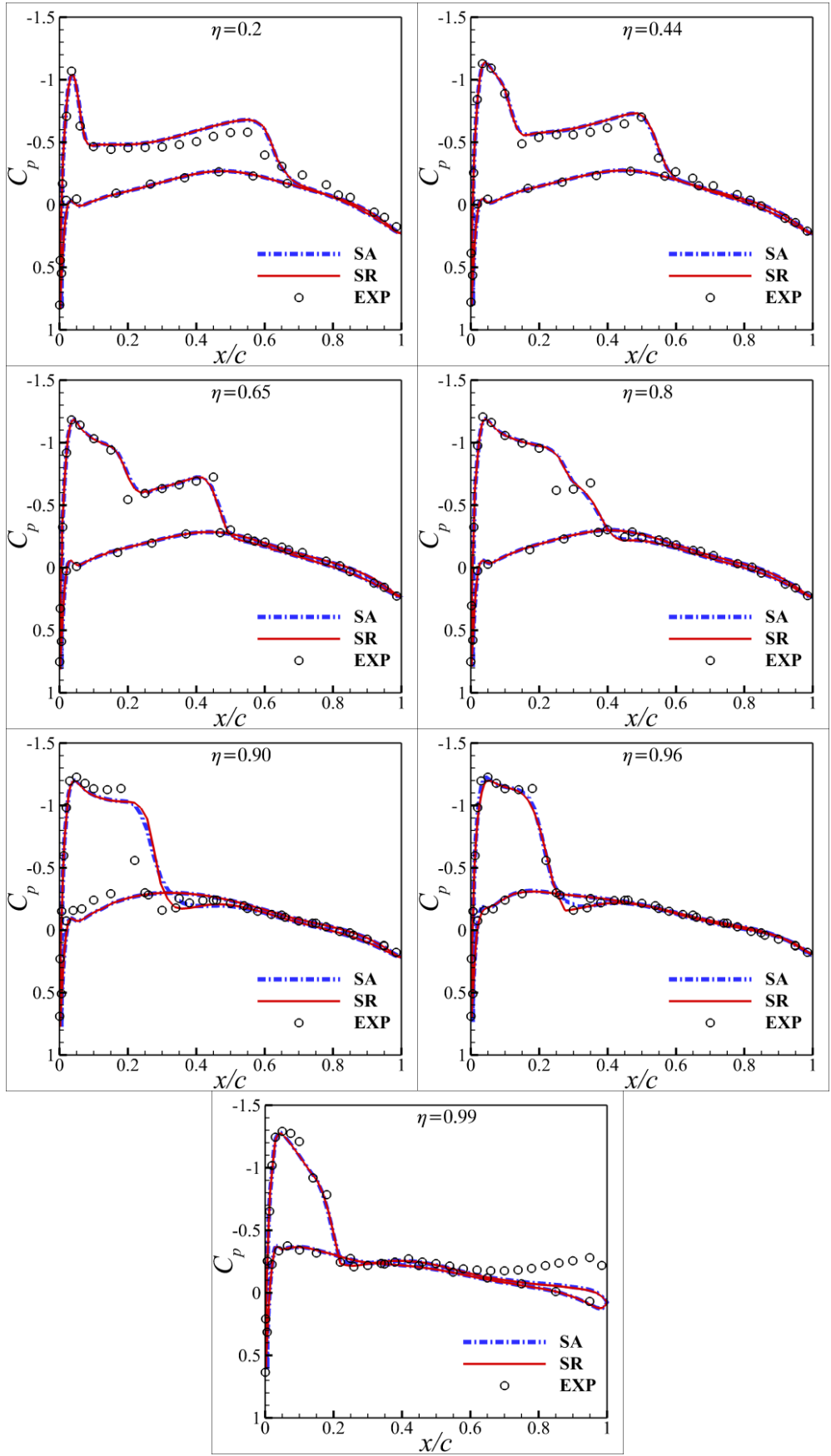

Figure 12 Comparison of wall pressure coefficient calculation results for the ONERA-M6 wing at Ma=0.84, AoA=3.06º, Re=11.72×10$^6$

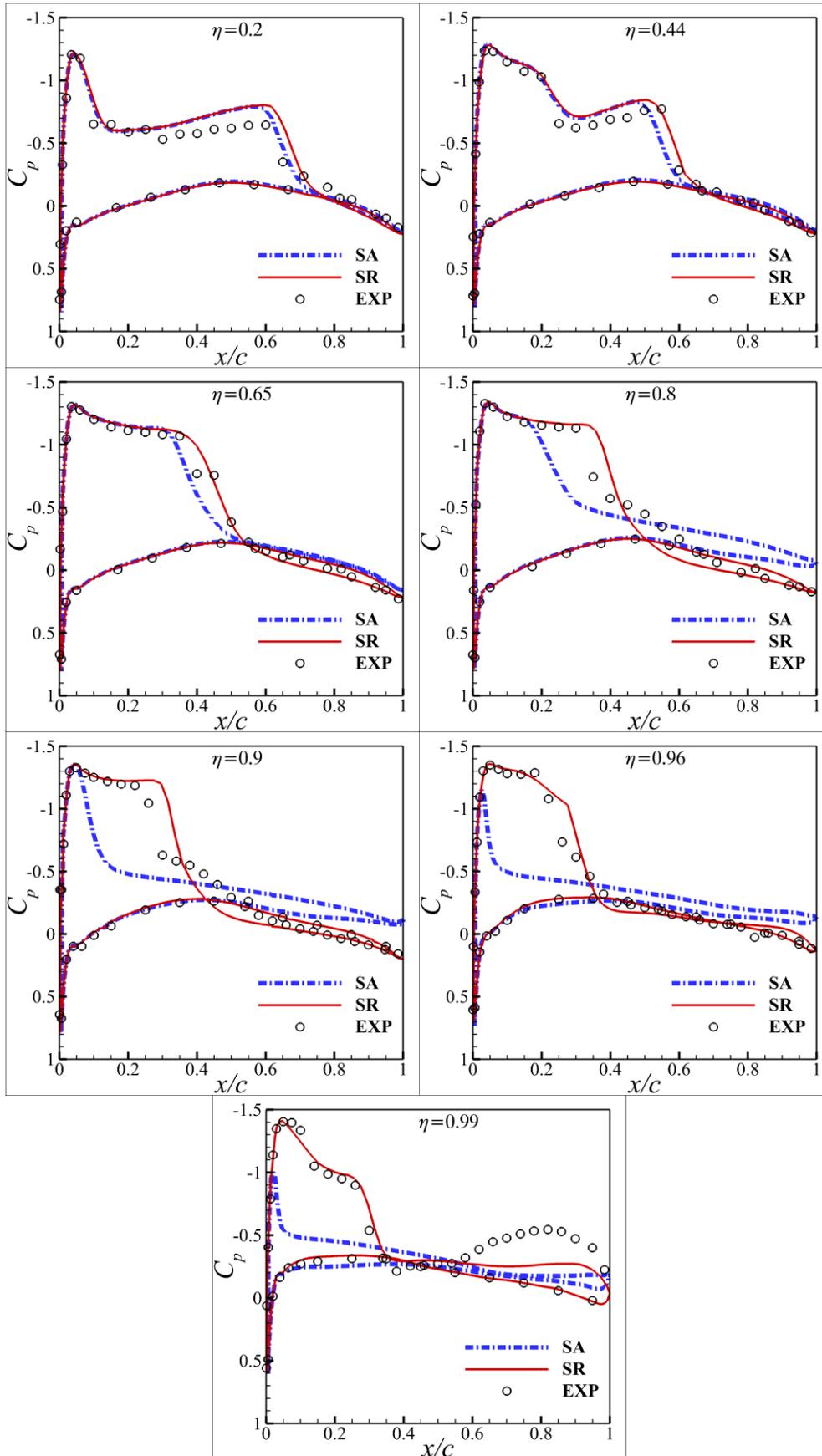

Figure 13 Comparison of wall pressure coefficient calculation results for the ONERA-M6 wing at Ma=0.84, AoA=5.06º, Re=11.72×10$^6$

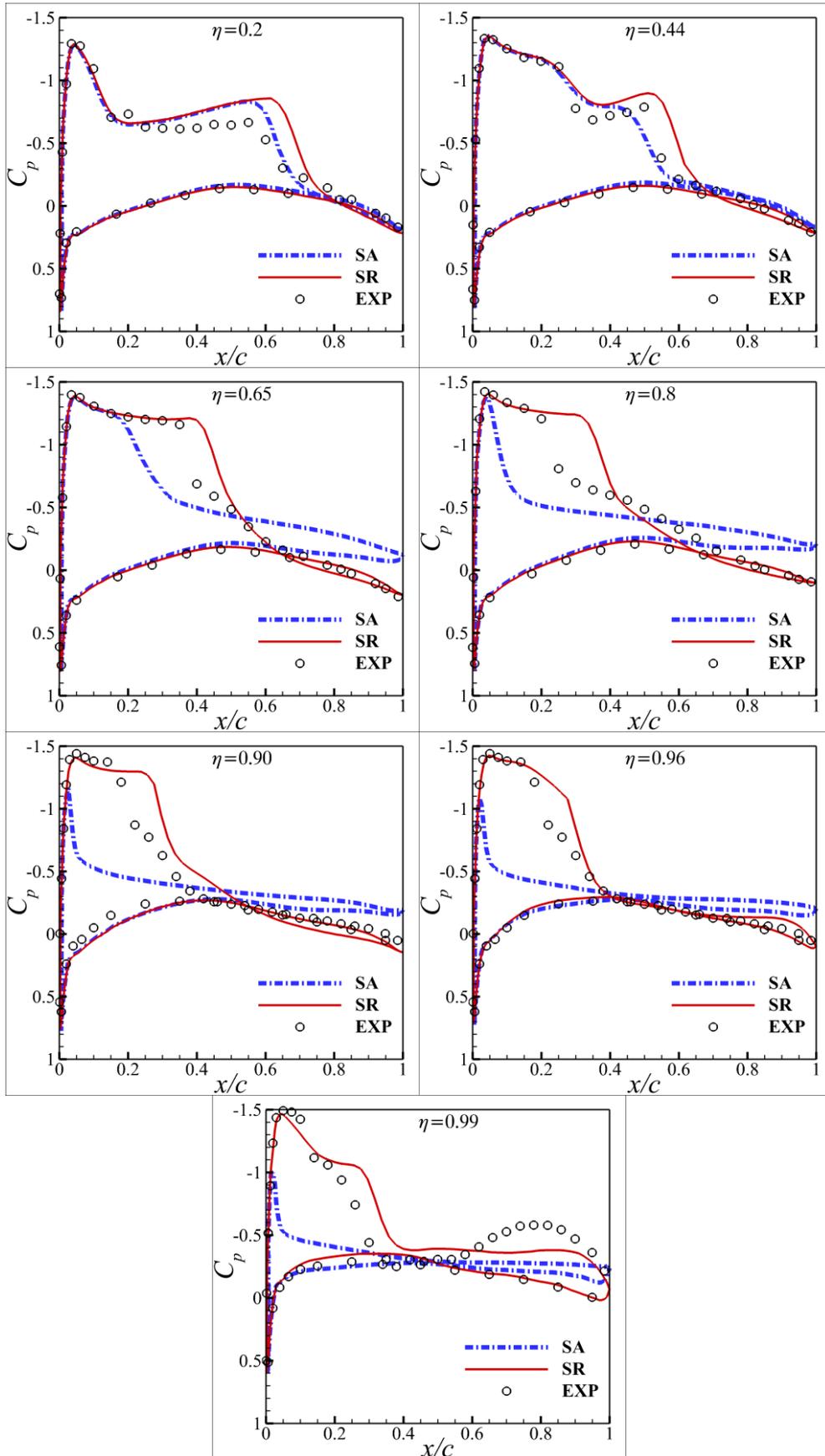

Figure 14 Comparison of wall pressure coefficient calculation results for the ONERA-M6 wing at Ma=0.84, AoA=6.06º, Re=11.72×10$^6$

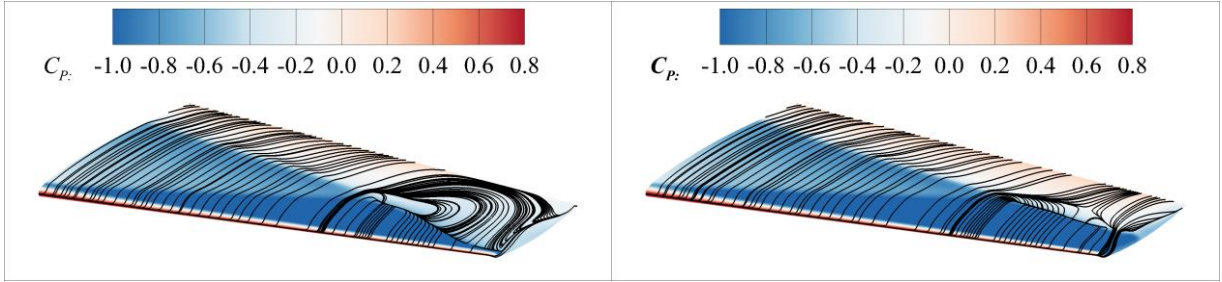

Figure 15 Streamline results on the M6 wing surface at AoA=5.06º. Left: SA model results; Right : corrected model results.

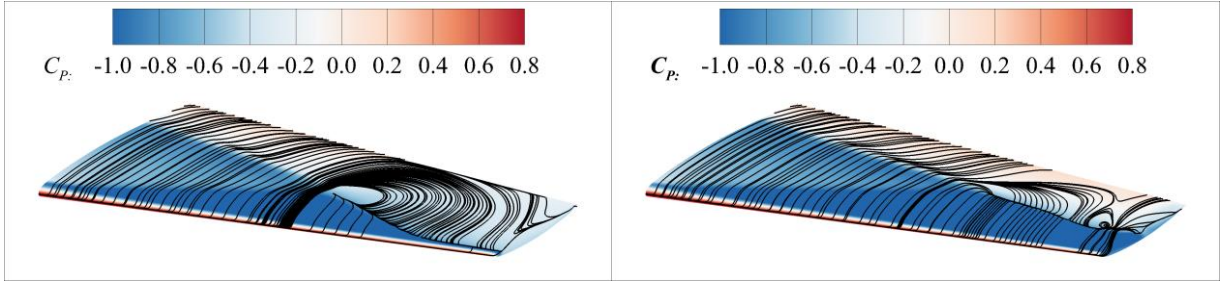

Figure 16 Streamline results on the M6 wing surface at AoA=6.06º. Left: SA model results; Right : corrected model results

Since the original experimental data in reference[49] does not provide specific concentrated force coefficient results, we instead compared the relative errors of the surface pressure coefficients at seven different sections under various inflow conditions. The error is calculated as follows:

$$Error = \frac{\|\hat{y} - y_{exp}\|_2}{\|y_{exp}\|_2}$$

Where $\hat{y}$ represents the computed value and $y_{exp}$ represents the experimental value. The error calculation results are shown in Figure 17. Compared to the original SA model, the corrected model reduces the average relative error of the computed results by 40.89% across three different inflow angles, with an average improvement in computational accuracy of 1.69 times. From another perspective, the corrected model maintains the same error level across all three angles of attack, with solution accuracy for separated flows being comparable to that for attached flows. In contrast, the classic SA model experiences a significant increase in computational error—about three times higher—in separated flow regions compared to attached flow regions. These results collectively demonstrate that the corrected model has strong generalization capability for three-dimensional flows and effective correction capability for shock-induced complex three-dimensional separated flows.

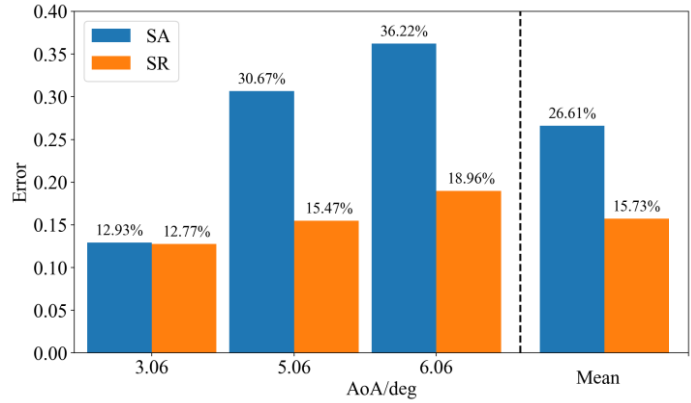

Figure 17 Comparison of relative errors in wall pressure coefficient calculations for different models under varying incoming flow angles of attack.

## 4.3 DU91-W2-250 Airfoil

This section presents the coupled computational results of the corrected model for the DU91-W2-250 airfoil. Since the training data for the corrected model includes assimilated data for the DU91-W2-250 airfoil at Ma=0.15, AoA=15º and Re=$3.0\times10^6$, the main focus here is to test the generalization ability of the model under different inflow conditions.

Figure 18 shows a comparison of the lift and drag coefficients for the training inflow conditions, where the green circles indicate the training data set, which corresponds to the data used for assimilation. Figure 19 presents the relative error statistics of the lift coefficient in the stall angle of attack range for the corresponding conditions. From the comparison in the figures, it is clear that, under the same Reynolds number and Mach number, the current method achieves high-precision generalization of the entire lift coefficient curve by learning from a single flow state data set. Compared to the traditional SA model, the corrected model improves the average computational accuracy by approximately 5.3 times, demonstrating the effectiveness and generalization capability of the modeling approach. For low-angle-of-attack separated flow, the results of the corrected model remain consistent with those of the SA model, indicating that the current model can selectively activate based on the flow state, allowing for high-precision simulation of both attached and separated flows. Figure 20 and Figure 21 compare the computational results at a higher Reynolds number, further supporting the conclusion that the corrected model maintains comparable accuracy to the SA model for attached flow, while significantly improving the computational accuracy for separated flows.

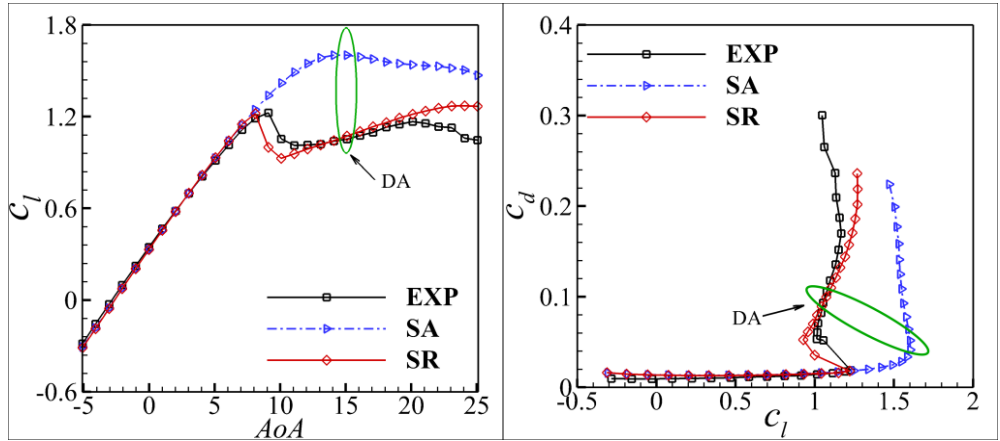

Figure 18 Comparison of lift and drag coefficient results for the DU91-W2-250 airfoil at Ma=0.15, Re=3.0×10$^6$.

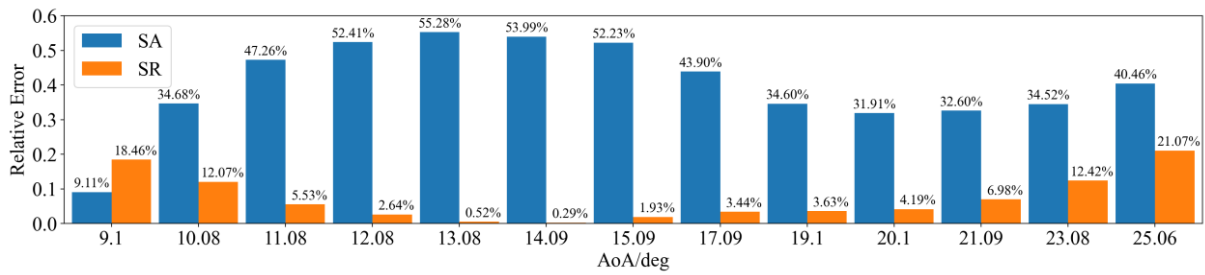

Figure 19 Comparison of relative errors in lift coefficients within the stall region for the DU91-W2-250 airfoil atMa=0.15, Re=3.0×10$^6$.

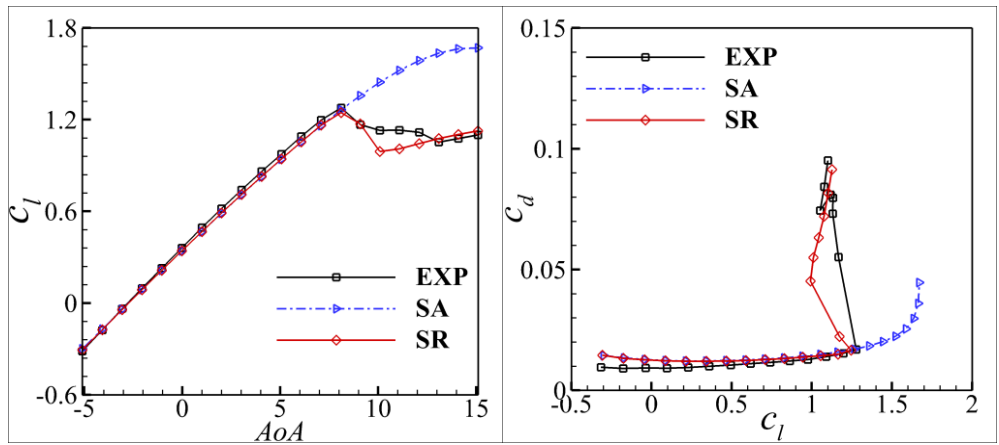

Figure 20 Comparison of lift and drag coefficient results for the DU91-W2-250 airfoil at Ma=0.15, Re=5.0×10$^6$.

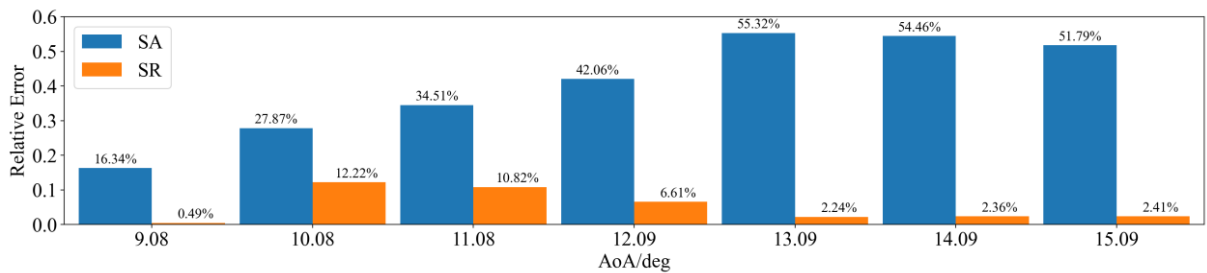

Figure 21 Comparison of relative errors in lift coefficients within the stall region for the DU91-W2-250 airfoil at Ma=0.15, Re=5.0×10$^6$.

In Figure 22, the left two columns of the dashed line compare the relative error of the lift coefficient at different Reynolds numbers, while the right side of the dashed line compares the average relative error of the lift coefficient for the three computed inflow states. In all computed states, the corrected model shows a significant improvement in computational accuracy. Overall, compared to the original SA model, the corrected model reduces the relative error by 84.00%, improving the computational accuracy by a factor of 6.25. Figure 23 presents a comparison of the surface pressure coefficient distribution for different angles of attack under the inflow conditions of Ma=0.2 and Re=3.0×10$^6$. In the entire stall angle region, the results of the corrected model are much closer to the experimental values, with smaller deviations and a separation point that is more consistent with experimental results. This is also corroborated by the results for the force coefficients, demonstrating the effectiveness of the current corrected model for simulating large-scale separated flow around airfoils.

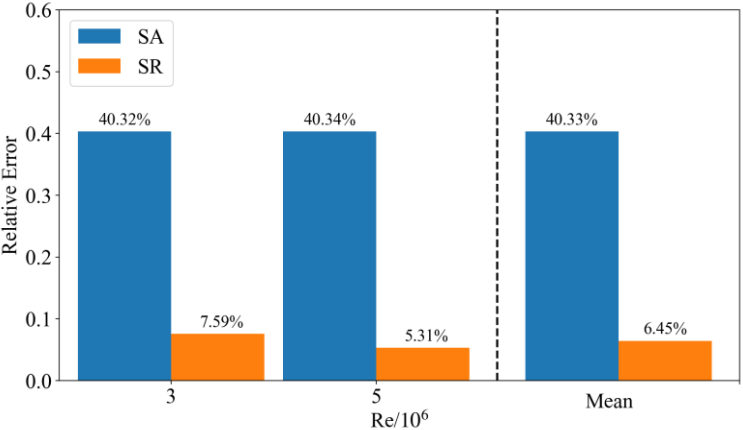

Figure 22 Comparison of average relative errors in lift coefficients within the stall region for the DU91-W2-250 airfoil under different Reynolds numbers.

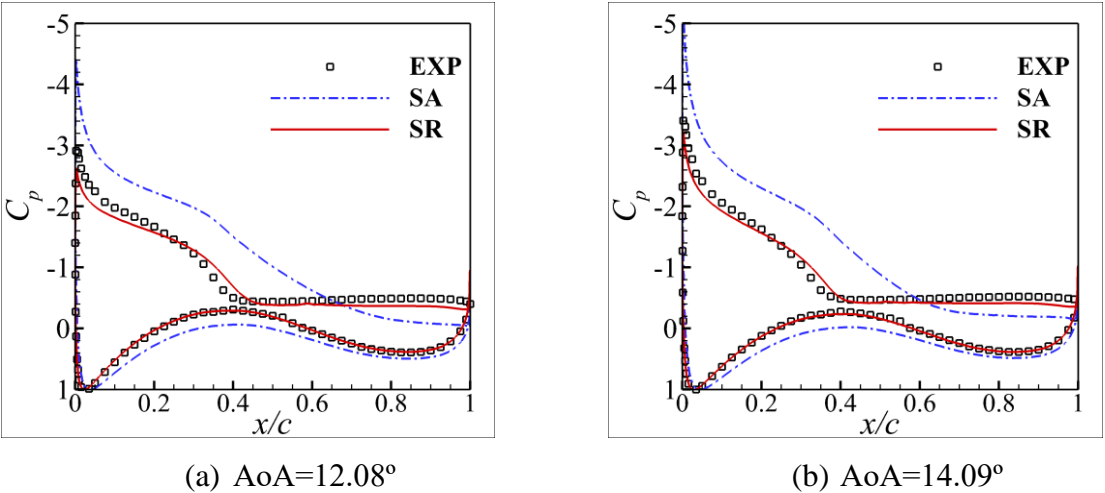

(a) AoA=12.08º       (b) AoA=14.09º

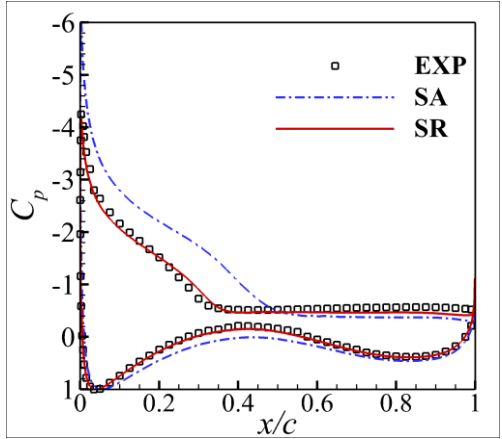
(c) AoA=17.09º

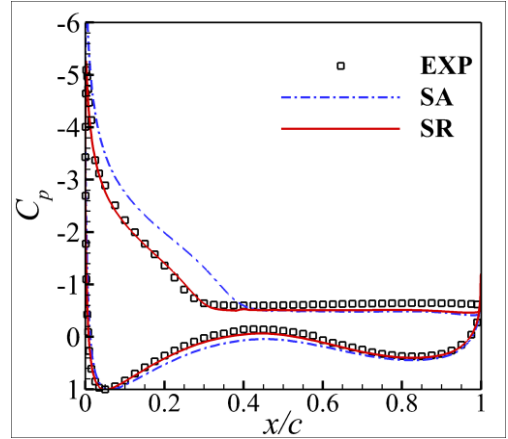
(d) AoA=20.1º

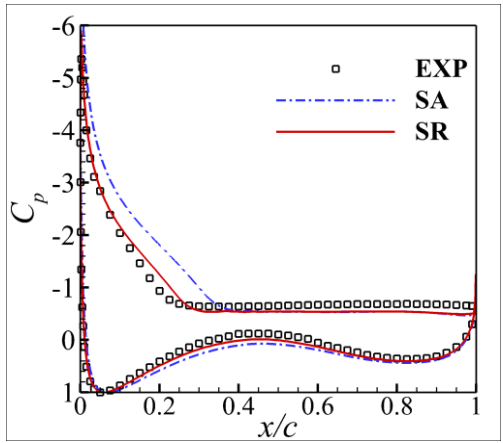
(e) AoA=22.09º

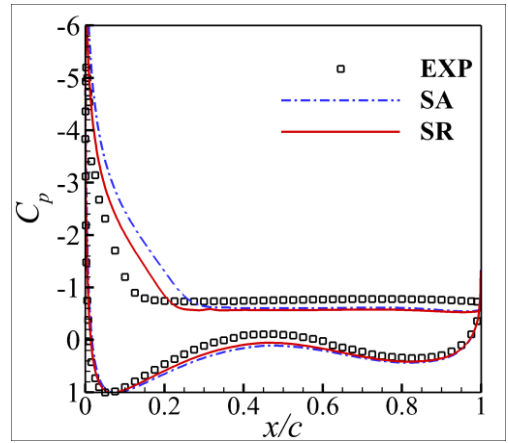
(f) AoA=25.06º

Figure 23 Comparison of wall pressure coefficient results for different angles of attack on the DU91-W2-250 airfoil at Ma=0.2, Re=3.0×10$^6$

## 4.4 S809 Airfoil

This section primarily presents the computational results of the symbolic regression-based corrected model applied to the S809 airfoil[50]. The lift coefficient curves, lift-to-drag polar curves, and the relative errors in the lift coefficient at the airfoil's stall angles for five typical operating conditions are compared, as shown in Figure 24 to Figure 33. In all test cases, compared to the SA model, the corrected model significantly improves computational accuracy in the high angle-of-attack separation region, while maintaining consistency with the SA model for attached flow. Figure 34 shows a comparison of the average relative error in the lift coefficient for the stall region under different Reynolds numbers. For each Reynolds number condition, the corrected model's lift coefficient average relative error is reduced by more than 55% compared to the original SA model. The last two columns on the far right of Figure 34 compare the average relative error of all the computed results across the test cases. Compared to the original SA model, the corrected model reduces the relative error by 66.79%, improving computational accuracy by a factor of 3.01. This demonstrates the generalization capability of the corrected model under varying shapes and operating conditions. The surface pressure coefficient distributions for different inflow angles at Re=2.0×10⁶ are shown in Figure 35 From the comparison in the figure, it is evident that for angles of attack greater than 10.2º , the corrected model shows significant improvement over the original SA model, and the surface pressure coefficient results are also closer to the experimental measurements. For the attached flow state at AoA=8.2º , both models yield results that are similar to the experimental data.

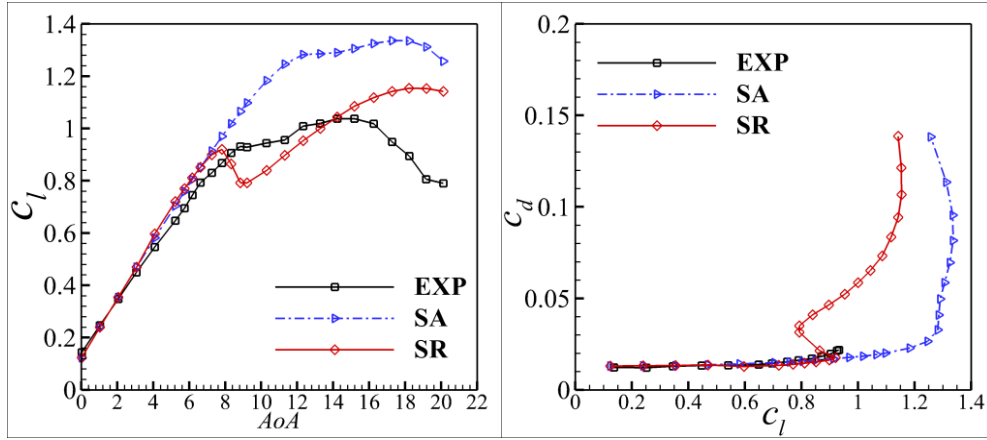

Figure 24 Comparison of lift and drag coefficient results for the S809 airfoil at Ma=0.15, Re=1.0×10⁶

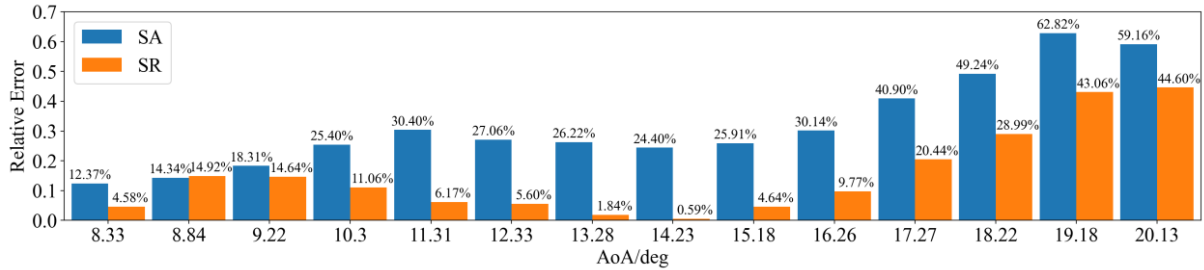

Figure 25 Comparison of relative errors in lift coefficients within the stall region for the S809 airfoil atMa=0.15 Re=1.0×10⁶

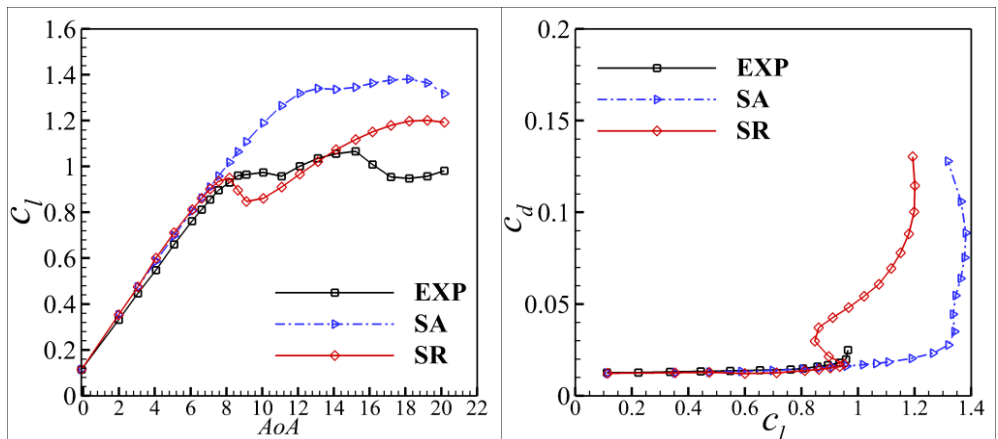

Figure 26 Comparison of lift and drag coefficient results for the S809 airfoil at Ma=0.15, Re=1.5×10⁶

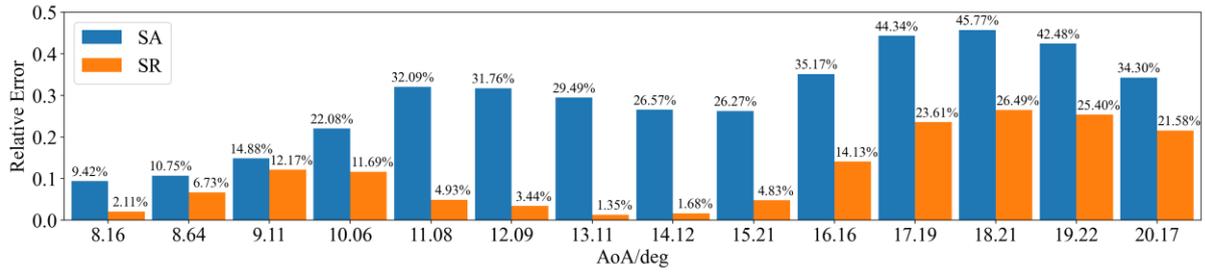

Figure 27 Comparison of relative errors in lift coefficients within the stall region for the S809 airfoil at Ma=0.15, Re=1.5×10$^6$

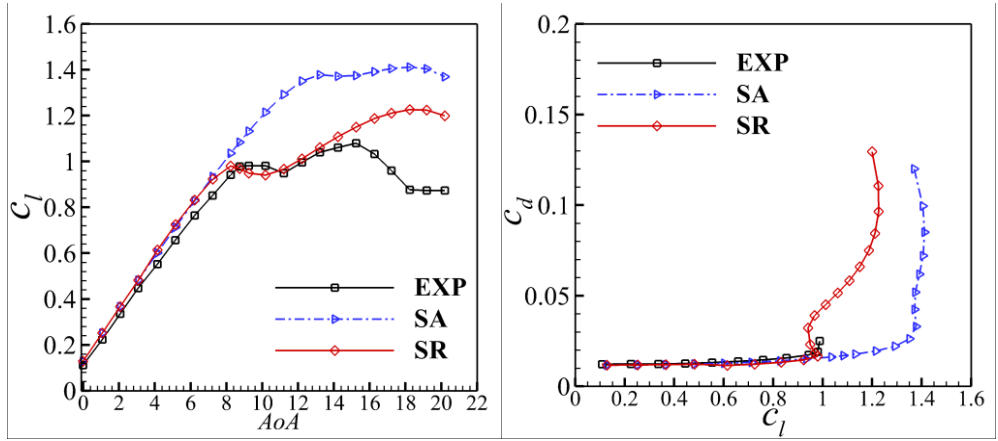

Figure 28 Comparison of lift and drag coefficient results for the S809 airfoil at Ma=0.15, Re=2.0×10$^6$

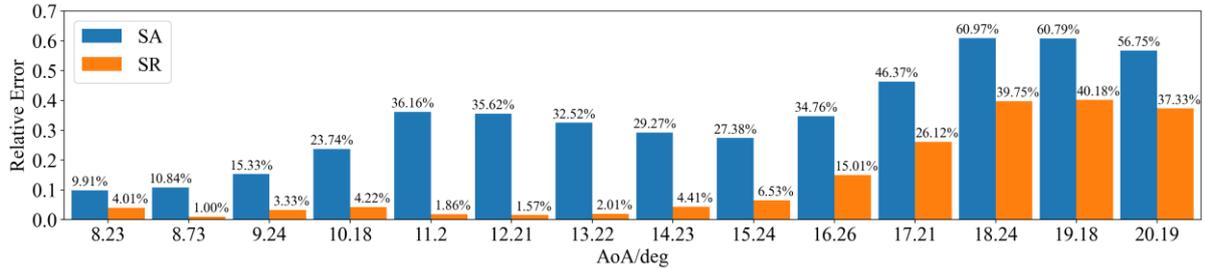

Figure 29 Comparison of relative errors in lift coefficients within the stall region for the S809 airfoil at Ma=0.15, Re=2.0×10$^6$

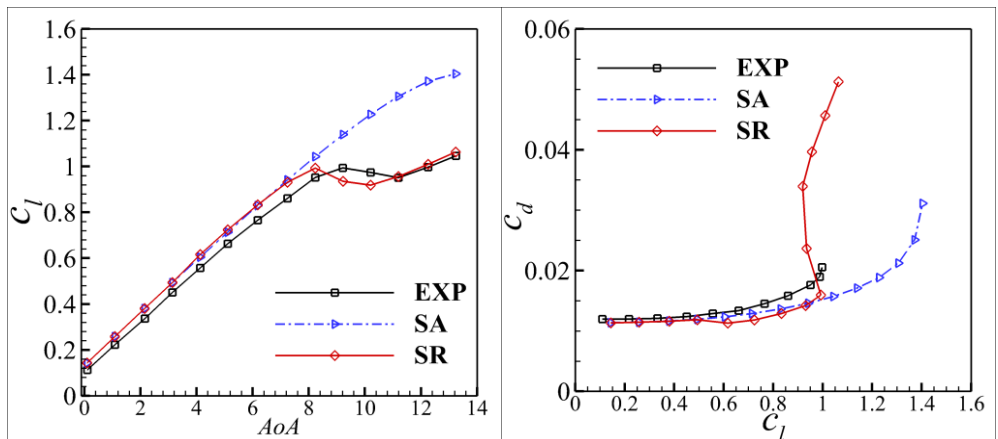

Figure 30 Comparison of relative errors in lift coefficients within the stall region for the S809 airfoil at Ma=0.15, Re=2.5×10$^6$

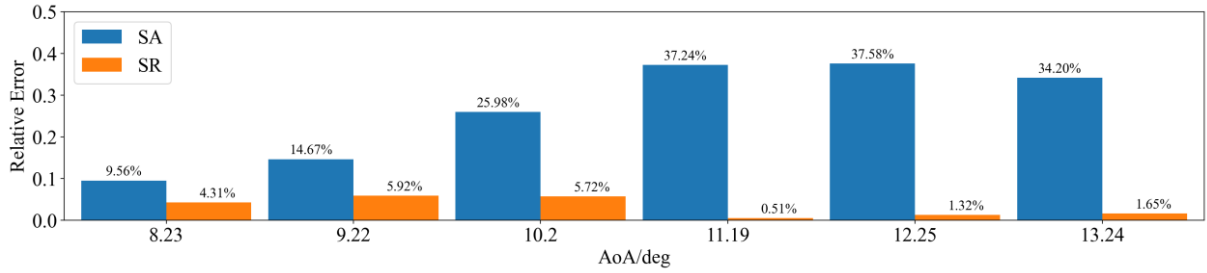

Figure 31 Comparison of relative errors in lift coefficients within the stall region for the S809 airfoil at Ma=0.15, Re=2.5×10$^6$

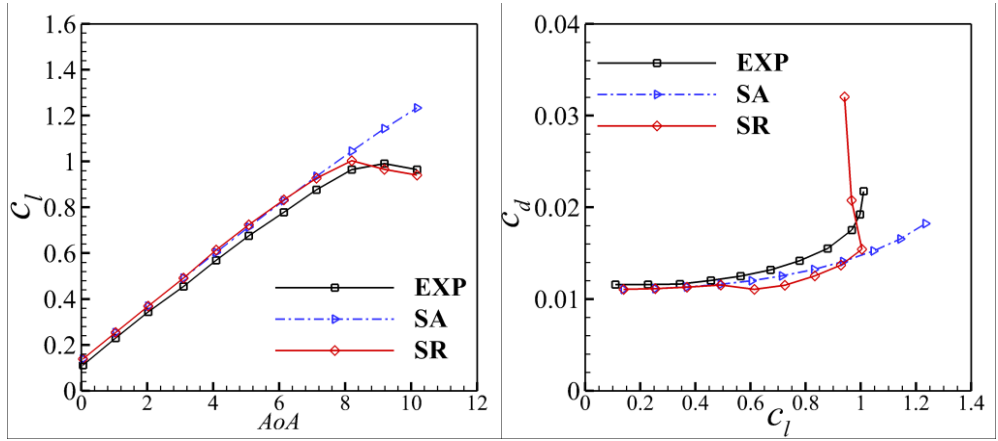

Figure 32 Comparison of relative errors in lift coefficients within the stall region for the S809 airfoil at Ma=0.15, Re=3.0×10$^6$

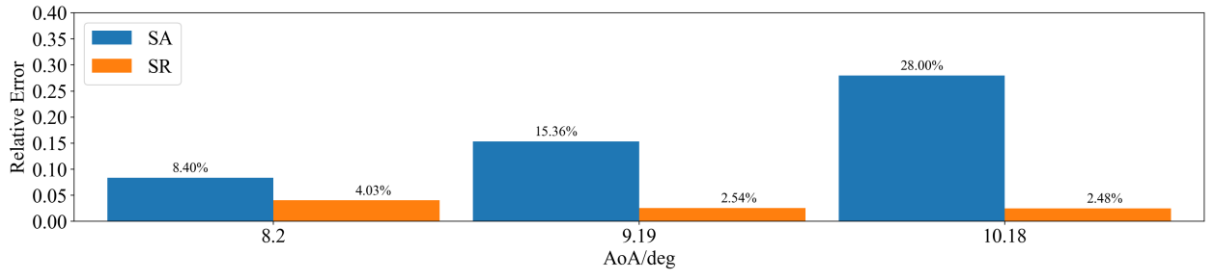

Figure 33 Comparison of relative errors in lift coefficients within the stall region for the S809 airfoil at Ma=0.15, Re=3.0×10$^6$

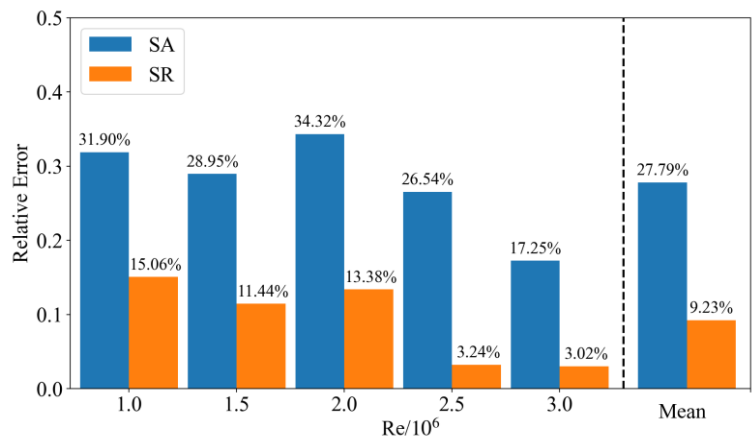

Figure 34 Comparison of average relative errors in lift coefficients within the stall region for the S809 airfoil under different Reynolds numbers.

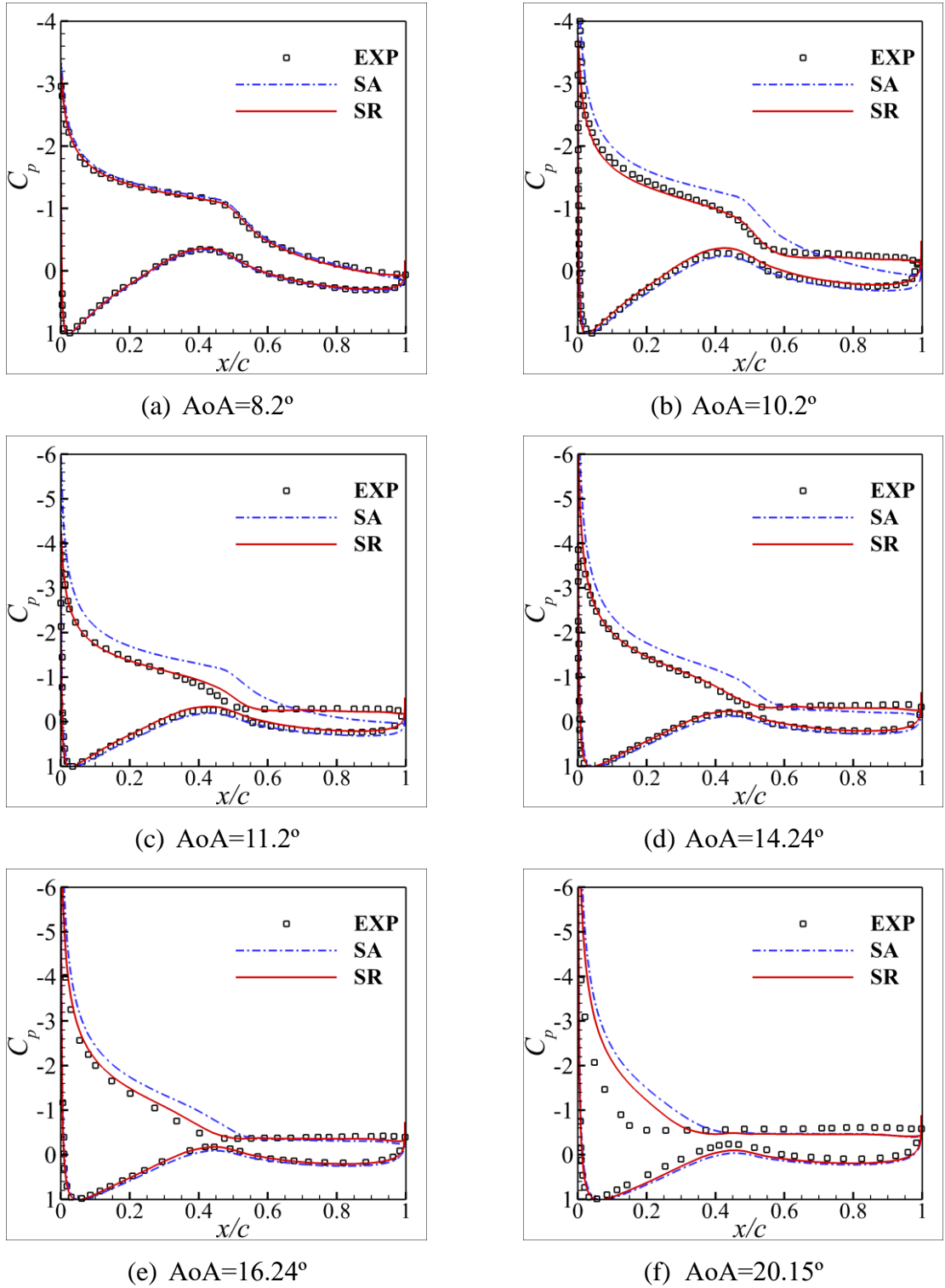

(a) AoA=8.2º   (b) AoA=10.2º
(c) AoA=11.2º   (d) AoA=14.24º
(e) AoA=16.24º   (f) AoA=20.15º

Figure 35 Comparison of wall pressure coefficient results for different angles of attack on the S809 airfoil at Ma=0.15, Re=2.0×10$^6$

## 4.5 S814 Airfoil

This section presents the computational results of the symbolic regression-based corrected model applied to the S814 airfoil[51] . The focus is on comparing the lift coefficient curves, lift-to-drag polar curves, and the relative errors of the lift coefficient in the stall region for four typical operating conditions, as shown in Figure 36 to Figure 45. The results demonstrate that, compared to the original SA model, the corrected model significantly improves computational accuracy in the high angle-of-attack separation region, while maintaining consistency with the SA model in the attached flow region.

Figure 46 shows a comparison of the average relative errors in the lift coefficient for the stall region under different Reynolds numbers. It can be seen that under all Reynolds number conditions, the corrected model's average relative error in the lift coefficient is reduced by more than 55% compared to the original SA model. Further statistical analysis of the overall average relative errors for all Reynolds numbers (the last two columns in Figure 46) reveals that the corrected model's relative error is reduced by 65.84%, improving computational accuracy by a factor of 2.92 compared to the original SA model. These results demonstrate that the symbolic regression-based corrected model exhibits excellent generalization capabilities under varying shapes and operating conditions, significantly improving the simulation accuracy of complex high-angle flows. In addition, a comparison of the surface pressure coefficient distributions for different inflow angles at Re=$1.5×10^6$ is shown in Figure 47. The corrected model's results are more consistent with the experimental data in predicting the lift coefficient after the stall angle, and the surface pressure coefficient distributions at high angles of attack also show better agreement with the experimental measurements.

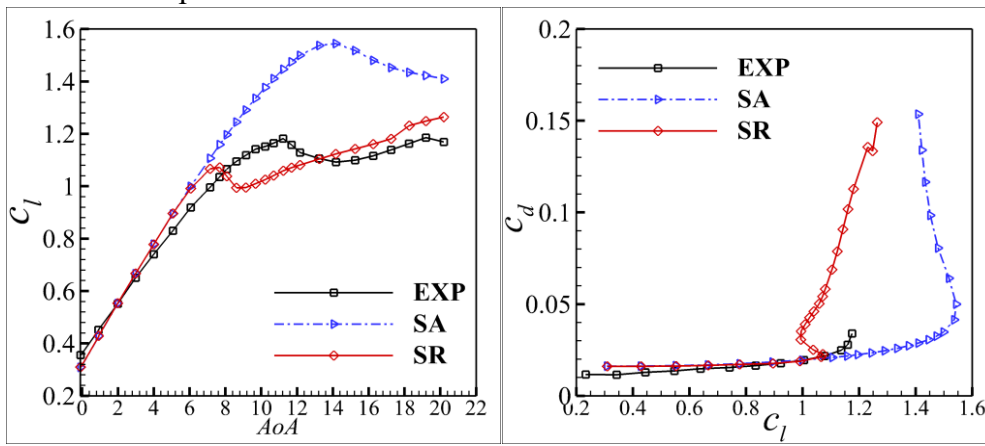

Figure 36 Comparison of lift coefficient and drag coefficient results for the S814 airfoil under flow conditions of  Ma=0.15, Re=$0.7×10^6$

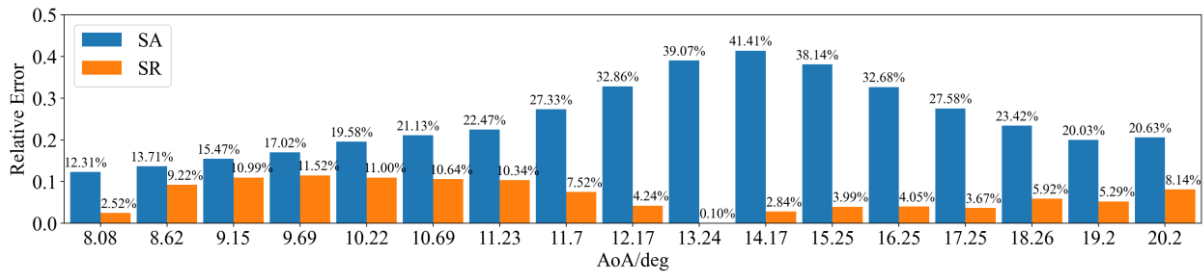

Figure 37 Comparison of relative errors in lift coefficient within the stall region for the S814 airfoil under flow conditions of Ma=0.15, Re=$0.7×10^6$

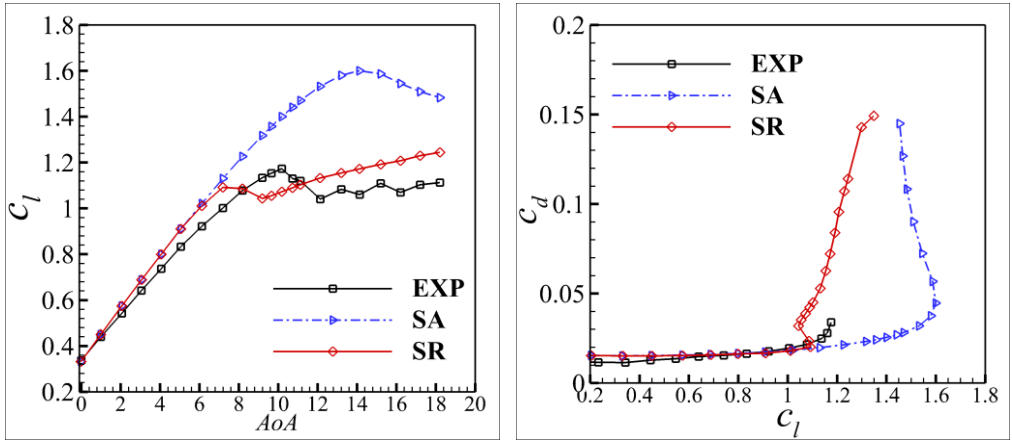

Figure 38 Comparison of lift coefficient and drag coefficient results for the S814 airfoil under flow conditions of Ma=0.15, Re=1.0×10$^6$

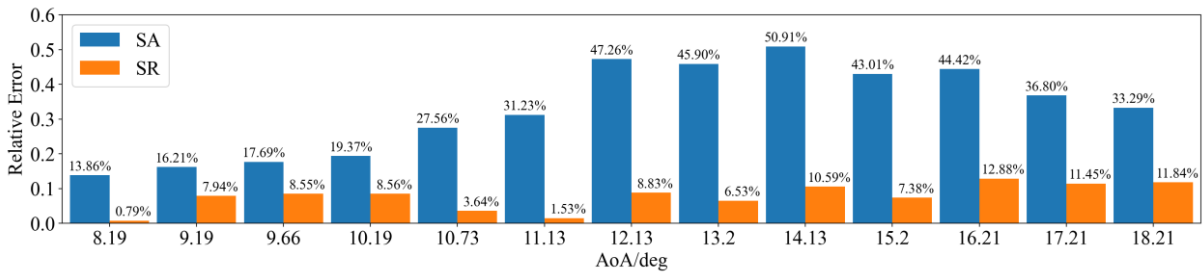

Figure 39 Comparison of relative errors in lift coefficient within the stall region for the S814 airfoil under flow conditions of Ma=0.15, Re=1.0×10$^6$

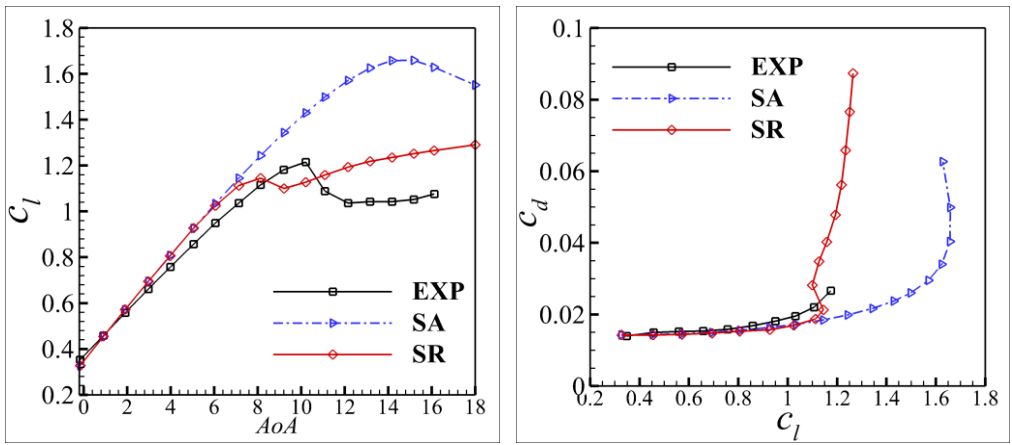

Figure 40 Comparison of lift coefficient and drag coefficient results for the S814 airfoil under flow conditions of Ma=0.15, Re=1.5×10$^6$

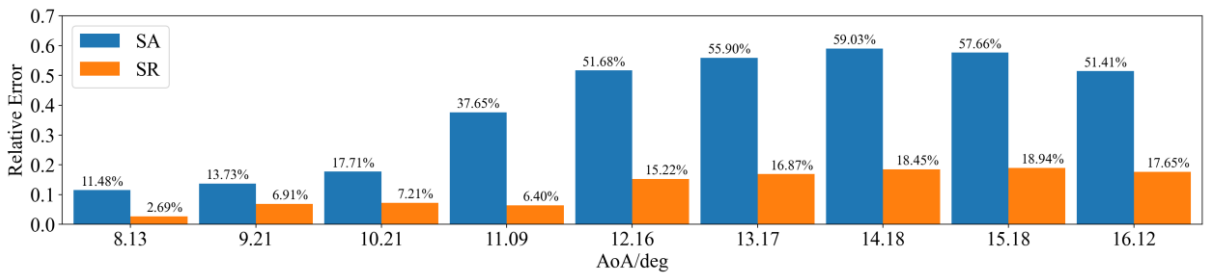

Figure 41 Comparison of relative errors in lift coefficient within the stall region for the S814 airfoil under flow conditions of Ma=0.15, Re=1.5×10$^6$

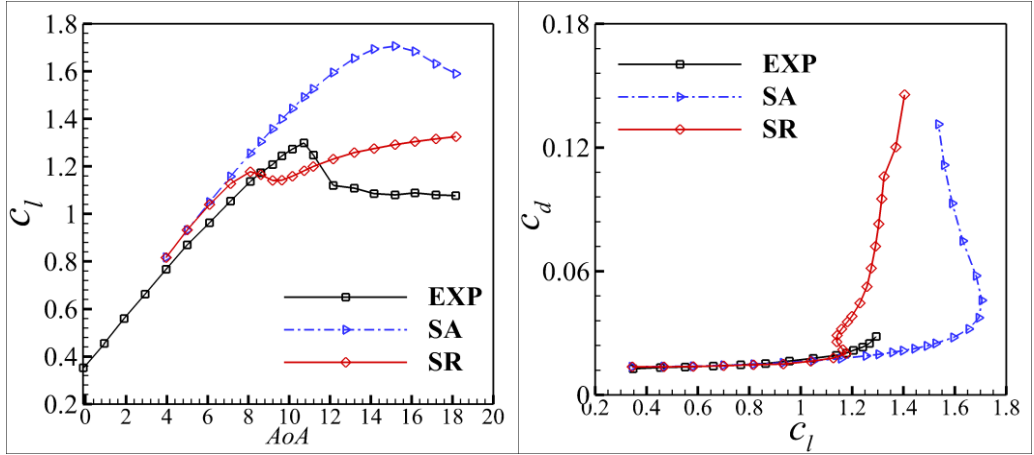

Figure 42 Comparison of lift coefficient and drag coefficient results for the S814 airfoil under flow conditions of Ma=0.15, Re=2.0×10$^6$

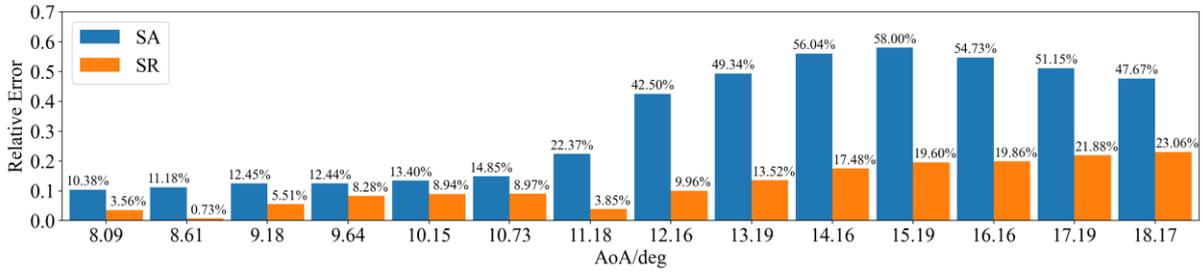

Figure 43 Comparison of relative errors in lift coefficient within the stall region for the S814 airfoil under flow conditions of Ma=0.15, Re=2.0×10$^6$

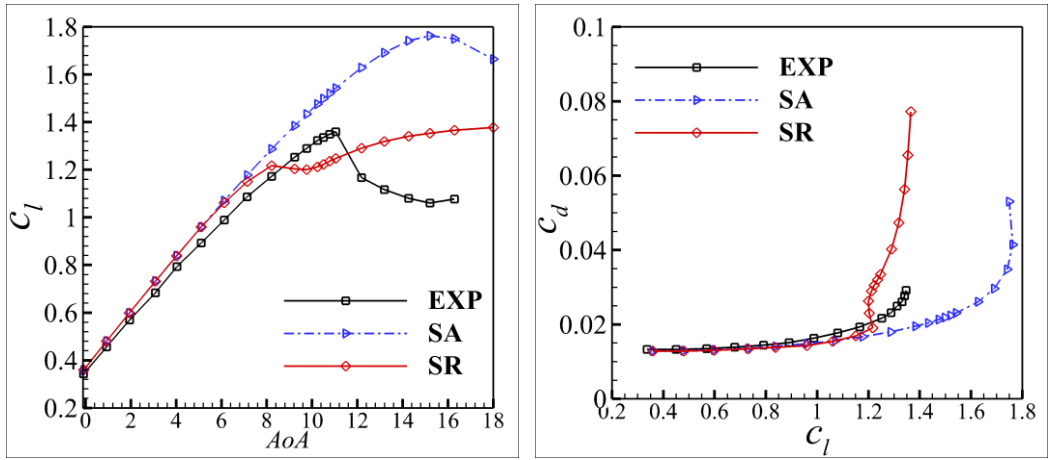

Figure 44 Comparison of lift coefficient and drag coefficient results for the S814 airfoil under flow conditions of Ma=0.15, Re=3.0×10$^6$

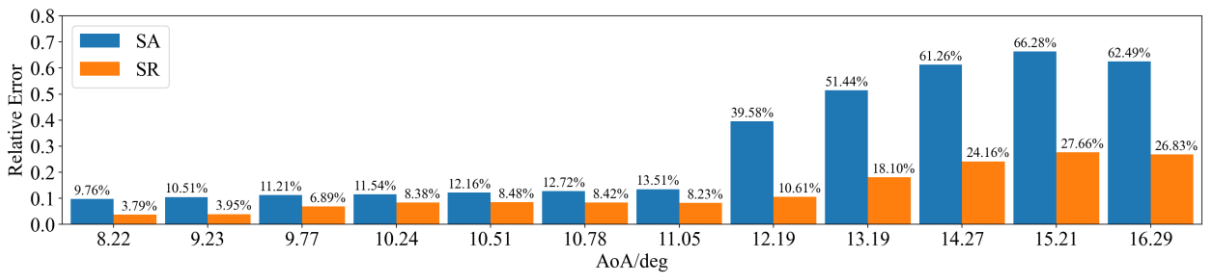

Figure 45 Comparison of relative errors in lift coefficient within the stall region for the S814 airfoil under flow conditions of Ma=0.15, Re=3.0×10$^6$
30

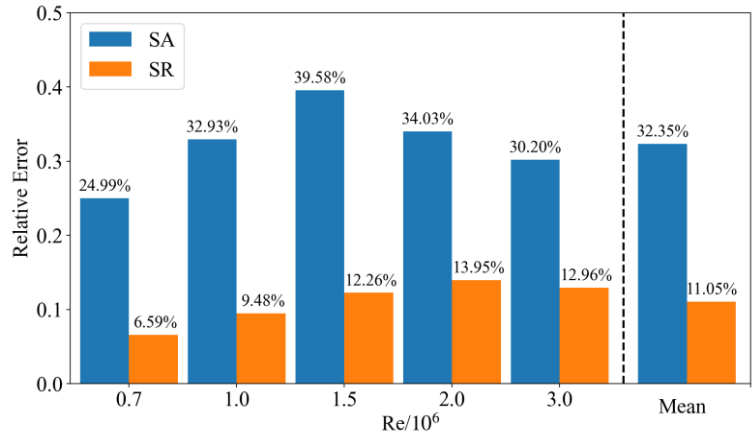

Figure 46 Comparison of average relative errors in lift coefficient within the stall region for the S814 airfoil at different Reynolds numbers.

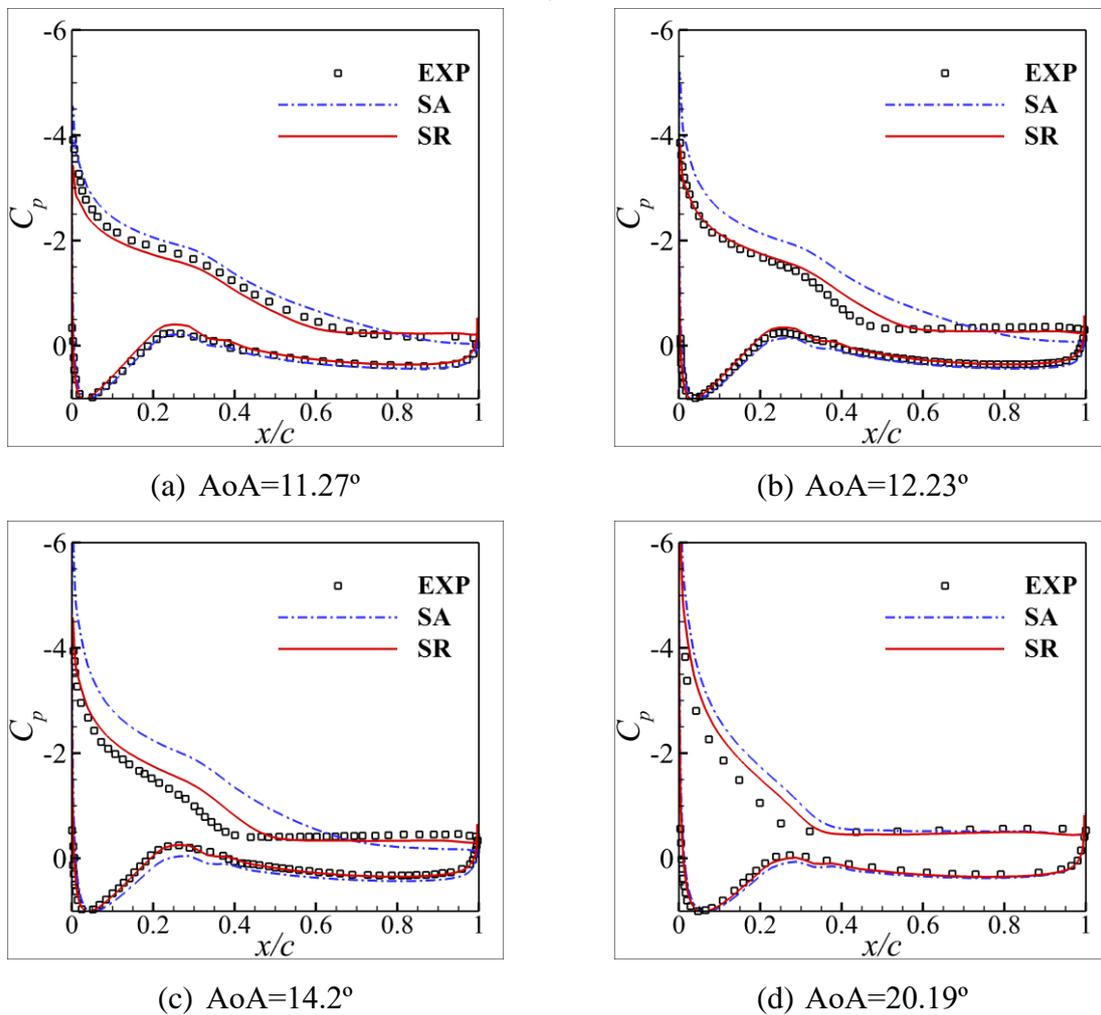

(a) AoA=11.27º

(b) AoA=12.23º

(c) AoA=14.2º

(d) AoA=20.19º

Figure 47 Comparison of wall pressure coefficients corresponding to different angles of attack for the S814 airfoil under flow conditions of Ma=0.15, Re=1.5×10$^6$

## 4.6 S805 Airfoil

This section presents the computational results of the symbolic regression-based corrected model applied to the S805 airfoil[52] . Firstly, a detailed comparison is made between the lift coefficient curves, lift-to-drag polar curves, and the relative errors of the lift coefficient at the stall angle for three typical operating conditions, as shown in Figure 48 to Figure 53 . The comparison results reveal that the corrected model provides a much better match with experimental data for the lift coefficient curves and lift-to-drag polar curves under all three typical conditions compared to the original SA model. In particular, in the high-angle-of-attack separation region, the corrected model more accurately captures the variations in the lift coefficient and lift-to-drag polar curves, showing a trend that is more consistent with the experimental measurements and significantly reducing prediction errors. Additionally, for the lift coefficient near the stall angle, the corrected model better reflects the nonlinear variations observed in the experimental data, further validating its predictive capability under complex flow conditions.

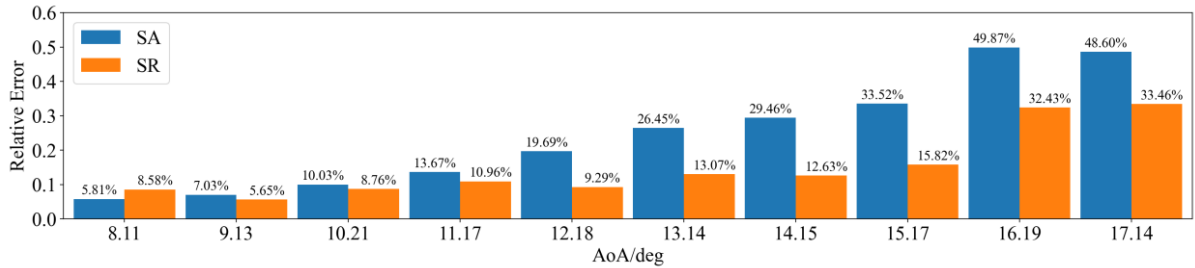

Figure 48 Comparison of relative errors in lift coefficient within the stall region for the S805 airfoil under flow conditions of Ma=0.15, Re=1.0×10$^6$

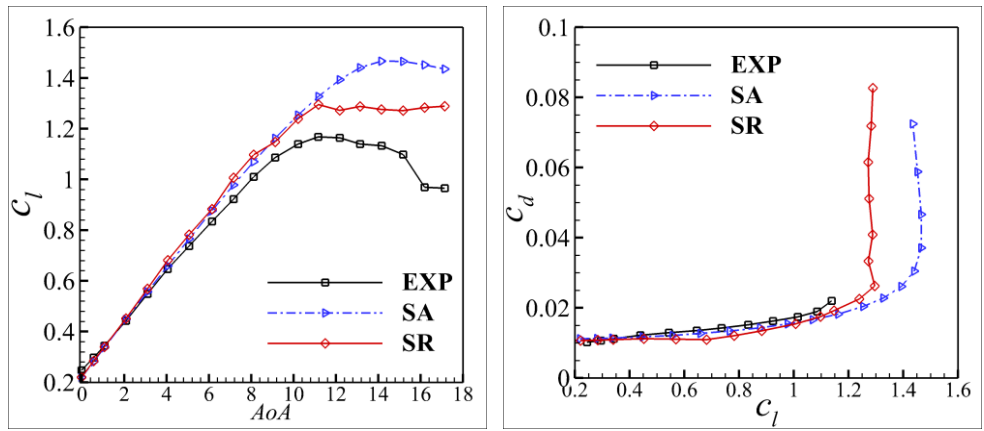

Figure 49 Comparison of lift coefficient and drag coefficient results for the S805 airfoil under flow conditions of Ma=0.15, Re=1.0×10$^6$

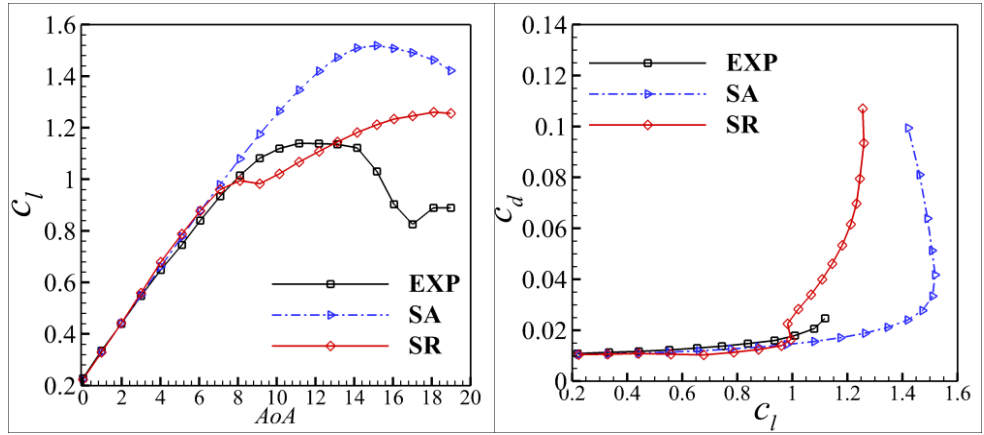

Figure 50 Comparison of lift coefficient and drag coefficient results for the S805 airfoil under flow conditions of Ma=0.15, Re=1.5×10$^6$

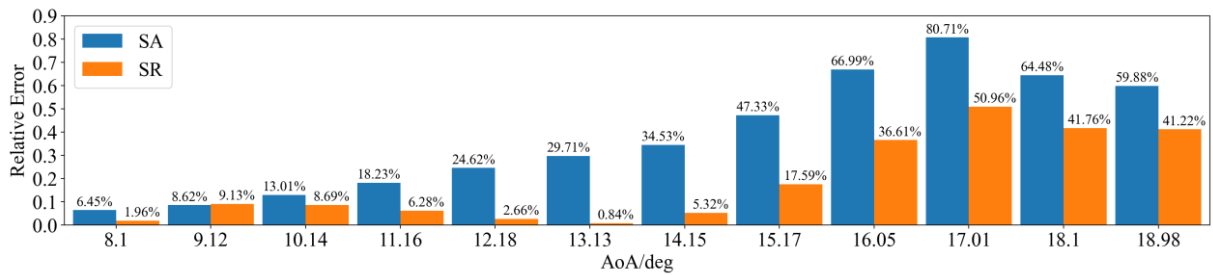

Figure 51 Comparison of relative errors in lift coefficient within the stall region for the S805 airfoil under flow conditions of Ma=0.15, Re=1.5×10$^6$

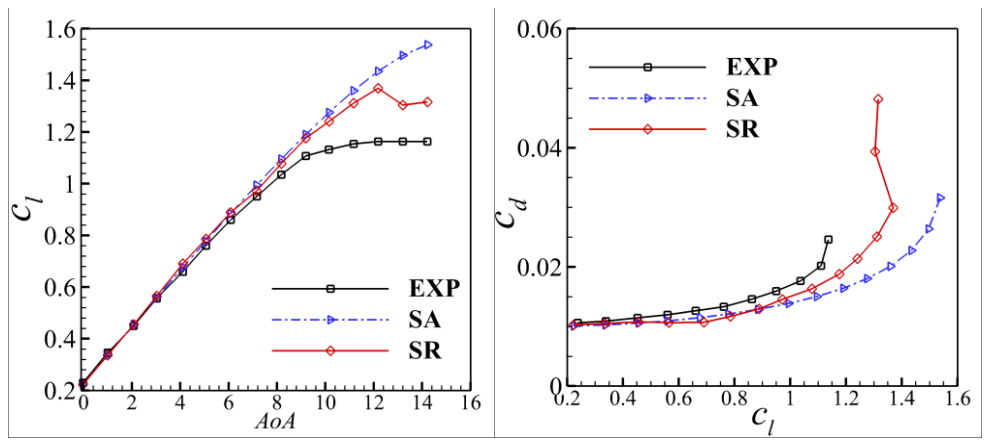

Figure 52 Comparison of lift coefficient and drag coefficient results for the S805 airfoil under flow conditions ofMa=0.15, Re=2.0×10$^6$

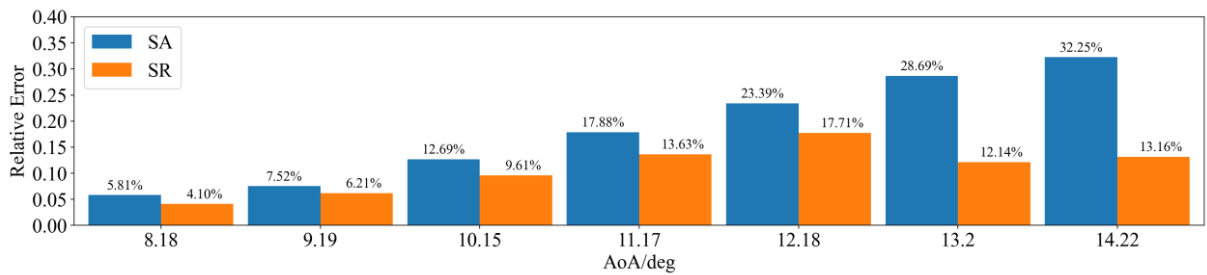

Figure 53 Comparison of relative errors in lift coefficient within the stall region for the S805 airfoil under flow conditions of Ma=0.15, Re=2.0×10$^6$

Further analysis was conducted on the wall pressure coefficient distribution at different inflow angles for the condition of Re=1.0×10$^6$, as shown in Figure 54. The results indicate that the corrected model provides a much closer match with the experimental data for the wall pressure coefficient distribution under high-angle-of-attack conditions, significantly reducing the deviations observed in the original SA model's predictions. In particular, after the stall angle, the corrected model more accurately captures the trend of the pressure gradient changes, especially in the key region of the airfoil's suction side, where it shows better alignment with the pressure distribution. In contrast, the original SA model's predictions under these conditions deviate more significantly from the experimental data and fail to accurately reflect the impact of separated flow on the wall pressure. This improvement in the corrected model further validates its predictive capability and accuracy under high-angle-of-attack separated flow conditions.

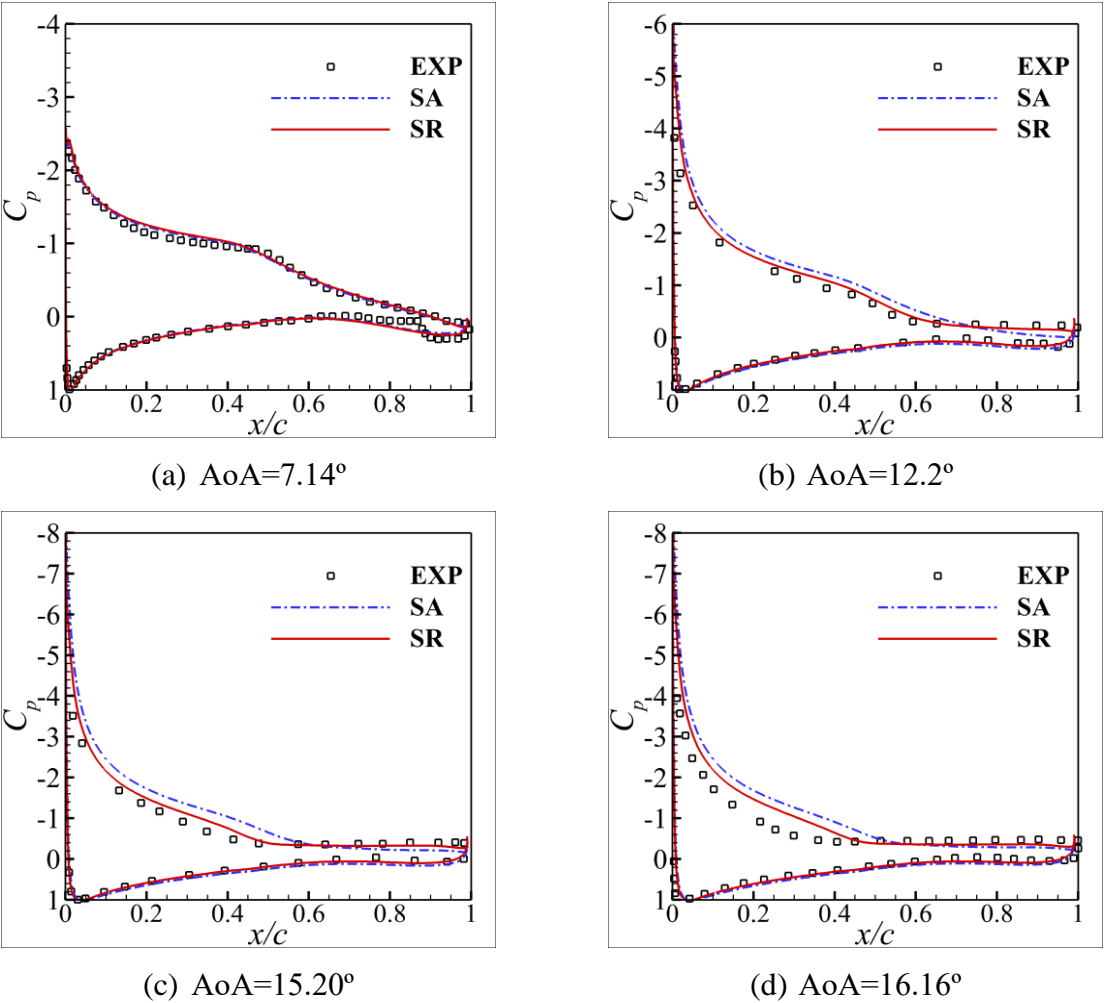

(a) AoA=7.14º  (b) AoA=12.2º

(c) AoA=15.20º  (d) AoA=16.16º

Figure 54 Comparison of wall pressure coefficients corresponding to different angles of attack for the S805 airfoil under flow conditions of Ma=0.15, Re=1.0×10$^6$

In addition, the comparison of the average relative errors in the lift coefficient for the stall region at different Reynolds numbers is shown in Figure 55. It can be seen that the corrected model significantly reduces the average relative errors across all Reynolds number conditions compared to the original SA model. For example, when considering the overall average relative error across all Reynolds number conditions, the corrected model reduces the error by 44.66% compared to the original SA model, with the calculation accuracy improving by a factor of 1.81. This result demonstrates that the corrected model not only significantly improves the prediction accuracy under individual conditions but also exhibits strong generalization capability.

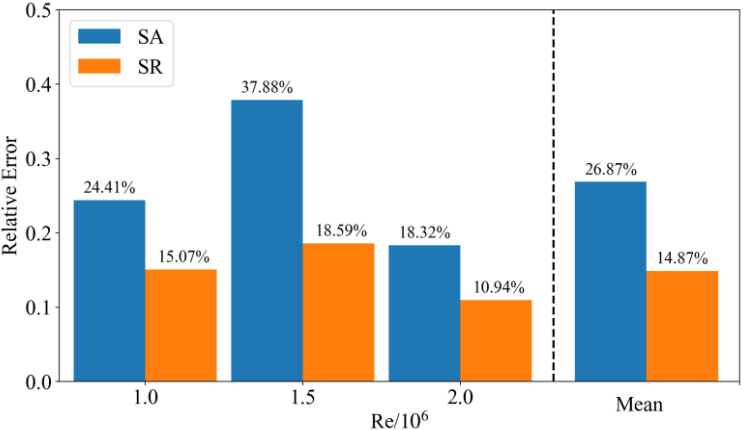

Figure 55 Comparison of average relative errors in lift coefficient within the stall region for the S805 airfoil at different Reynolds numbers.

In summary, the symbol regression-based corrected model demonstrates excellent predictive capability and generalization performance under varying shapes and operating conditions, particularly showing significant potential in simulating high Reynolds number separation flows. This provides an important basis for further improving RANS turbulence models and enhancing their engineering applicability.

## 4.7 Curved Backward-Facing Step and Hump

This section presents the test results of the proposed corrected model on two-dimensional Curved Backward-Facing Step (CBFS) and hump test cases. Both of these cases involve flow separation due to abrupt changes in geometry, with fixed separation points that are typically difficult for RANS methods to accurately predict the reattachment points, which differs from the previously tested airfoil and wing cases. The objective of this section is to verify the generalization ability of the proposed model for such separation cases. Figure 56 and Figure 57 show the shape and mesh diagrams for the CBFS and hump test cases, where the Mach number for the hump case is 0.1 and the Reynolds number is $9.36 \times 10^5$, and the Mach number for the CBFS case is 0.01 with a Reynolds number of $1.37 \times 10^4$。

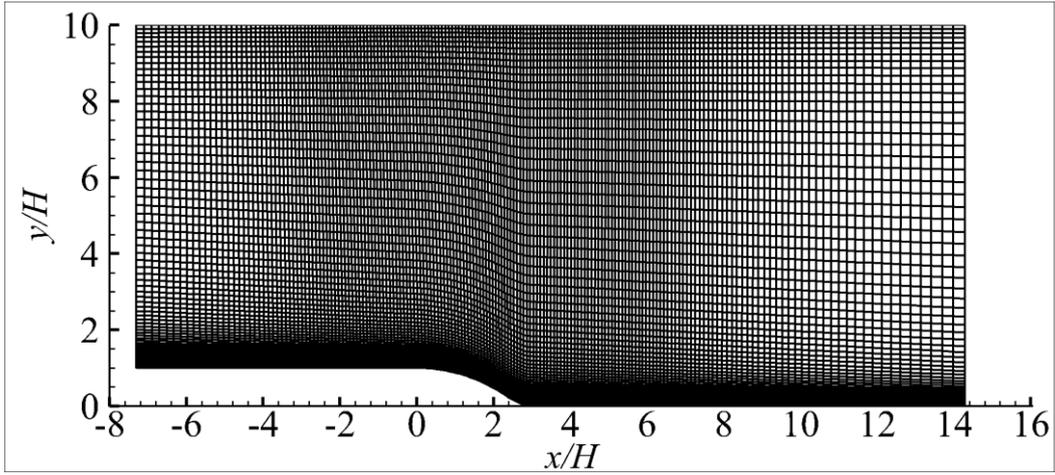
Figure 56 Schematic diagram of the CBFS test case geometry and computational mesh.

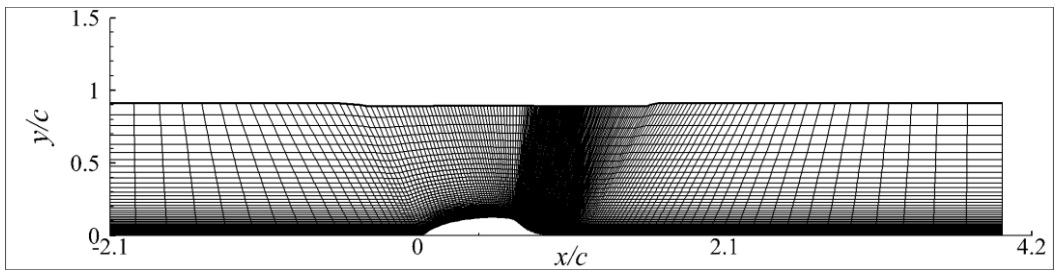
Figure 57 Schematic diagram of the hump test case geometry and computational mesh.

Figure 58 and Figure 59 show the comparison of the computed results for the wall pressure coefficient and friction drag coefficient for the CBFS and hump test cases using different models. From the figures, it can be observed that for this type of flow, the corrected model's results are nearly identical to those of the original SA model. While there was no significant improvement, there was also no noticeable degradation in the results. This is primarily due to the significant differences in the flow separation mechanisms between these two cases and the airfoil flow, and because the training data for the corrected model did not include this type of flow data, making it difficult for the model to effectively correct for this kind of flow. This indicates that there is still room for improvement in the unified modeling of different separation flow types. Future research could explore including multiple types of separation flows in the training data set to develop models with stronger generalization capabilities, which is one of the key directions for further work in this study.

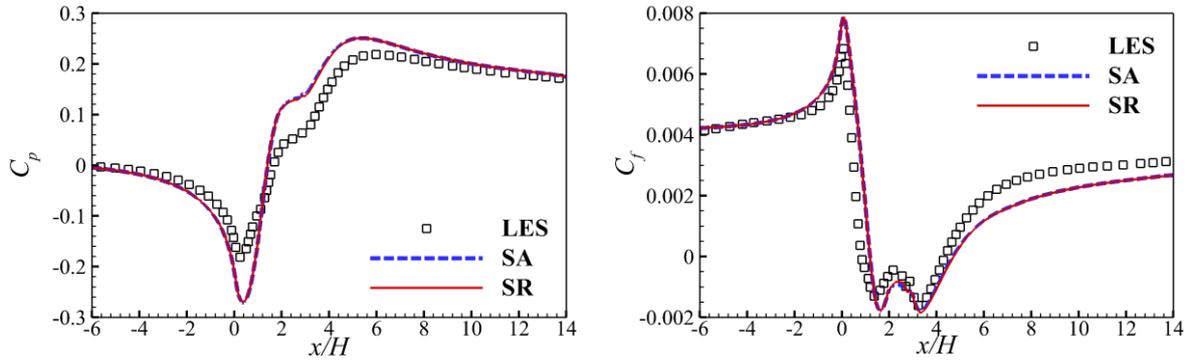

(a) Wall pressure coefficient  (b) Wall friction drag coefficient

Figure 58 Comparison of wall pressure coefficient and friction drag coefficient results for the CBFS test case.

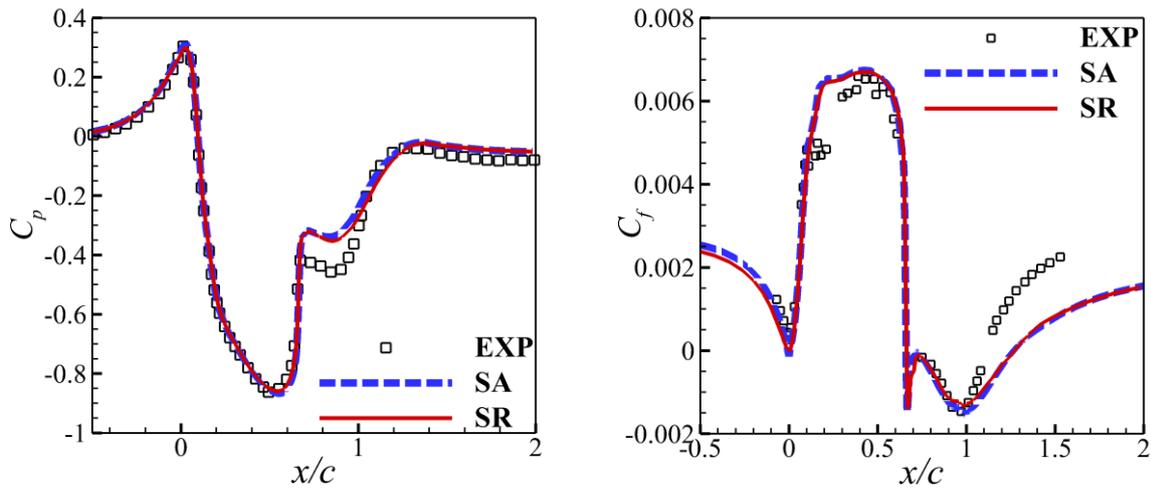

(a) Wall pressure coefficient  (b) Wall friction drag coefficient

Figure 59 Comparison of wall pressure coefficient and friction drag coefficient results for the hump test case.

## 4.8 2D Zero Pressure Gradient Flat Plate

The classic SA model's parameters are calibrated for simple flows, such as zero-pressure-gradient flat plates, ensuring the SA model's good predictive ability for such attached flows. However, modifications to the turbulence model may lead to ill-posed behavior for these types of problems, making it necessary to examine the performance of the corrected model in attached flows. Since the correction proposed in this study affects the turbulence model's production terms, influencing the distribution of eddy viscosity and ultimately the velocity profile, this section uses the zero-pressure-gradient flat plate as an example to verify the corrected model's predictive ability for such simple attached flows.

Figure 60 compares the computed results from different models for the zero-pressure-gradient flat plate case, including the friction drag coefficient distribution (left) and velocity profile distributions at different stations (right). The comparison of results in the figure shows that the corrected model retains the ability to predict the turbulent boundary layer velocity profile. The velocity profiles at two different flow directions match well with the baseline SA model, as shown in the right image. Furthermore, as shown in the left image, the corrected model's prediction of the flat plate friction coefficient also agrees closely with the baseline SA model, indicating that the corrected model maintains the baseline SA model's accuracy for attached flows.

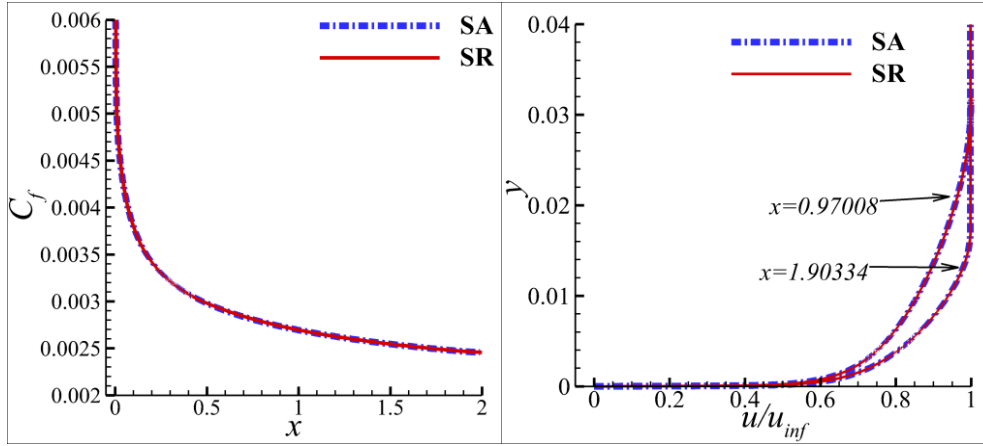

Figure 60 Comparison of computational results for different models in the zero-pressure-gradient flat plate test case, Left figure: Friction drag coefficient, Right figure: Velocity profiles at different stations.

## 5 Conclusions

Supported by high-quality data from data assimilation, this study selects two typical high-Reynolds number wing separation flows as research objects: (1) SC1095 thin airfoil separation at transonic large angles of attack, and (2) DU91-W2-250 thick airfoil separation at large angles of attack. First, the sample dataset suitable for the symbolic regression algorithm was constructed through preprocessing techniques such as data sample reduction and feature construction. Then, the symbolic regression algorithm was used to obtain algebraic expressions for the correction field of the SA model's production terms. The resulting expressions establish a relationship between the local time-averaged flow field characteristics and the corresponding correction coefficient β, which is physically reasonable: the correction only applies near the wall and not in the far-field region, with a value range of [0, 2], ensuring the non-negativity of the production term coefficient.

To ensure that this formula is consistent with the standard SA model for flat plates and airfoil attached flows, the formula was further modified in this chapter to ensure its reasonableness under favorable pressure gradient and zero pressure gradient conditions. Finally, the corrected formula was embedded into the CFD solver for post-validation tests on airfoils, wings, curved backward-facing steps, humps, flat plates, and other cases, with a comparison to the results of the standard SA model. The results show that, for attached flow conditions, the corrected model's performance is consistent with the standard SA model. For airfoil separation flow, the corrected model significantly improves the results, with the average relative error in the lift coefficient for the airfoil stall region reduced by more than 50% compared to the SA model. The corrected model also demonstrates good generalization ability, showing effective corrections in multiple different airfoil cases and 3D wing cases under various conditions, indicating the promising prospects of the data-driven turbulence modeling method for high-Reynolds-number complex separation flows proposed in this chapter.

The corrected expressions based on symbolic regression in this study work well for separation flows caused by adverse pressure gradients in airfoil shapes. In the future, further improvements in modeling data richness can be made, along with optimizations to the symbolic regression algorithm. This could involve adding physical constraints to increase the model's representational capacity, enhancing the model's generalization ability, and achieving effective corrections for other forms of separation flow.

# 基于符号回归的 SA 模型修正及其在分离流动中的应用

孙旭翔 [1,2,3]，单湘淋 [1,2,3]，刘溢浪 [1,2,3]，张伟伟 [1,2,3,*]

1. 西北工业大学 航空学院，西安 710072；
2. 西北工业大学 流体力学智能化国际联合研究所，西安 710072；
3. 飞行器基础布局全国重点实验室，西安 710072；

*通讯作者：张伟伟  邮箱: aeroelastic@nwpu.edu.cn

**摘要**：本文主要围绕高雷诺数分离流动的数值模拟问题，提出了一种基于数据驱动的方法来改进 SA 湍流模型的预测能力。首先，通过对两种典型的翼型大迎角分离流动进行了数据同化工作，获得了高可信度的流场数据集。在此基础上，采用符号回归方法建立了白箱模型，用于对 SA 模型的生成项进行修正。为了验证修正模型的有效性，本文选取了多种典型翼型和机翼（如 SC1095 翼型、DU91-W2-250 翼型、ONERA-M6 机翼等）作为测试算例，从亚音速到跨声速，从几十万 Re 数到千万 Re 数，从小迎角到大迎角等状态，对比分析了修正模型与标准 SA 模型的计算结果。结果表明，修正模型能够有效改善分离流动的预测精度，在保持附着流动预测能力的同时，显著提升了对分离涡分布和流动分离点位置的再现能力，对失速迎角下的升力预测平均相对误差降低了 69.2%，计算精度提升 3 倍以上。此外，通过零压力梯度平板算例的验证，进一步证明了修正模型在湍流边界层速度剖面和摩擦阻力系数分布上的良好预测性能。本文的研究结果为高雷诺数分离流动的数值模拟提供了新的思路和方法，有助于提升工程实践中对复杂流动现象的准确模拟能力。

**关键词**：湍流建模；符号回归；分离流动；数据驱动。

# 1 引言

湍流是流体运动的普遍形态，广泛存在于各种系统之中[1]。在工程领域中，许多应用都依赖于对湍流流动的预测，例如：飞机设计、大气流动分析、燃气轮机引擎以及反应流等。尽管过去几十年来计算能力有了显著提升，但对于大多数实际应用场景而言，完整求解纳维-斯托克斯（Navier-Stokes, NS）方程仍然难以实现。因此，工程师们不得不通过模化湍流来应对这一挑战。

在用于模化湍流的各种方法中，雷诺平均纳维-斯托克斯（Reynolds-Averaged Navier-Stokes, RANS）方法作为一种广泛应用的技术脱颖而出。RANS 方法基于将流动变量分解为时间平均的均值量和脉动量这一概念。这种分离使得能够求解描述均值流的方程，而未被解析的湍流脉动则通过湍流模型进行建模。其在预测流体流动行为中的关键作用已在广泛的工程应用中得到体现，涵盖空气动力学[2]，传热[3]，燃烧[4]等领域.

湍流建模既是 RANS 框架也是计算流体动力学（Computational Fluid Dynamics，CFD）[5]中的关键要素之一。在构建 RANS 方程的封闭模型方面，研究者们付出了巨大努力，并由此发展出多种著名的湍流模型。其中具有代表性的包括一阶方程 Spalart–Allmaras（SA）模型[6]、两方程模型 $k-\epsilon$[7]、$k-\omega$[8] 及其变体[5]。这些模型通常基于一套简化的典型流动实验数据标定的参数。然而，由于每个模型都是针对特定流动工况进行专门开发，且其参数缺乏明显的普适性。另一方面，尽管计算能力有所提升，直接数值模拟（Direct Numerical Simulation, DNS）和大涡模拟（Large Eddy Simulation, LES）等高保真度数值模拟方法在航空航天应用中常见的高雷诺数流动中的应用仍然受到限制[9]。因此，进一步提高 RANS 方法的准确性仍是一个具有研究价值的领域[10]。

随着机器学习技术的发展，研究者开始利用机器学习和数据驱动方法来增强湍流模型[11, 12]。例如，使用来自 LES/DNS 的高保真度数据训练一个数据驱动的模型，以替代或改进传统的湍流模型[13-18]。Ling 和 Templeton[19, 20] 构建了以伽利略不变量形式存在的模型不变量，并利用机器学习从传统模型计算中预测不合理的涡粘性区域。Zhu 等人[17]开发了一个基于单层神经网络的纯粹数据驱动湍流黑箱代数模型，并实现了与求解器的耦合计算。他们的结果显示，所构建的模型在准确性和计算效率上均表现良好，同时对不同流动条件和几何形状表现出良好的泛化能力。这验证了此类替代模型的可行性。

在高保真数据难以获取的情况下（例如高雷诺数流动），基于数据同化的流场反演方法已逐渐成为一种有效的手段[21-24]。Duraisamy[25]、Singh[26-28] 和 Parish 等人[29] 提出了结合流场反演与机器学习（Flowfield Inversion with Machine Learning, FIML）的建模框架。在这一框架中，空间分布的校正系数被引入到湍流模型中，并通过伴随方法进行优化，随后使用神经网络对其进行建模。基于 FIML 框架，可在流场反演过程中引入表征非平衡湍流效应的先验物理约束[30, 31]，同时可将对湍流各向异性校正引入非线性涡粘模型中[32]。Wu 等人以曲线后台阶为研究对象，针对 SST 模型中 ω 方程的破坏项进行了流场反演修正[22]。基于反演获得的高保真流场数据，他们采用符号回归（SR）方法开发了一个显式代数模型，将局部流动特征映射至校正场。该模型在不同测试算例中均展现出优于传统 SST 模型的性能，具有良好的泛化能力。此外，该研究团队的其他工作[33, 34]也进一步凸显了结合流场反演与符号回归在数据驱动湍流建模中的广泛应用潜力。

与基于伴随方法的流场反演不同，Yang 等人[35]采用了集合卡尔曼滤波（Ensemble Kalman Filter, EnKF）来对 $k-\omega-\gamma-A_r$ 转捩模型中的系数进行反演，并依据反演数据训练了神经网络和随机森林模型。一些研究还利用数据同化方法对经典湍流模型的系数进行了修正[36, 37]。例如，Li 等人[37] 确定了 $k-\omega$ 模型的系数，以使速度误差相对于高保真模拟结果最小化。随后，优化问题通过基于梯度的技术或集合方法来解决。近期，作为建模与数据同化方法的结合，已有研究尝试直接利用场反演方法依据观测数据构建神经网络模型[38-41]。

在数据驱动的建模范式中，模型的泛化性是非常重要的性能。泛化性是指模型在未见数据集上的预测能力，是衡量模型性能的重要指标。由于湍流问题本身的复杂性，当前的数据驱动湍流模型的泛化能力有限，通常只在与训练数据相近的外形、工况上具有较好的预测能力，这与模型选择、训练数据规模、特征选择、物理规律嵌入等因素密切相关。Wu 等人[34]针对 FIML 框架提出了模型泛化性能的四个等级，分别是（1）模型对于相似外形的泛化性；（2）模型对简单附着流动的泛化性；（3）模型对具有相同的物理机制但是完全不同的外形、工况的泛化性，比如模型在逆压梯度主导的分离流动数据上进行了训练，在其他由逆压梯度主导的分离流动上表现出色；（4）模型在具有完全不同的分离特征、几何形状和来流工况的一系列测试算例上都表现良好，比如模型既可以准确预测逆压梯度引起的分离，也能准确预测因钝体几何引起的分离。他们认为当前的数据驱动湍流模型多数停留在第一级的泛化能力，仅有少数模型[22,42]具有第三级的泛化能力，并强调了第二级泛化能力对于数据驱动模型的重要性。McConkey 等人[43]的研究结论也支持这一观点，他们在由周期山、bump、曲线后台阶、方管、收缩扩张管道构成的数据集上对随机森林、神经网络、XGBoost[44]三个模型的泛化能力进行了测试，并建议使用 XGBoost 模型进行湍流建模工作，因为该模型具有良好的泛化性能和调优成本。他们还发现所构建的模型可以很好地泛化到与训练数据相似的流动中，但缺乏对未见流动类型的泛化能力，因此他们认为机器学习方法最适合为给定的流动类型开发专门的模型，这也是工业流动中经常遇到的问题。Li 等人[45]对 TBNN、FIML、PIML 三个框架的泛化性进行了比较研究，结果表明 FIML 具有更好的泛化性能。2022 年 NASA 湍流建模研讨会的结果表明：当前大多数机器学习模型仅能泛化到与训练集相似的流动状态。SA 模型的提出者 Spalart 也在此次会议的特邀报告上指出了当前机器学习湍流模型的诸多问题，特别是在泛化性与可解释性方面，当前的研究结果并没有产生一个"通用型"湍流模型，很多模型在湍流平板上甚至不能合理预测边界层。尽管目前存在诸多不足，他并不完全排除机器学习可能带来的范式变化，但也强调这需要克服许多深层次的问题。他建议机器学习在目前阶段更多地充当"辅助角色"，而不是独立设计完整模型的"架构师"。综合来看，数据驱动湍流模型的泛化性是研究者普遍面临的困难和挑战，需要更深入细致的分析和探讨。

上述研究表明了通过数据驱动方法构建或改进湍流模型的潜力；然而，大多数相关工作集中于低雷诺数简单几何分离流或高雷诺数附着流。目前对于工业应用中常见的高雷诺数复杂分离流的关注仍显不足。本文特别针对这一方面，旨在开发一种强泛化性的数据驱动模型，以提高经典 SA 湍流模型对高雷诺数分离流的模拟精度。

本文接下来的安排如下：第 2 节介绍了主要的数值计算方法。第 3 节阐述了符号回归建模的关键方面，包括采样点选取以及输入输出特征的构建。第 4 节展示了修正模型在多个测试用例中的计算结果，并将其与实验数据和 SA 模型的结果进行了对比。最后，

第 5 节进行总结与展望。

## 2 数值方法

守恒形式的 RANS 方程定义如下：

$$\frac{\partial \rho}{\partial t} + \frac{\partial (\rho u_i)}{\partial x_i} = 0,$$

$$\frac{\partial (\rho u_i)}{\partial t} + \frac{\partial (\rho u_i u_j)}{\partial x_j} = -\frac{\partial p}{\partial x_i} + \frac{\partial}{\partial x_j}(\tau_{ij}^l + \tau_{ij}^t),$$

$$\frac{\partial (\rho E)}{\partial t} + \frac{\partial (\rho E u_j)}{\partial x_j} = -\frac{\partial (p u_j)}{\partial x_i} + \frac{\partial}{\partial x_j}\left[\left(\tau_{ij}^l + \tau_{ij}^t\right)u_i\right] + \frac{\partial}{\partial x_j}\left(q_j - c_p \rho \overline{T'u_j'}\right)$$

其中 $\tau_{ij}^t$ 被称为雷诺应力项，考虑使用经典的涡粘假设对其进行封闭：

$$\tau_{ij}^t = \mu_t \left(\frac{\partial \overline{u}_i}{\partial x_j} + \frac{\partial \overline{u}_j}{\partial x_i}\right) - \frac{2}{3}\rho k \delta_{ij}$$

$$k = \frac{1}{2}\overline{u_k' u_k'} = \frac{1}{2}\left(\overline{u'^2} + \overline{v'^2} + \overline{w'^2}\right)$$

其中 $\mu_t$ 是涡粘系数，$k$ 是湍动能项。本文所使用的湍流模型为 SA 模型[46]，其形式如下：

$$\mu_t = \rho \hat{v} f_{v1}$$

$$\frac{D\hat{v}}{Dt} = C_{b1}(1-f_{t2})\hat{S}\hat{v} + \frac{1}{\sigma}\left[\frac{\partial}{\partial x_j}\left((v+\hat{v})\frac{\partial \hat{v}}{\partial x_j}\right) + C_{b2}\frac{\partial \hat{v}}{\partial x_i}\frac{\partial \hat{v}}{\partial x_i}\right] - (C_{w1}f_w - \frac{C_{b1}}{\kappa^2}f_{t2})(\frac{\hat{v}}{d})^2$$

模型中的生成项、破坏项和输运项分别定义如下：

$$P = C_{b1}(1-f_{t2})\hat{S}\hat{v},$$

$$D = (C_{w1}f_w - \frac{C_{b1}}{\kappa^2}f_{t2})(\frac{\hat{v}}{d})^2$$

$$T = \frac{1}{\sigma}\left[\frac{\partial}{\partial x_j}\left((v+\hat{v})\frac{\partial \hat{v}}{\partial x_j}\right) + C_{b2}\frac{\partial \hat{v}}{\partial x_i}\frac{\partial \hat{v}}{\partial x_i}\right]$$

其中 $d$ 是壁面距离，模型中的相关参数与函数定义如下：

$$f_{v1} = \frac{\chi^3}{\chi^3 + c_{v1}^3}, \chi = \frac{\hat{v}}{v}, \hat{S} = \Omega + \frac{\hat{v}}{\kappa^2 d^2}f_{v2}, \Omega = \sqrt{2W_{ij}W_{ij}}$$

$$f_{v2} = 1 - \frac{\chi}{1+\chi f_{v1}}, f_w = g\left[\frac{1+c_{w3}^6}{g^6 + c_{w3}^6}\right]^{1/6}, g = r + c_{w2}(r^6 - r),$$

$$r = \min\left[\frac{\hat{v}}{\hat{S}\kappa^2 d^2}, 10\right], f_{t2} = c_{t3}\exp\left(-c_{t4}\chi^2\right), W_{ij} = \frac{1}{2}\left(\frac{\partial u_i}{\partial x_j} - \frac{\partial u_j}{\partial x_i}\right),$$

$$c_{b1} = 0.1355, \sigma = 2/3, c_{b2} = 0.622, \kappa = 0.41,$$

$$c_{w2} = 0.3, c_{w3} = 2, c_{v1} = 7.1, c_{t3} = 1.2, c_{t4} = 0.5, c_{w1} = \frac{c_{b1}}{\kappa^2} + \frac{1+c_{b2}}{\sigma}$$

SA 模型对附着流、小分离流动具有较好的精度,但不能准确地捕捉大分离流动。本文的研究旨在通过使用基于数据驱动的 SA 模型生产项校正来解决这一缺陷:

$$\frac{D\hat{v}}{Dt} = \beta(\boldsymbol{x})P(\hat{v},\boldsymbol{U}) - D(\hat{v},\boldsymbol{U}) + T(\hat{v},\boldsymbol{U})$$

# 3 符号回归建模

本文对两种典型分离流动进行了同化工作,得到了对应的高可信度流场数据与 SA 生成项修正系数空间分布。这两种典型流动分别为:(1)SC1095 翼型在来流状态 Ma=0.6, AoA=9.17°, Re=4.9×10⁶ 下的分离流动。标准 SA 模型在此类型的流动中预测的流动分离点靠前,从而使得在激波与逆压梯度的影响下分离涡占据上翼面 80%左右的面积,升力迅速下降,数值模拟得到的升力系数较实验值偏小。在数据同化过程中通过对 SA 模型生成项修正系数空间分布的调整,使得流动分离点靠后,一定程度上抑制了流动分离,减小了分离涡的大小,从而使得数值模拟结果与实验结果吻合。同化前后的速度场对比如图 1 所示。(2)DU91-W2-250 翼型在来流状态 Ma=0.15, AoA=15.0º, Re=3.0×10⁶ 下的分离流动。该算例是一个典型厚翼型的后缘分离流动,标准 SA 模型在此类型的流动中预测的流动分离点靠后,分离涡较小,从而使得在翼型失速迎角范围通过数值模拟得到的升力系数较实验值偏大。数据同化通过减小上翼面前缘 $\beta$ 的方式增强了流动分离,使得分离点靠前,分离涡变大,计算结果与实验结果吻合。该算例同化前后的速度场与流线对比见图 2。这两类流动是非常具有代表性的翼型分离流动,因此取这两个算例同化得到的流场数据作为数据驱动模型的样本数据集。

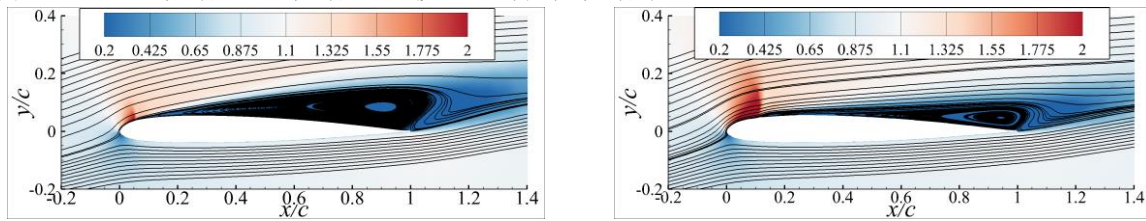

(a) SA 模型计算得到的流场　　　　　(b) 数据同化得到的流场

图 1 SC1095 翼型 Ma=0.6, AoA=9.17°, Re=4.9×10⁶ 状态下流场对比,左图为基于标准 SA 模型计算得到的流场和流线分布,右图为通过同化 SA 生成项系数分布得到的最优流场和流线分布。经过同化,上翼面的流动分离被抑制,分离点靠后,分离涡减小。

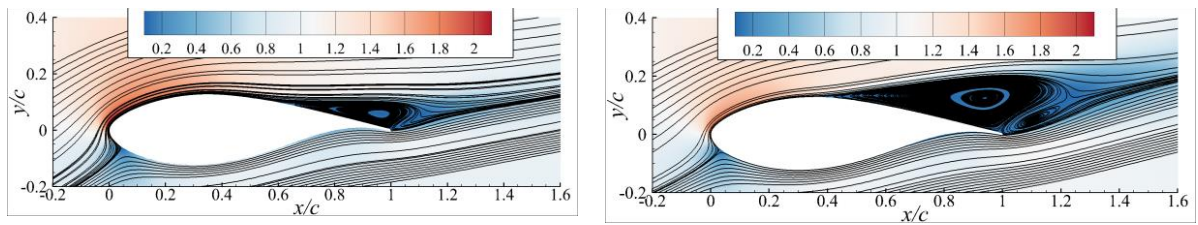

(a) SA 模型计算得到的流场　　　　　(b) 数据同化得到的流场

图 2 DU91-W2-250 翼型 Ma=0.15, AoA=15.0º, Re=3.0×10⁶ 来流状态下的流场对比,左图为基于标准 SA 模型计算得到的流场和流线分布,右图为通过同化 SA 生成项系数分布得到的最优流场和流线分布。经过同化,上翼面的流动分离被增强,分离点靠前,分离涡增大。

本章使用符号回归方法构建当地流场特征到修正系数 $\beta$ 之间的模型。与神经网络类模型相比，基于符号回归方法的模型数据量需求偏小[22, 33]，因此有必要对流场样本进行下采样，精简流场样本，选取具有代表性的样本点用于符号回归建模。在本章所要构建的关系式中，模型输入是时均流动特征，输出是对应网格单元上的生成项修正系数，其空间分布如图 3 所示。通过其空间分布可以看到，在流场的大多数区域 $\beta$ 值接近于 1，即不对生成项进行修正。只有在壁面附近的 $\beta$ 值才有较大的变化，因此训练样本所需要的数据点主要在壁面附近随机选取 2000 个，总共 4000 个数据点用于符号回归模型训练。表 1 列举了用于建模的特征及其表达式，在建模过程中通常需要对特征进行一定的数值缩放与转换使得特征量级基本一致，这样有助于模型的训练，参考已有文献的工作[13, 33]，对部分特征进行了变换，变换公式列举在表 1 最后一列。

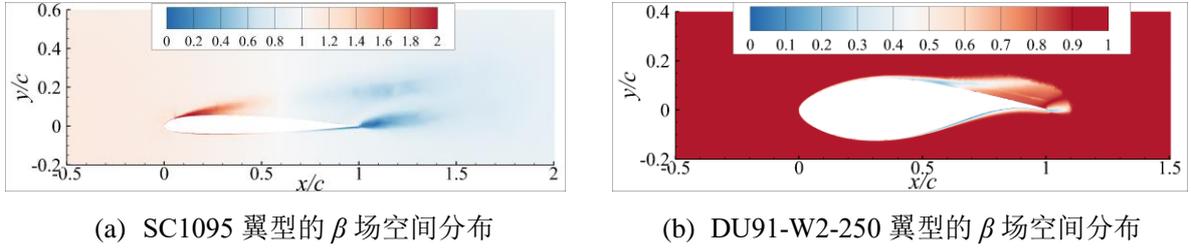

(a) SC1095 翼型的 $\beta$ 场空间分布　　(b) DU91-W2-250 翼型的 $\beta$ 场空间分布

图 3 训练样本集中不同翼型同化得到的 $\beta$ 场分布情况.

表 1 用于符号回归建模的输入特征表达式与变换公式

| 特征 | 表达式 | 变换公式 |
| --- | --- | --- |
| $x_0$ | $\sqrt{2W_{ij}W_{ij}}, W_{ij}=\dfrac{1}{2}\left(\dfrac{\partial U_i}{\partial x_j}-\dfrac{\partial U_j}{\partial x_i}\right)$ | $\log(x+1)$ |
| $x_1$ | $\sqrt{2S_{ij}S_{ij}}, S_{ij}=\dfrac{1}{2}\left(\dfrac{\partial U_i}{\partial x_j}+\dfrac{\partial U_j}{\partial x_i}\right)$ | $\log(x+1)$ |
| $x_2$ | $P\rho^{-\gamma}-1$ | — |
| $x_3$ | $\dfrac{x_0^2-x_1^2}{x_0^2+x_1^2}$ | — |
| $x_4$ | $S_{ij}S_{ji}$ | — |
| $x_5$ | $W_{ij}W_{ji}$ | — |
| $x_6$ | $d$ | — |
| $x_7$ | $\sqrt{\dfrac{\partial P}{\partial x_i}\dfrac{\partial P}{\partial x_i}}$ | $\log(x+\varepsilon)$ |
| $x_8$ | $1-\tanh(8r_d)^3\ \ r_d=\dfrac{\nu_l+\nu_t}{\kappa^2 d^2\sqrt{U_{ij}U_{ij}}}$ | — |

在表 1 所列举的特征中，$x_0$ 和 $x_1$ 分别是旋变率张量与应变率张量的模。特征 $x_2$ 表征流场的熵[17, 21]。$x_3$ 是无量纲 $Q$ 准则。$x_4$ 和 $x_5$ 分别是基于旋变率张量和应变率张量所

定义的不变量[1, 14, 22]。$x_6$表示网格格心到壁面的最小距离。$x_7$表征了压力梯度的大小，变换公式中的$\varepsilon$是一个为了保证计算稳定的小量，实际计算时取为$10^{-9}$。$x_8$是用于边界层识别的特征[47]，该表达式是在 DDES 方法中被使用来保证边界层内部依然使用 RANS 方法求解的保护函数，因此在边界层中为 0，外部为 1。所有用于特征计算的基本流量变量都是无量纲的，无量纲方式如下：

$$x_i = \frac{\overline{x}_i}{L}, \rho = \frac{\overline{\rho}}{\rho_\infty}, U_i = \frac{\overline{U}_i}{V_\infty}, P = \frac{\overline{P}}{\rho_\infty a_\infty^2}, \mu = \frac{\overline{\mu}}{\mu_\infty}$$

其中下标"∞"表示自由来流参数，上标 "-" 表示有量纲的值。综上所述，本研究要构建的代数关系表述如下：

$$\beta = f(x_0,...,x_8)+1$$

在$\beta$等于 1 的时候表示采用的是原始的 SA 模型形式，不进行任何修正。由于在流场的绝大部分$\beta$的取值都和 1 接近，因此在所要构建的代数关系式中先验性的加上 1，近壁面附近的差量通过符号回归模型构建。

本文通过 PySR 库来构建符号回归模型，预设的操作算符见表 2，损失函数定义为预测值与目标值之间误差的平方和，具体如下：

$$loss = \sum_i (Y_{i,pred} - Y_{i,target})^2$$

表 2 符号回归使用的操作符

| 操作符类型 | 操作符 |
| --- | --- |
| 一元算符 | $\exp(x_i)$， $\tanh(x_i)$ |
| 二元算符 | $x_i + x_j, x_i - x_j, x_i * x_j, x_i / x_j$ |

符号回归模型得到的模型虽然有提到的若干优势，但是相应的也存在模型表示能力不足、对于特征要求严格、在较大规模数据集上难以有效建模等劣势。本文在构建模型时所使用的数据集为 4000 个数据点，且多为逆压梯度下的流场数据，缺乏顺压梯度、零压力梯度场景下的数据，这会在一定程度上限制模型的泛化性以及修正结果，特别是会使得在流场某些区域或附着流状态下错误激活产生修正，导致最终计算结果出现偏差甚至错误。在尝试扩大数据集规模与流场数据类型重新建模时，发现建模过程难以进行。为了避免修正公式的错误激活，有必要根据相关物理先验对得到的公式进行相关修正，这种后验修正的过程是将物理先验与数据驱动模式相结合的过程，为当前多篇文献所采用[22, 33, 34, 48]。经过修正之后，该模型被嵌入到 CFD 求解器中，用于双向耦合求解各类流动，在下面的算例中经由模型修正计算得到的结果以"SR"表示。

# 4 模型验证

本小节主要给出带有 SR 修正的 SA 湍流模型在不同外形、来流状态上的计算结果，以及与标准 SA 模型结果、实验结果的对比。测试算例主要包括 SC1095 翼型、ONERA-M6 机翼、DU-91-W2-250 翼型、S809 翼型、S814 翼型、S805 翼型、曲线后台阶、驼峰以及零压力梯度平板，不同翼型的形状示意图见图 4。其中 SC1095 翼型和 ONERA-M6 机翼主要考察修正模型对于跨声速状态激波诱导分离流动的修正效果，DU91-W2-250 翼型和 S 系列风力机翼型主要考察修正模型在大迎角分离流动上的修正效果。在这两类算例中，前一种主要表现为激波引起的大范围分离流动，需要适当抑制分离涡的大小、将分离点向后修正；后一种则正好相反，需要通过修正使得流动分离点靠前，从而使得计算结果与实验结果更加吻合。曲线后台阶和驼峰算例用于考察修正模型在不同分离机制算例上的表现。零压力梯度平板算例则用于考察修正模型在简单外形附着流动上的效果，修正模型在该算例上应与经典 SA 模型保持一致的结果。在所有的测试算例中，修正模型中的可调节参数均保持一致。

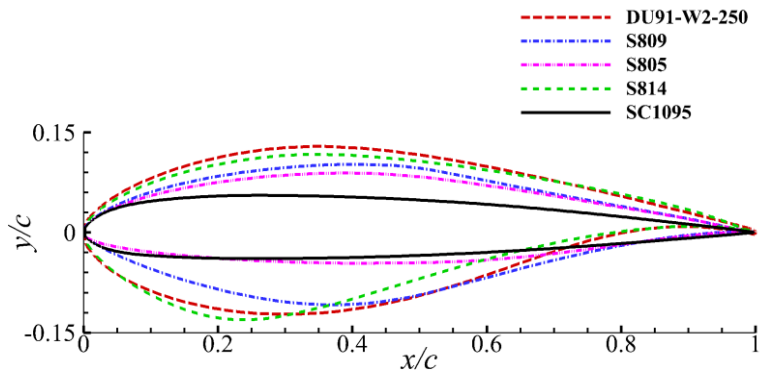

图 4 测试翼型外形示意图

## 4.1 SC1095 翼型

本小节给出修正模型在 SC1095 翼型上的耦合计算结果。由于修正模型的训练数据中包含了 SC1095 翼型在 Ma=0.6，AoA=9.17º，Re=$4.9\times10^6$ 状态下的同化数据，因此这里主要测试修正模型在不同来流状态上的泛化能力。图 5 给出了在 Ma=0.6、Re=$4.9\times10^6$ 来流状态下升、阻力系数的结果对比，其中绿色实线与黑色箭头标记了模型的训练数据对应的状态，黑色实线表示实验结果，红色实线表示经过带 SR 修正的 SA 模型的结果，蓝色虚线表示原始 SA 模型结果，本章后面的图中若不做特殊说明均遵循此类表示。根据图中结果对比可以看到，经过修正的模型可以有效改善在失速迎角区域的计算结果，升阻力系数与实验值均更加吻合。在来流迎角较小的状态下，流动为附着流，修正模型的表现则与标准 SA 模型基本一致，这说明模型仅在分离流动下激活，在附着流上保持了与标准 SA 模型的一致性。

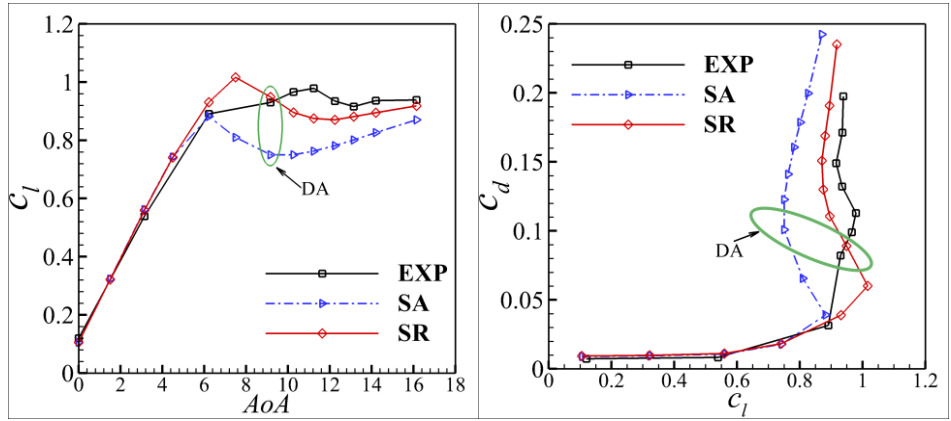

图 5 SC1095 翼型 Ma=0.6, Re=4.9×10⁶ 来流状态下升力系数、阻力系数对比

图 6 中给出了迎角 AoA=6.22º、9.17º 时壁面压力系数的对比结果，结果表明经过修正的模型可以更加准确地预测表面压力系数分布结果。图 7 中给出了 SC1095 翼型在 Ma=0.6, AoA=9.17º, Re=4.9×10⁶ 下两种模型求解得到的速度场以及流线示意图。可以看到修正模型的计算结果激波位置更加靠后，且有效抑制了流动的分离，减小了分离涡大小，使得计算结果与实验结果更加吻合。图 8 中壁面摩擦阻力系数的对比结果也表明修正模型抑制了流动分离，流动分离点靠后。以上结果均表明修正模型对于激波诱导的分离流动具有良好的修正左右。

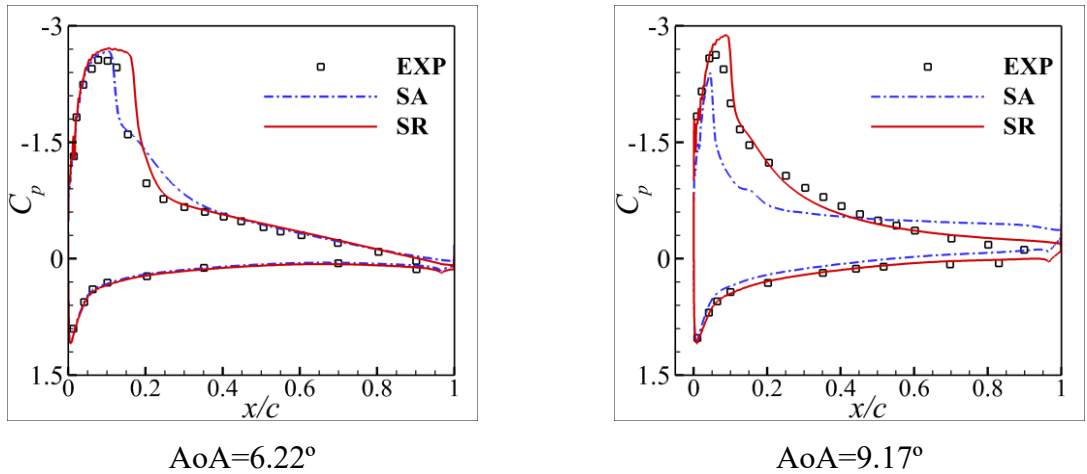

AoA=6.22º　　　　　　　　　　　　　AoA=9.17º

图 6 SC1095 翼型 Ma=0.6, Re=4.9×10⁶ 来流状态下不同迎角对应的壁面压力系数结果对比

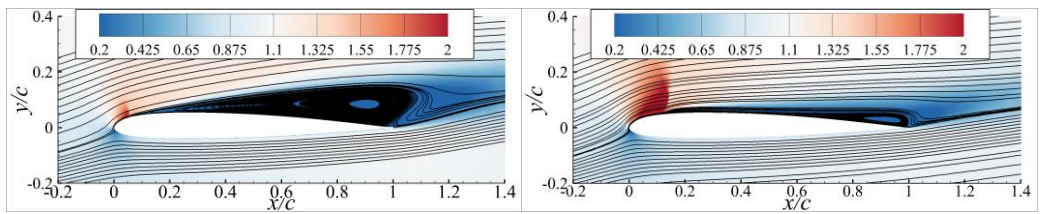

图 7 SC1095 翼型在 Ma=0.6, AoA=9.17º, Re=4.9×10⁶ 下不同方法求解得到的流场对比，左图：SA 模型计算结果，右图：带 SR 修正的模型计算结果

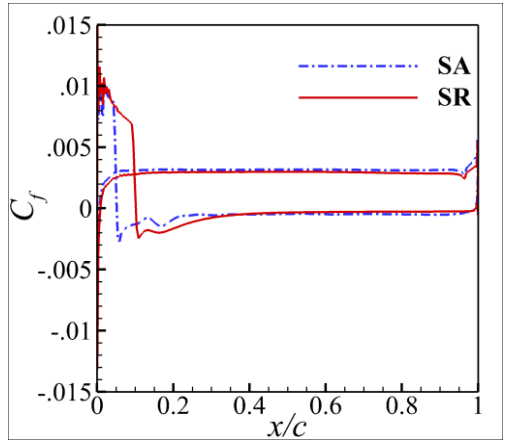

图 8 SC1095 翼型在 Ma=0.6, AoA=9.17º, Re=4.9×10$^6$ 下不同方法求解的摩擦阻力系数分布对比

图 9 给出了在失速迎角下经典 SA 模型的升力系数计算结果与修正模型计算结果相比实验数据的相对误差。可以看到，相比于 SA 模型，修正模型在不同迎角下的模拟结果均显著降低。对于整个失速区域的升力预测，相较于 SA 模型（相对误差 15.94%），SR 模型的升力相对误差显著降低至 5.32%，计算误差降低了 66.62%，计算精度提升至原来的 3.0 倍。表明了当前修正模型在分离区域具有良好的适用性，可以显著提升 RANS 方法对于分离流动的模拟精度。

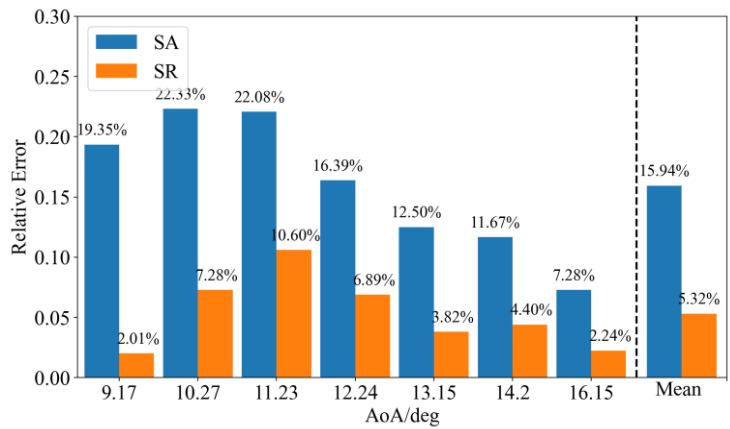

图 9 SC1095 翼型在 Ma=0.6, Re=4.9×10$^6$ 状态下失速迎角下升力系数相对误差对比

图 10 给出了在 Ma=0.5, Re=4.34×10$^6$ 的来流状态下基于两个模型计算得到的升阻力系数的结果对比，结果显示在失速迎角区域修正模型的计算结果与实验值高度一致，说明模型具有良好的变状态泛化能力。图 11 进一步给出了该状态失速迎角下升力系数的相对误差结果，可以看到，相比于 SA 模型，修正模型在不同迎角下的预测结果均显著降低。对于整个失速区域的升力预测，相较于 SA 模型（相对误差 12.28%），SR 模型的升力相对误差显著降低至 1.56%，计算误差降低了 87.30%，计算精度提升至原来的 7.87 倍。

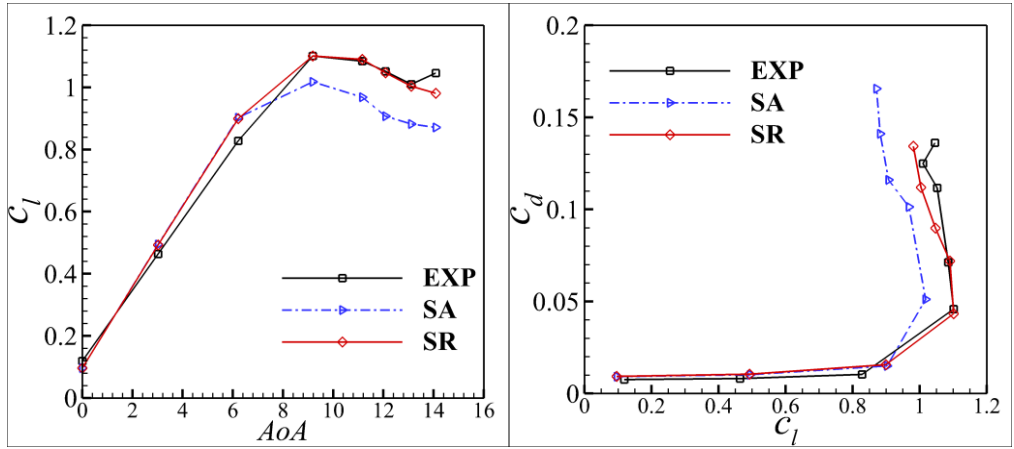

图 10 SC1095 翼型 Ma=0.5, Re=4.34×10⁶ 来流状态下升力系数、阻力系数对比

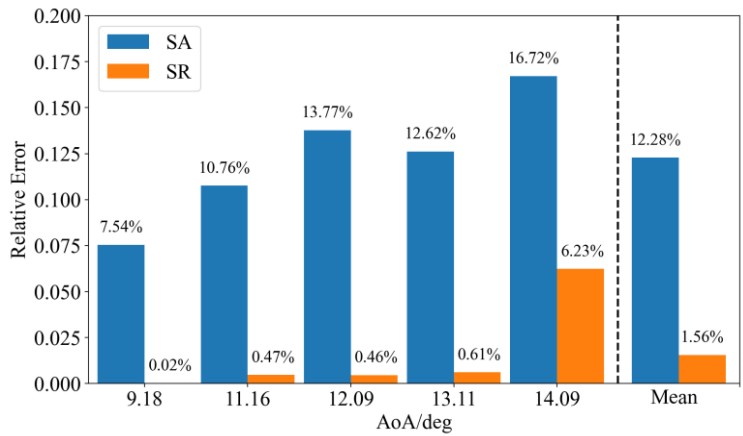

图 11 SC1095 翼型在 Ma=0.5, Re=4.34×10⁶ 状态下失速迎角下升力系数相对误差对比

## 4.2 ONERA-M6 机翼

本小节主要给出模型在 ONERA-M6 机翼算例上三个典型计算状态的结果对比，来流马赫数为 0.84，来流迎角分别为 3.06º、5.06º、6.06º，基于平均气动弦长的雷诺数为 11.72×10⁶。在三个迎角中，第一个对应的是附着流，其余两个对应的均为激波诱导的分离流动。该算例主要考察修正模型在三维附着流与分离流动上的表现。图 12 至图 14 分别给出了不同来流迎角下机翼不同截面位置处的的壁面压力系数分布对比，可以看到在 AoA=3.06º 附着流动下（图 12）两个模型的计算结果十分接近，说明修正模型在三维附着流上也可以保证与 SA 模型结果的一致。而在后面两个分离流动中（图 13 和图 14）修正模型的计算结果显著优于原始 SA 模型计算结果，说明在这两个典型分离流场景下开启了修正并起到了积极作用。结合分离流动下不同模型计算的表面流线对比结果（图 15 和图 16）可以看到，在两个分离流动中，靠近翼根的两个截面（$\eta$=0.2/0.44）流动为附着流动，因此两个模型的压力系数计算结果仍保持一致。其余靠近翼梢的截面中均出现了激波诱导的分离流动，在这些截面上修正模型均进行了有效修正，对激波位置的预测相比 SA 模型更为精确，因此对于激波后分离流动的压力变化的预示也更为精确，得到了与实验结果更为吻合的计算结果。

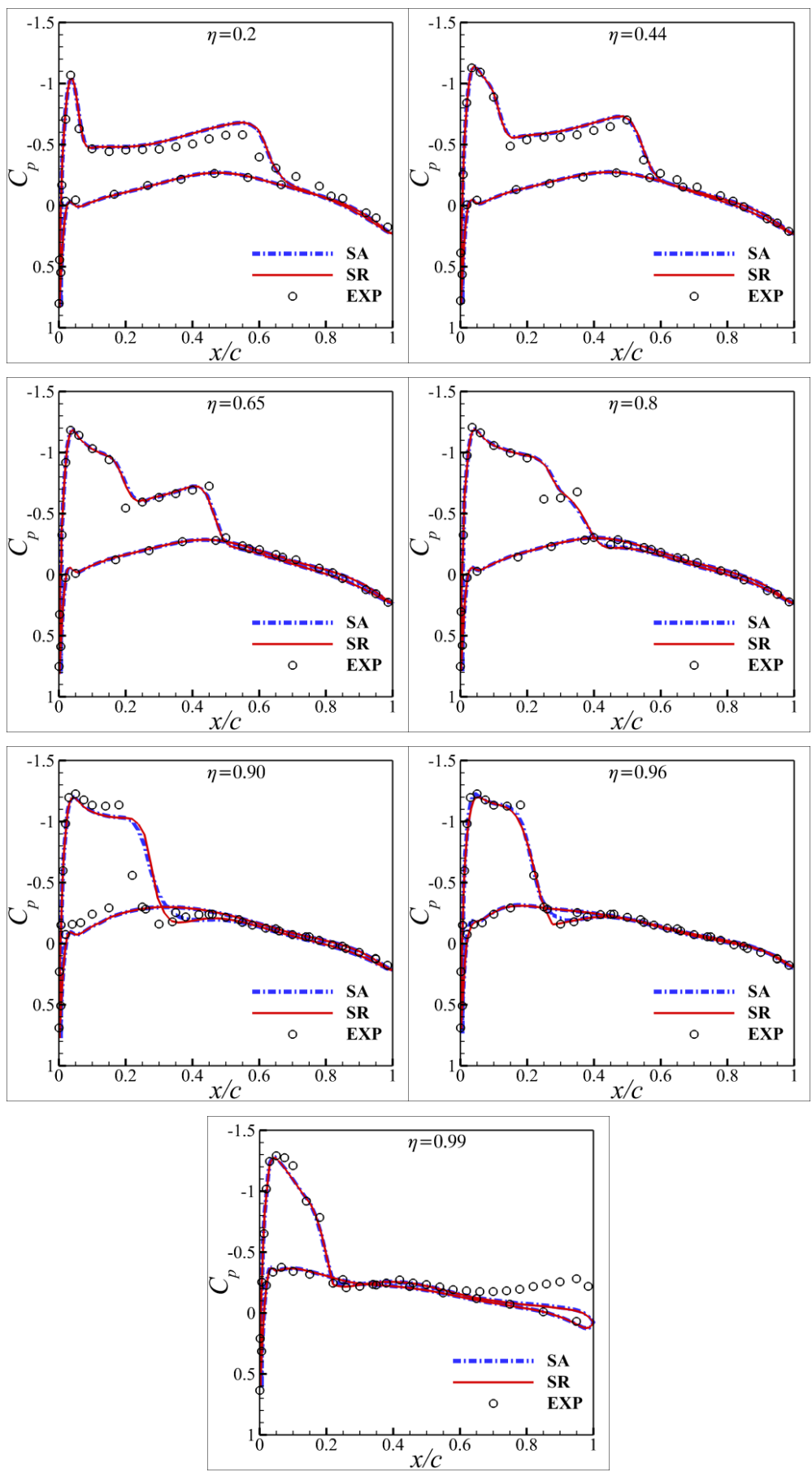

图 12 ONERA-M6 机翼 Ma=0.84, AoA=3.06º, Re=11.72×10$^6$ 壁面压力系数计算结果对比

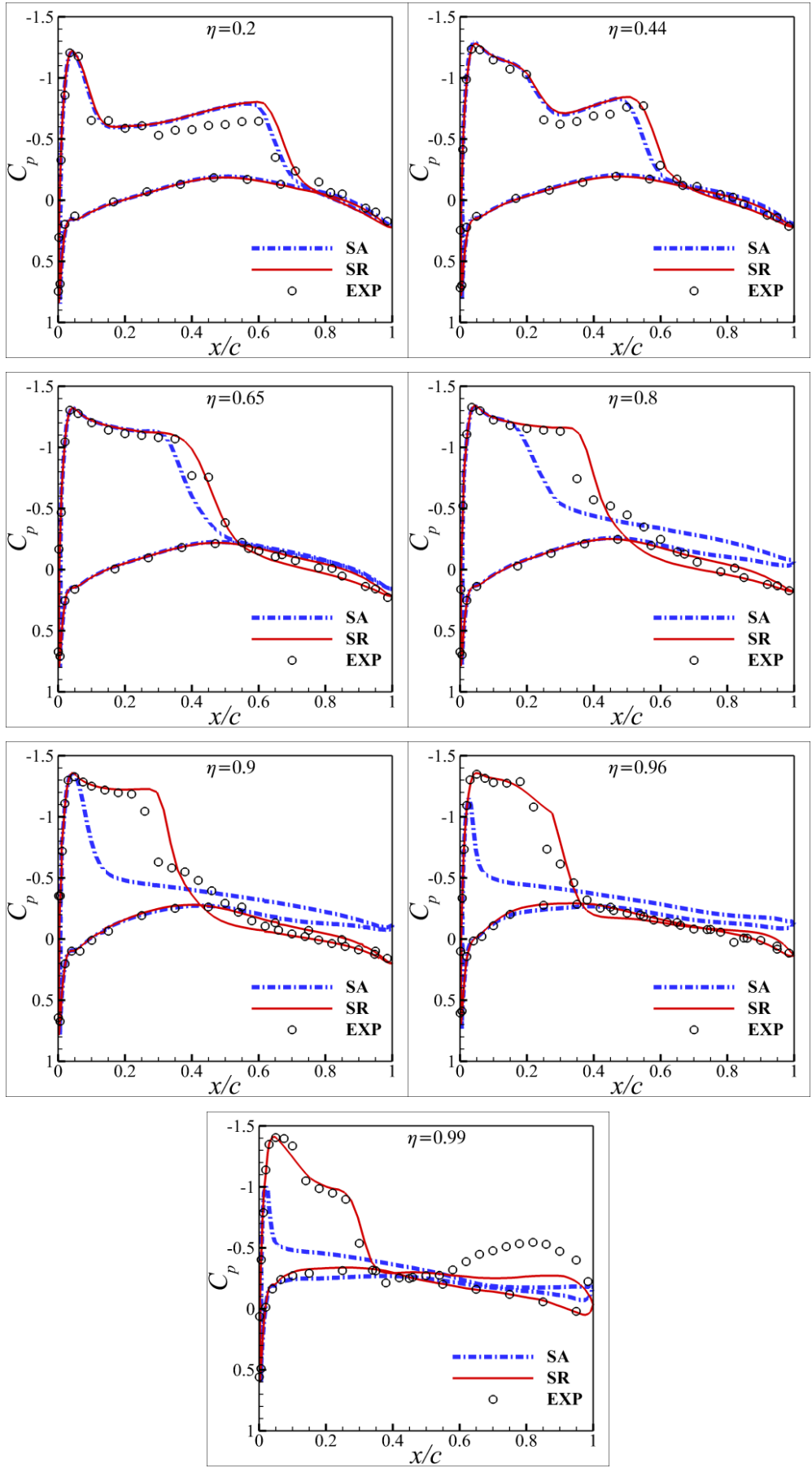

图 13 ONERA-M6 机翼 Ma=0.84, AoA=5.06º, Re=11.72×10⁶ 壁面压力系数计算结果对比

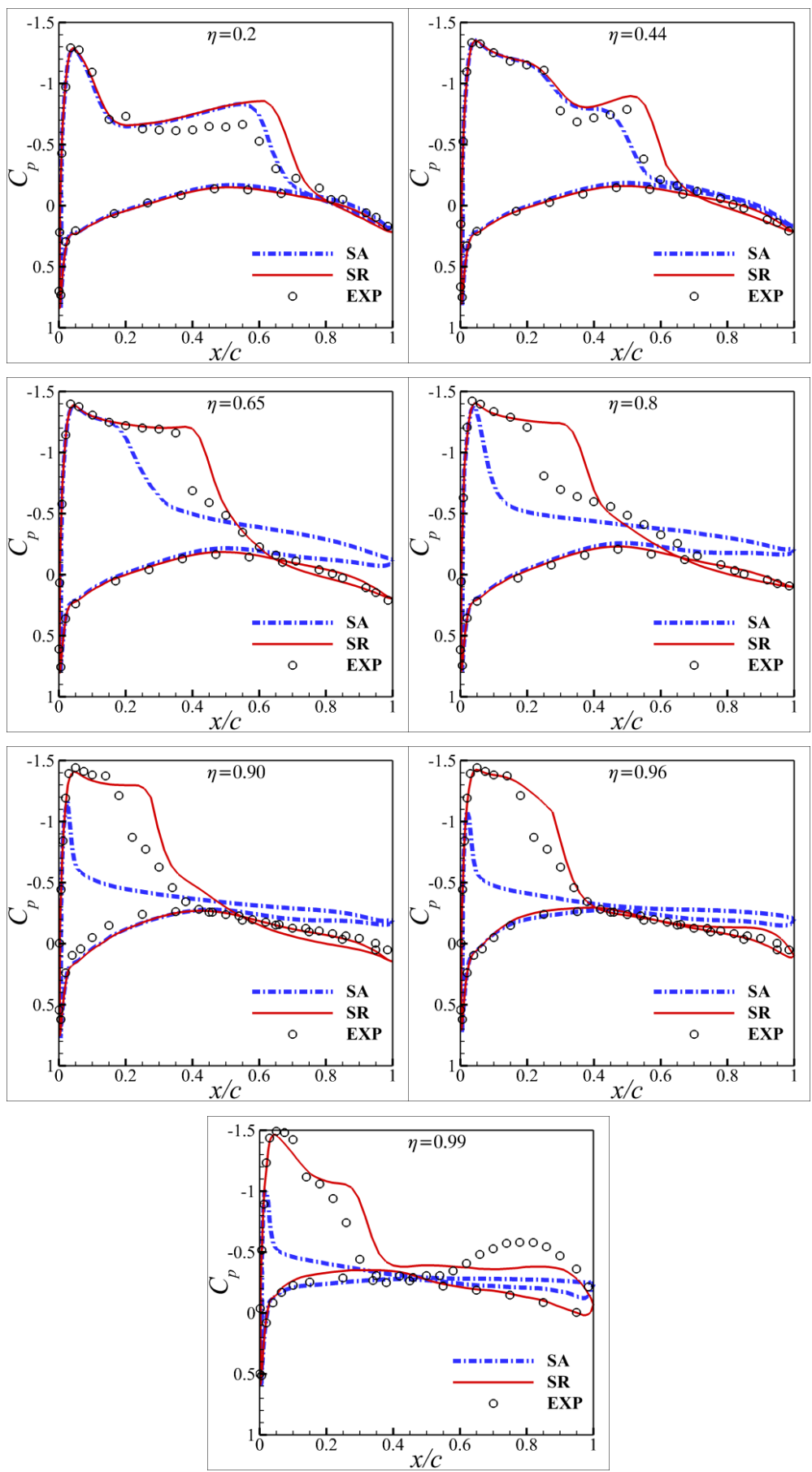

图 14 ONERA-M6 机翼 Ma=0.84, AoA=6.06º, Re=11.72×10⁶ 壁面压力系数计算结果对比

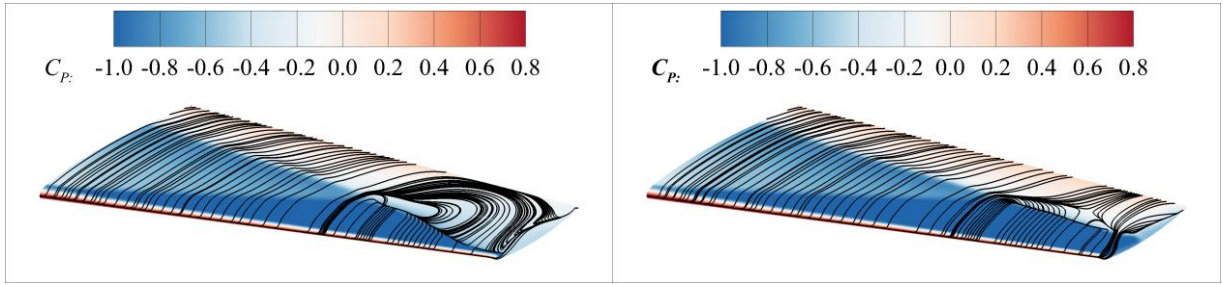

图 15 AoA=5.06º 时 M6 机翼表面流线结果，左图：SA 模型结果，右图：修正模型结果

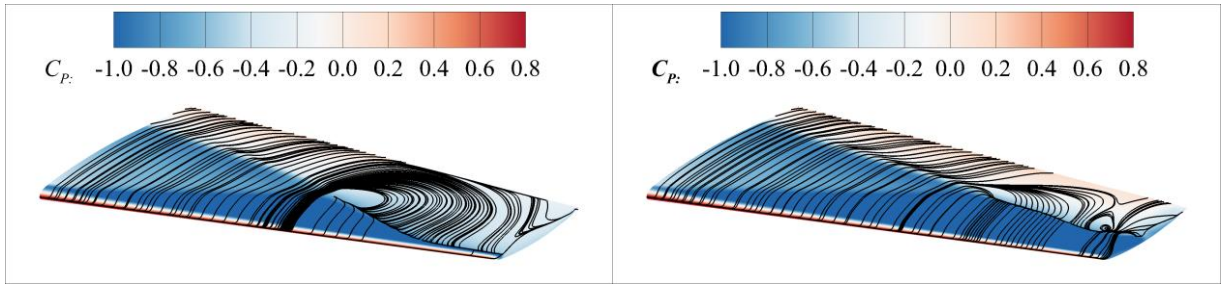

图 16 AoA=6.06º 时 M6 机翼表面流线结果，左图：SA 模型结果，右图：修正模型结果

由于原始实验文献[49]中没有给出具体的集中力系数结果，这里计算对比了不同来流状态下七个截面处的壁面压力系数相对误差，误差计算方式如下：

$$Error = \frac{\|\hat{y} - y_{exp}\|_2}{\|y_{exp}\|_2}$$

其中 $\hat{y}$ 表示计算值，$y_{exp}$ 表示实验值。误差计算结果如图 17 所示，相比于原始 SA 模型，修正模型在三个不同来流迎角状态上计算结果平均相对误差降低 40.89%，平均计算精度提高 1.69 倍。另一个角度来看，修正后的计算结果在三个迎角上均保持了同样的误差水平，对于分离流动的求解精度与附着流相当。而经典 SA 模型在分离流动上的计算误差相比附着流显著提升 3 倍左右。以上结果均表明所构建的修正模型对于三维流动具有较好的泛化能力，在激波诱导的三维复杂分离流动上也具有良好的修正能力。

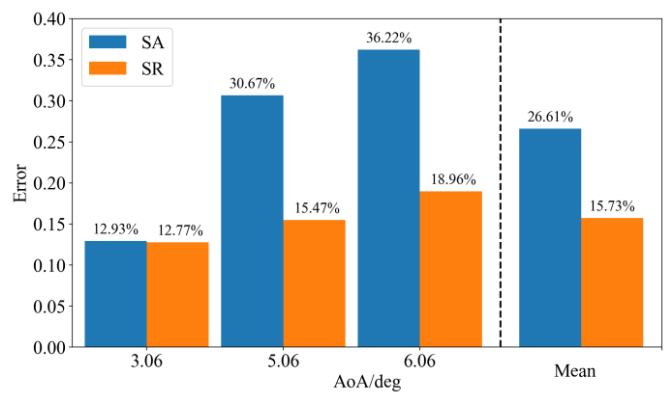

图 17 不同来流迎角下不同模型计算得到的壁面压力系数相对误差对比

## 4.3 DU91-W2-250 翼型

本小节给出修正模型在 DU91-W2-250 翼型上的耦合计算结果。由于修正模型的训练数据中包含了 DU91-W2-250 翼型在 Ma=0.15，AoA=15º，Re=3.0×10⁶ 状态下的同化数据，因此这里主要测试在不同来流状态上的泛化能力。

图 18 给出了包含训练来流状态下升力系数、阻力系数的结果对比，其中绿色圆圈标记了模型的训练数据集，即就是用于数据同化的状态。图 19 是对应状态下翼型失速迎角范围内的升力系数相对误差统计结果。结合图中的结果对比可以看到，在相同雷诺数与马赫数下，当前的方法可以仅靠对于单个状态流动数据的学习建模而实现对于整条升力系数曲线的高精度泛化，相比传统 SA 模型，修正模型的平均计算精度提高约 5.3 倍，表明了当前建模方法的有效性与泛化性。而对于小攻角分离流动，修正模型的结果则依然与 SA 模型保持一致，表明当前模型对于不同流动状态可以进行选择性激活，实现对于附着流和分离流的整体高精度模拟。图 20 与图 21 是在更高雷诺数上的计算结果对比，该结果依然支持本文已有的结论，即修正模型在附着流动上可以保持与 SA 模型相当的精度，同时在分离流动上的计算精度显著提升。

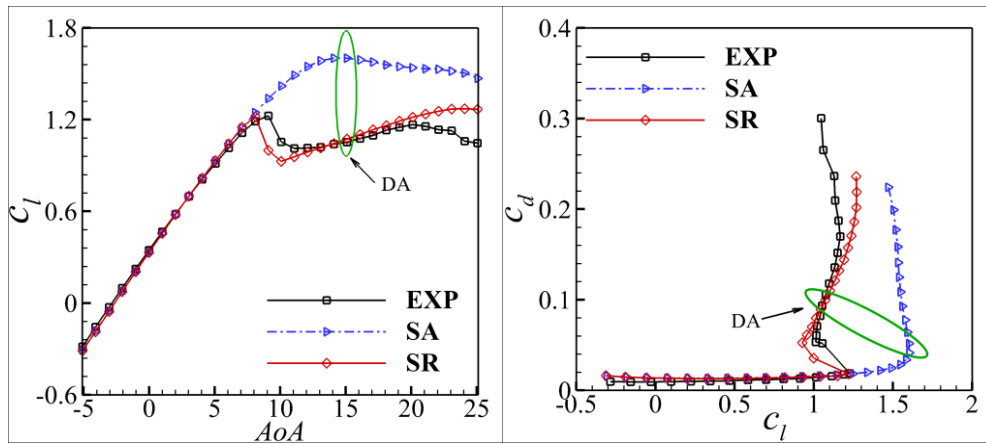

图 18 DU91-W2-250 翼型 Ma=0.15, Re=3.0×10⁶ 来流状态下升力系数、阻力系数结果对比

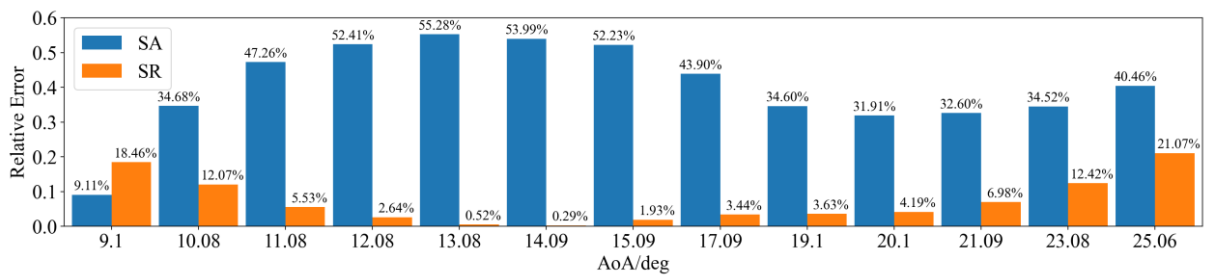

图 19 DU91-W2-2509 翼型 Ma=0.15, Re=3.0×10⁶ 来流状态下失速区域的升力系数相对误差对比

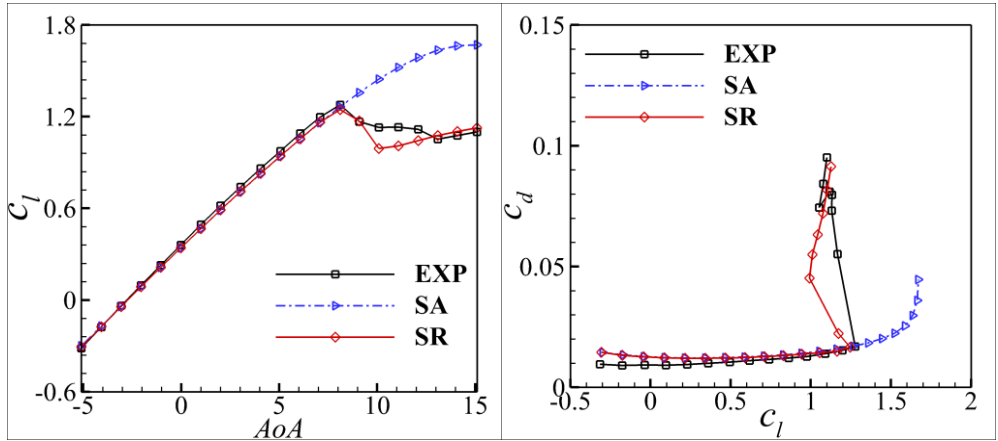

图 20 DU91-W2-250 翼型 Ma=0.15, Re=5.0×10$^6$ 来流状态下升力系数、阻力系数结果对比

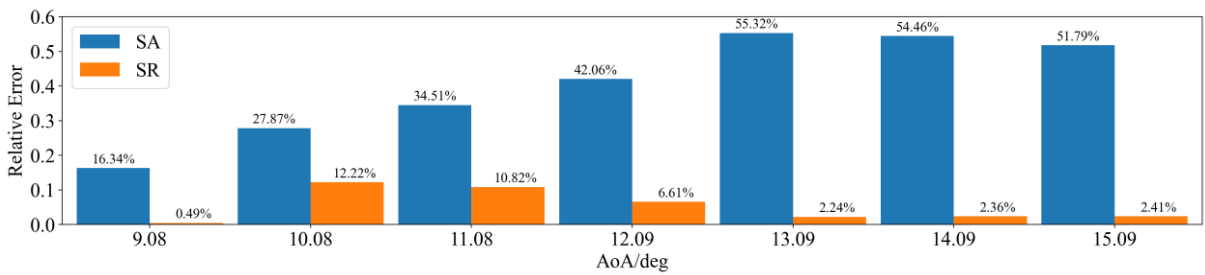

图 21 DU91-W2-2509 翼型 Ma=0.15, Re=5.0×10$^6$ 来流状态下失速区域的升力系数相对误差对比

图 22 中虚线左侧两列对比了不同雷诺数下升力系数的相对误差，虚线右侧则对比了所计算的三个来流状态下升力系数的平均相对误差。在所有计算的状态下，修正模型计算精度均有大幅提升。整体而言，相比于原始 SA 模型，修正之后的模型计算结果相对误差降低了 84.00%，计算精度提升至原来的 6.25 倍。图 23 DU91-W2-250 翼型 Ma=0.2, Re=3.0×10$^6$ 来流状态下不同迎角对应的壁面压力系数结果对比

给出了 Ma=0.2, Re=3.0×10$^6$ 来流状态下不同迎角对应的壁面压力系数分布结果对比，在整个失速迎角区域修正模型的结果均与实验值更加接近，偏差更小，与实验结果的分离位置更为接近，也与力系数的结果互为印证，表明了当前修正模型对于翼型大分离流动模拟的有效性。

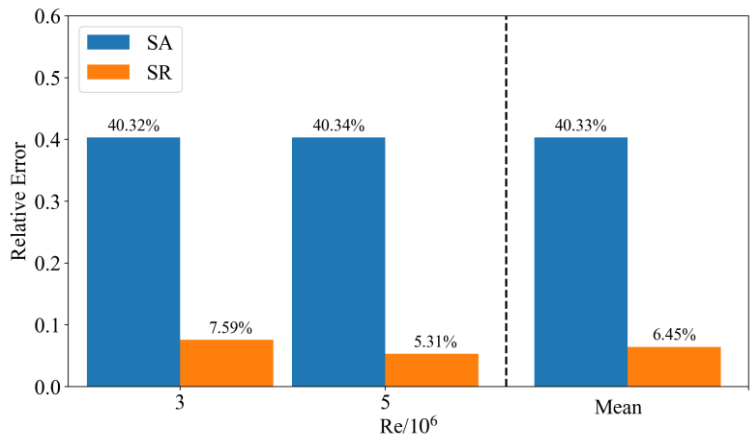

图 22 DU91-W2-250 翼型不同雷诺数下失速区域平均升力系数相对误差对比

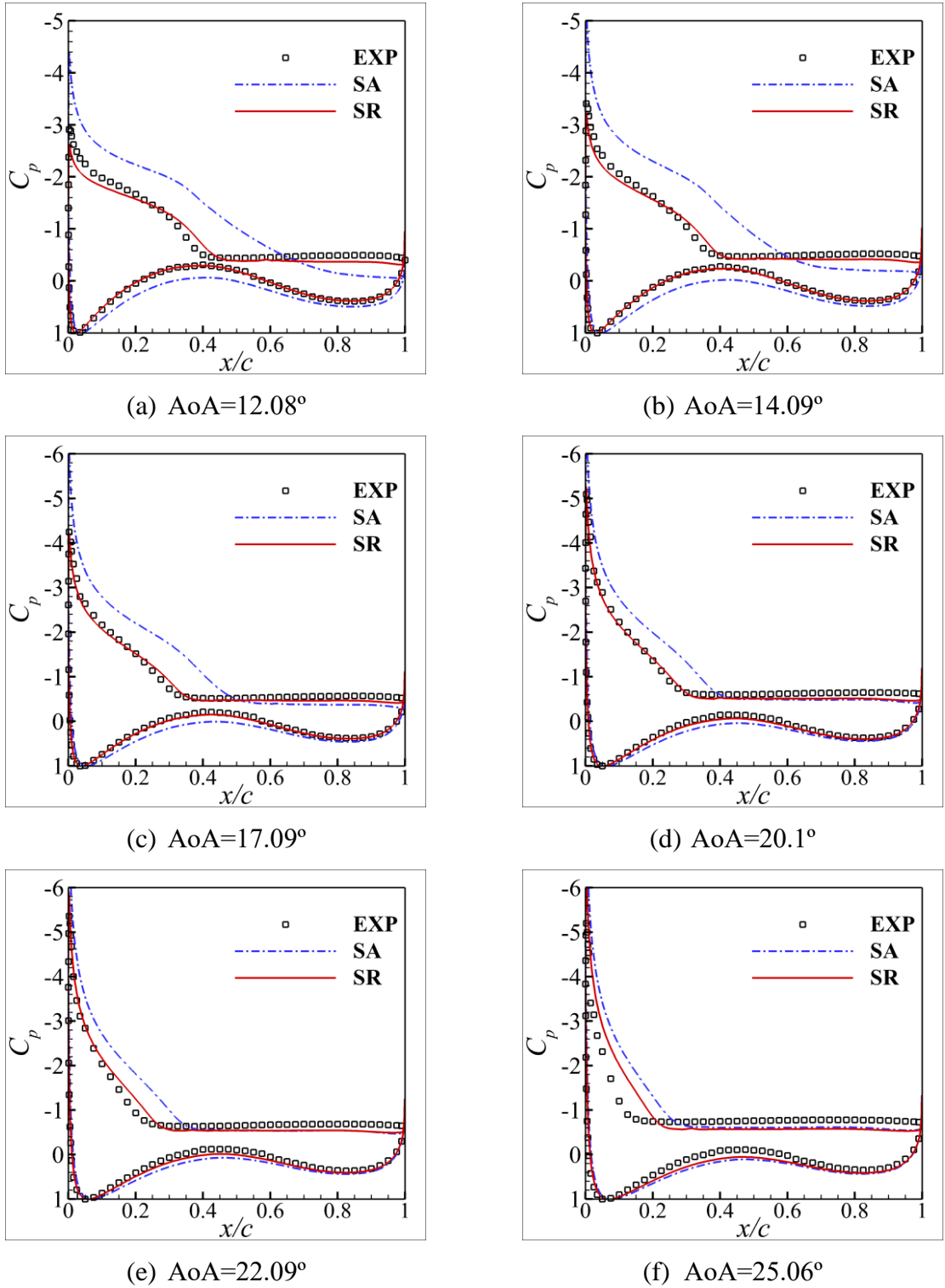

(a) AoA=12.08º  (b) AoA=14.09º
(c) AoA=17.09º  (d) AoA=20.1º
(e) AoA=22.09º  (f) AoA=25.06º

图 23 DU91-W2-250 翼型 Ma=0.2, Re=3.0×$10^6$ 来流状态下不同迎角对应的壁面压力系数结果对比

## 4.4 S809 翼型

本小节主要给出基于符号回归的修正模型在 S809 翼型[50]上的计算结果。对比了五个典型工况下数值模拟与试验测量得到的升力系数曲线与升阻极曲线以及在翼型失速

迎角下的升力系数相对误差，分别如图 24 至图 34 所示。在所有的测试算例中，相比于 SA 模型，修正模型在大迎角分离区的计算精度均有显著提升，而在附着流均与 SA 模型保持了一致。不同雷诺数下失速区域升力系数的平均相对误差对比见图 34，每个雷诺数状态下修正模型计算得到的升力系数平均相对误差相比于原始 SA 模型计算结果均降低了 55%以上。图 34 中最右边两列统计对比了当前所有算例计算结果的平均相对误差，相比于原始 SA 模型，修正之后的模型计算结果相对误差降低了 66.79%，计算精度提升至原来的 3.01 倍。表明了本文所开发的修正模型在变外形、变工况状态下的泛化性能。在 Re=2.0×10⁶ 情况下不同来流迎角对应的壁面压力系数分布如图 35 所示。由图中的结果对比可以看到，在迎角大于 10.2º 的部分修正模型较原始 SA 模型结果均有明显提升，壁面压力系数计算结果与实验测量结果也更为符合。对于 AoA=8.2º 的附着流状态两种模型结果均与实验结果相近。

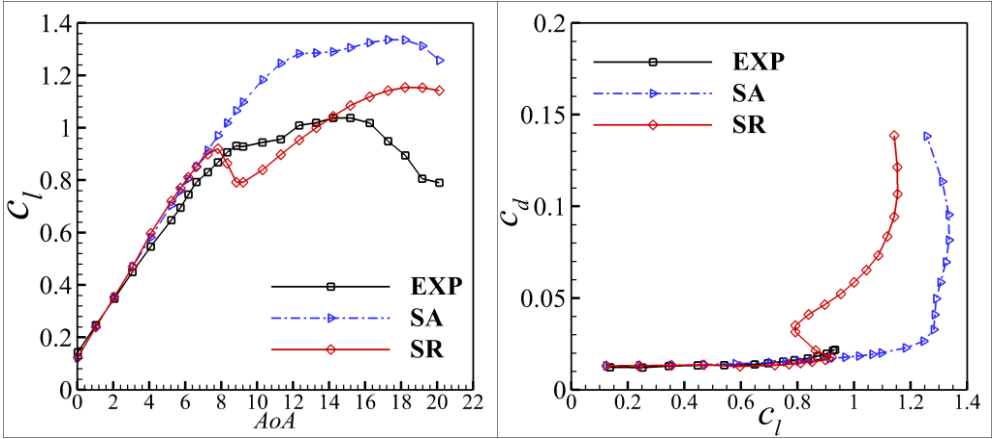

图 24 S809 翼型 Ma=0.15, Re=1.0×10⁶ 来流状态下升力系数、阻力系数结果对比

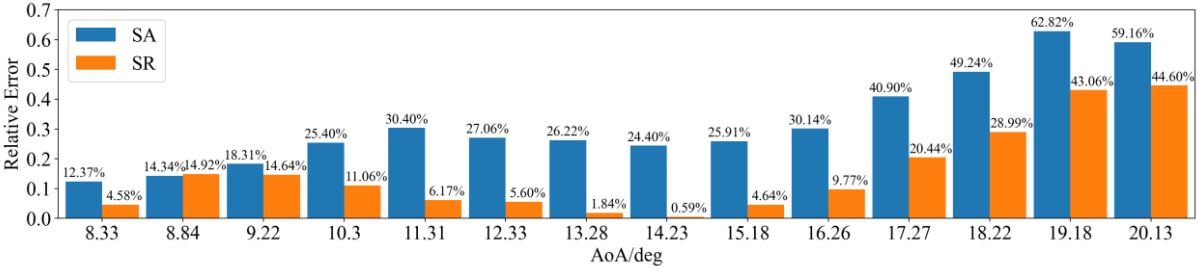

图 25 S809 翼型 Ma=0.15 Re=1.0×10⁶ 来流状态下失速区域的升力系数相对误差对比

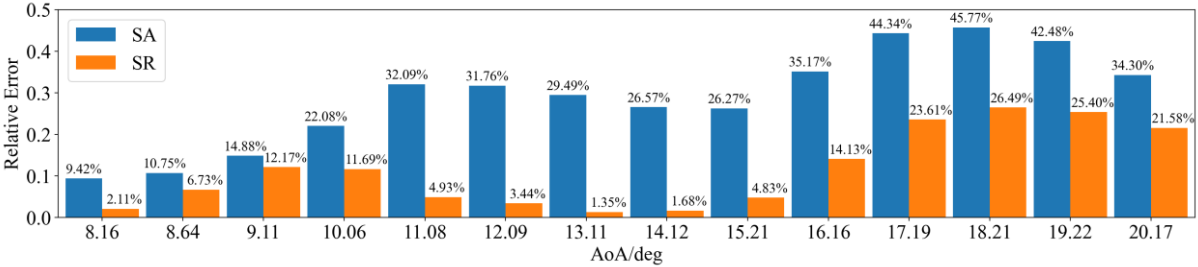

图 26 S809 翼型 Ma=0.15, Re=1.5×10⁶ 来流状态下失速区域的升力系数相对误差对比

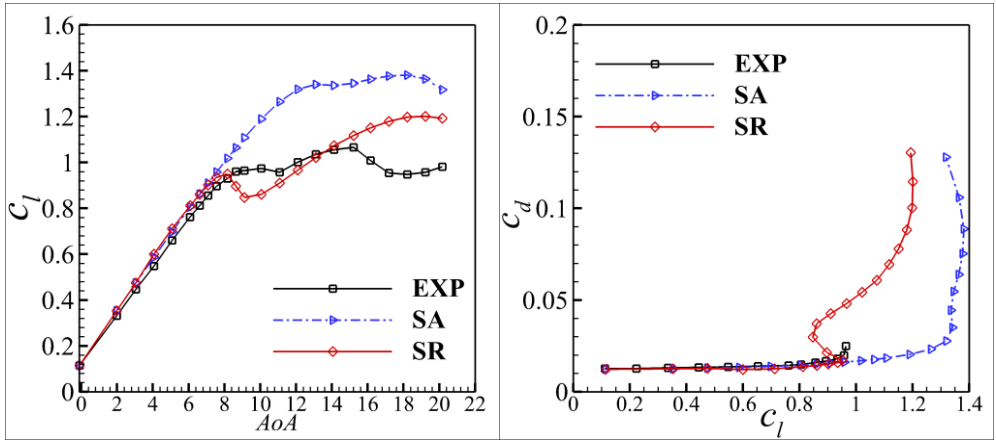

图 27 S809 翼型 Ma=0.15, Re=1.5×10$^6$ 来流状态下升力系数、阻力系数结果对比

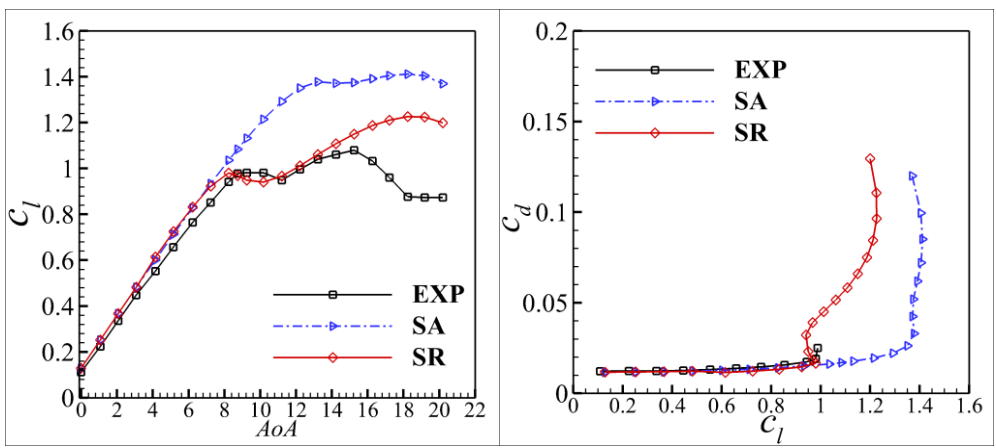

图 28 S809 翼型 Ma=0.15, Re=2.0×10$^6$ 来流状态下升力系数、阻力系数结果对比

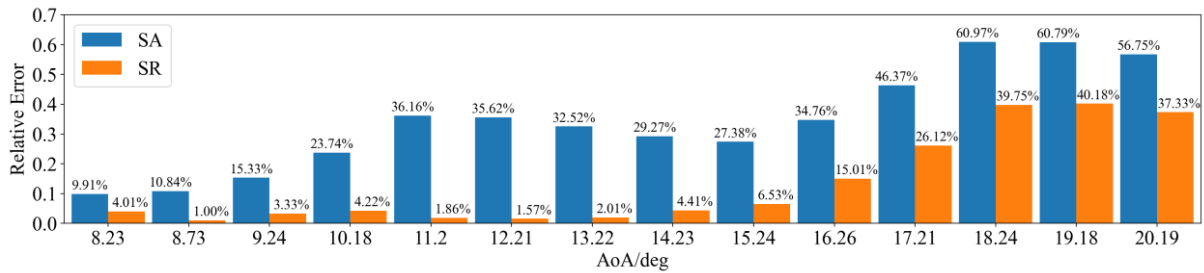

图 29 S809 翼型 Ma=0.15, Re=2.0×10$^6$ 来流状态下失速区域的升力系数相对误差对比

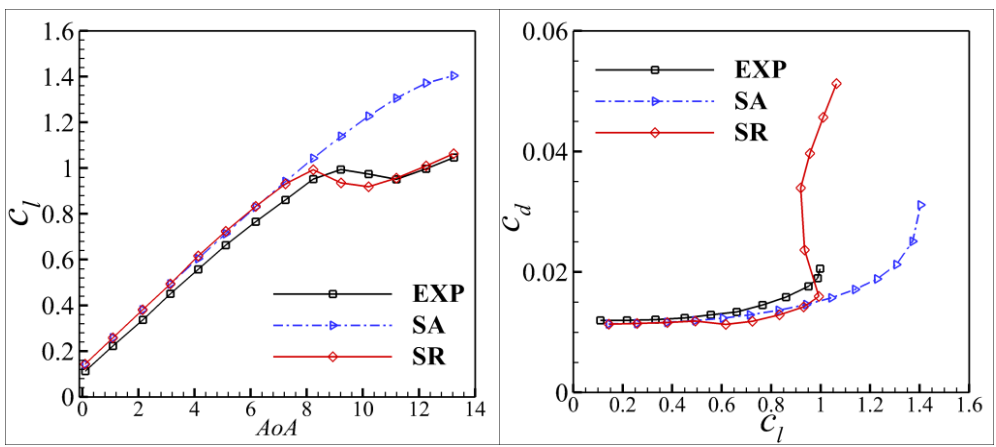

图 30 S809 翼型 Ma=0.15, Re=2.5×10$^6$ 来流状态下失速区域的升力系数相对误差对比

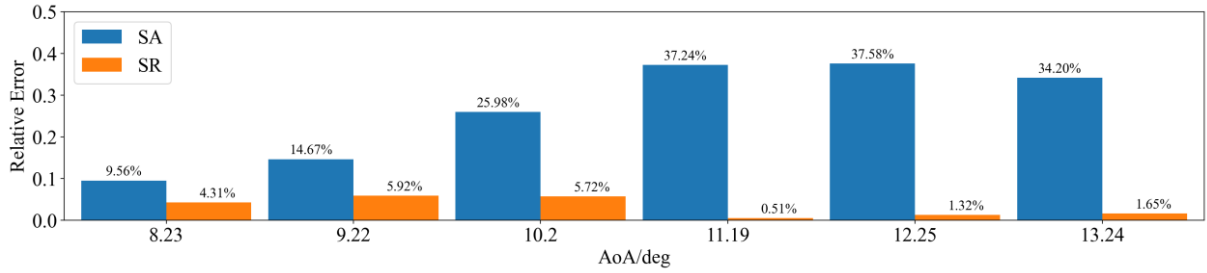

图 31 S809 翼型 Ma=0.15, Re=2.5×10$^6$ 来流状态下失速区域的升力系数相对误差对比

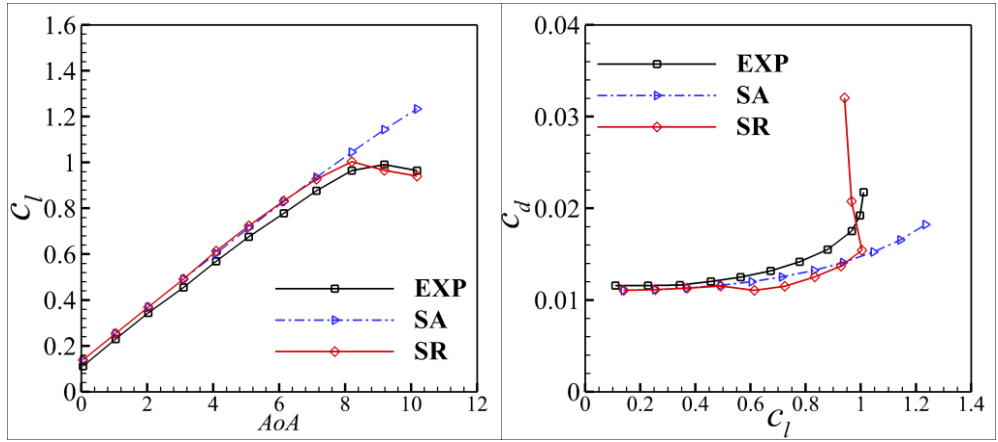

图 32 S809 翼型 Ma=0.15, Re=3.0×10$^6$ 来流状态下失速区域的升力系数相对误差对比

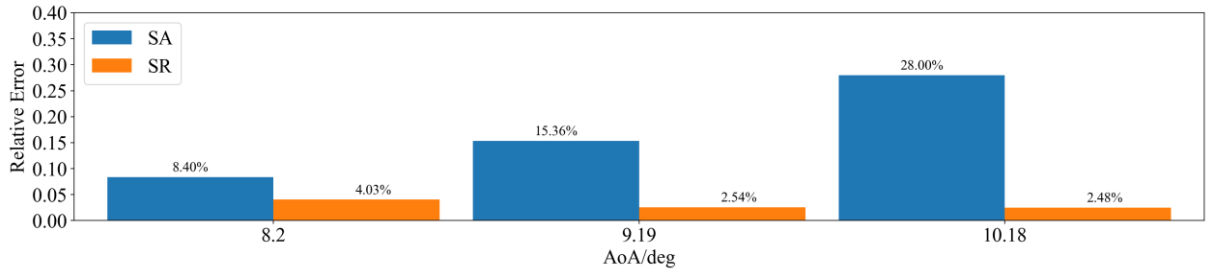

图 33 S809 翼型 Ma=0.15, Re=3.0×10$^6$ 来流状态下失速区域的升力系数相对误差对比

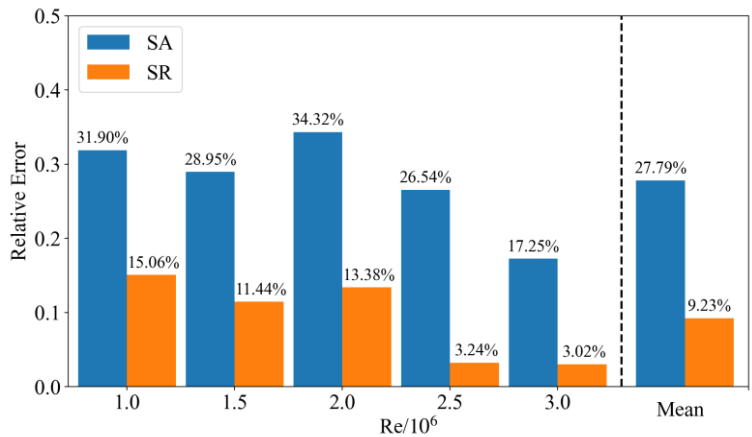

图 34 S809 翼型不同雷诺数下失速区域平均升力系数相对误差对比

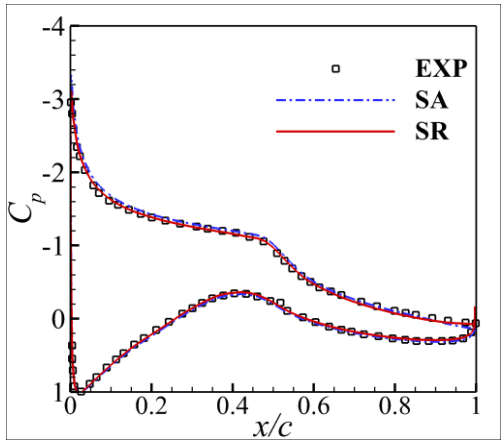
(a) AoA=8.2º

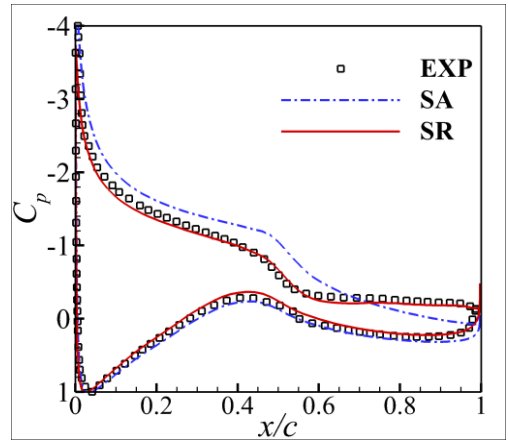
(b) AoA=10.2º

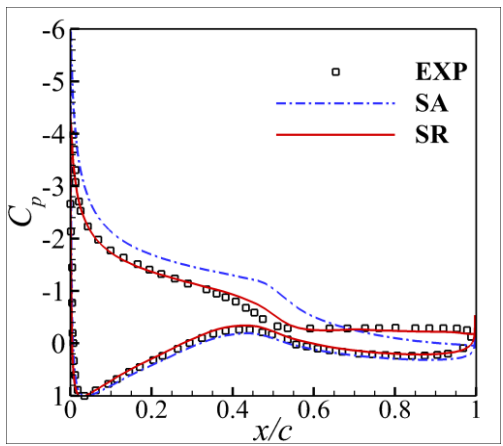
(c) AoA=11.2º

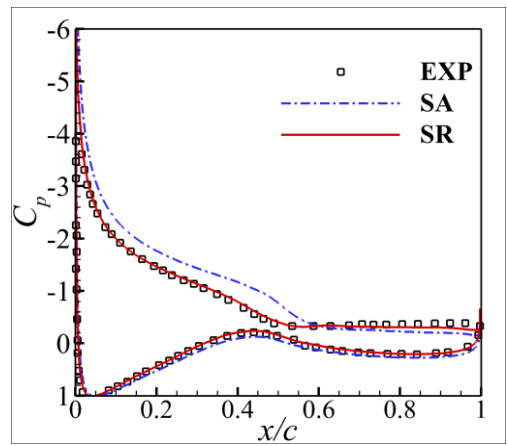
(d) AoA=14.24º

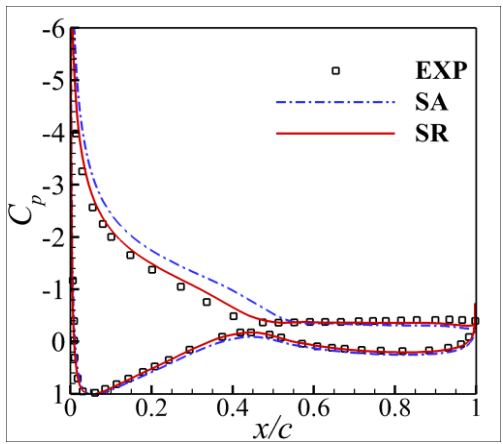
(e) AoA=16.24º

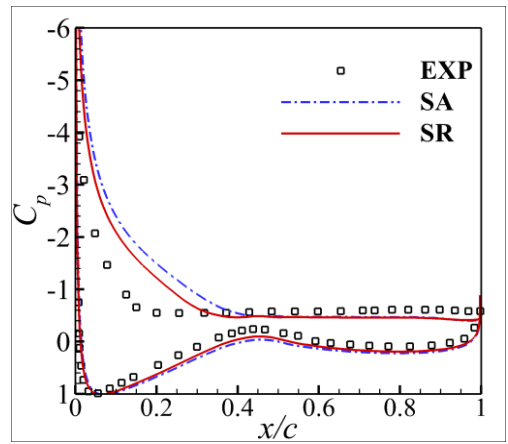
(f) AoA=20.15º

图 35 S809 翼型 Ma=0.15, Re=2.0×10$^6$ 来流状态下不同迎角对应的壁面压力系数结果对比

## 4.5 S814 翼型

本小节展示了基于符号回归的修正模型在 S814 翼型[51]上的计算结果。重点对比了四个典型工况下数值模拟与实验测量得到的升力系数曲线、升阻极曲线，以及翼型失速

区域升力系数的相对误差，分别如图 36 至图 45 所示。结果表明，相较于原始 SA 模型，修正模型在大迎角分离流区域的计算精度显著提高，而在附着流区域与 SA 模型保持了一致性。不同雷诺数下失速区域升力系数的平均相对误差对比见图 46，显示了不同雷诺数下失速区域的升力系数平均相对误差对比。可以看出，在各雷诺数条件下，修正模型的升力系数平均相对误差相较原始 SA 模型降低了 55%以上。进一步统计所有雷诺数的总体平均相对误差（图 46 右两列）修正模型较原始 SA 模型的相对误差降低了 65.84%，计算精度提升至原来的 2.92 倍。综上结果，基于符号回归的修正模型在变外形和变工况条件下表现出了优异的泛化能力，显著提升了对大迎角复杂流动的模拟精度。

此外，对 Re=1.5×10$^6$ 下不同来流迎角的壁面压力系数分布进行了对比分析，如图 47 所示。修正模型的计算结果在失速迎角之后的升力系数预测上与实验数据更加吻合，在大迎角条件下的壁面压力系数分布也与实验测量结果更为一致。

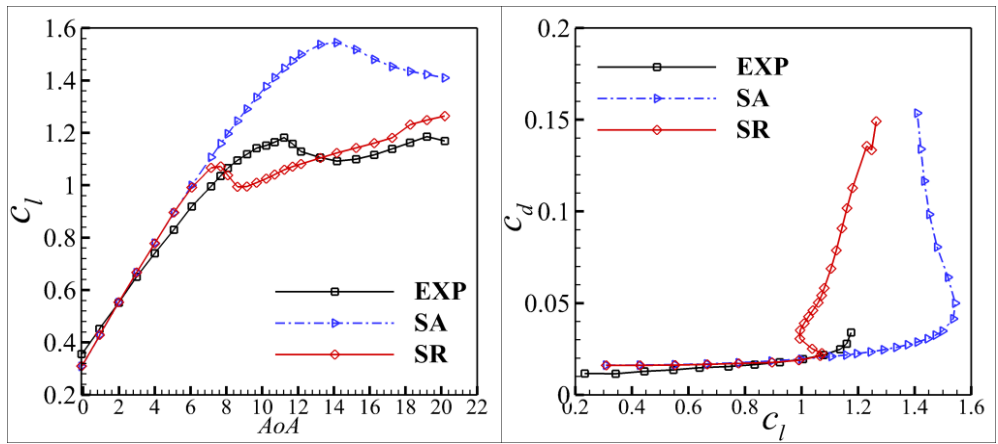

图 36 S814 翼型 Ma=0.15, Re=0.7×10$^6$ 来流状态下升力系数、阻力系数结果对比

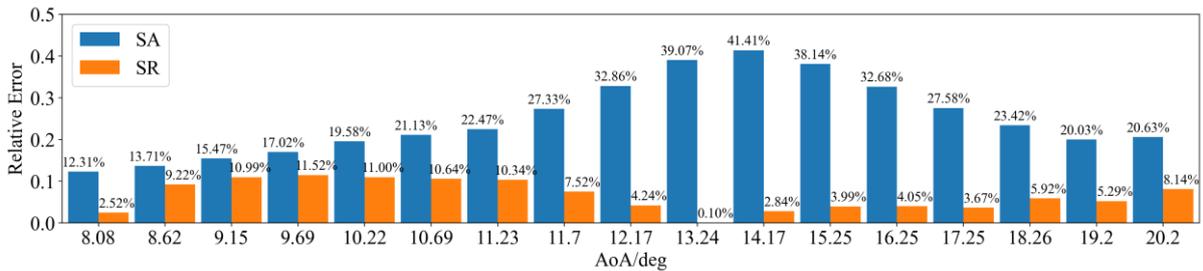

图 37 S814 翼型 Ma=0.15, Re=0.7×10$^6$ 来流状态下失速区域的升力系数相对误差对比

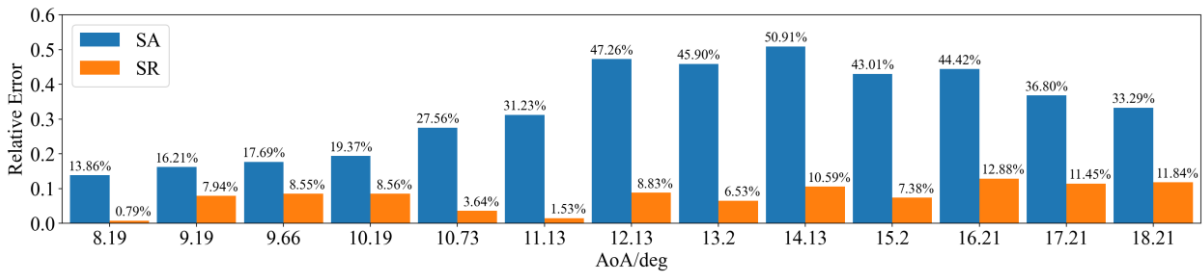

图 38 S814 翼型 Ma=0.15, Re=1.0×10$^6$ 来流状态下失速区域的升力系数相对误差对比

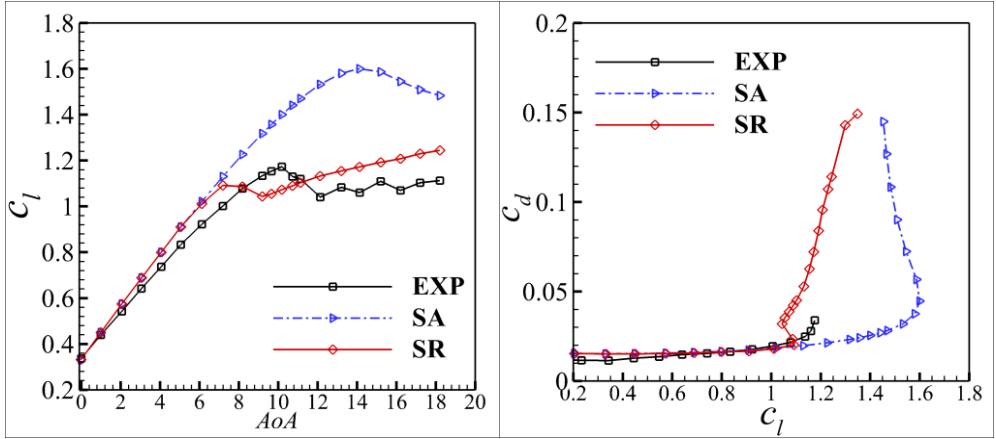

图 39 S814 翼型 Ma=0.15, Re=1.0×10$^6$ 来流状态下升力系数、阻力系数结果对比

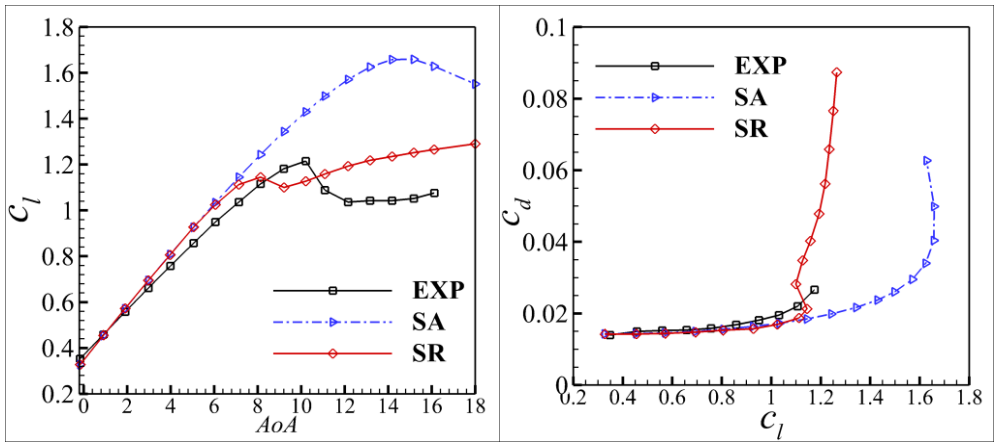

图 40 S814 翼型 Ma=0.15, Re=1.5×10$^6$ 来流状态下升力系数、阻力系数结果对比

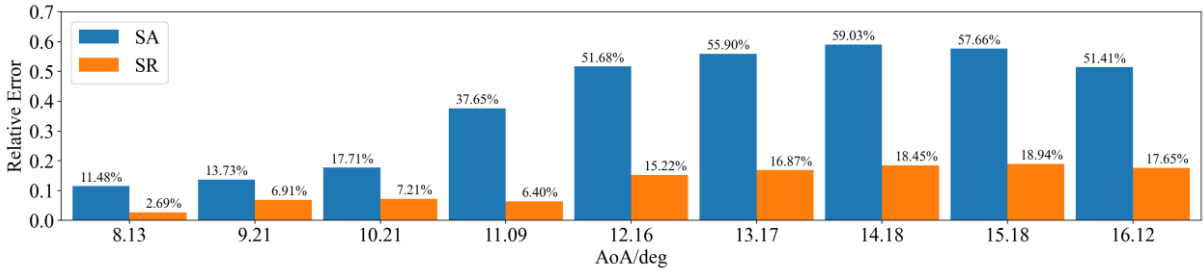

图 41 S814 翼型 Ma=0.15, Re=1.5×10$^6$ 来流状态下失速区域的升力系数相对误差对比

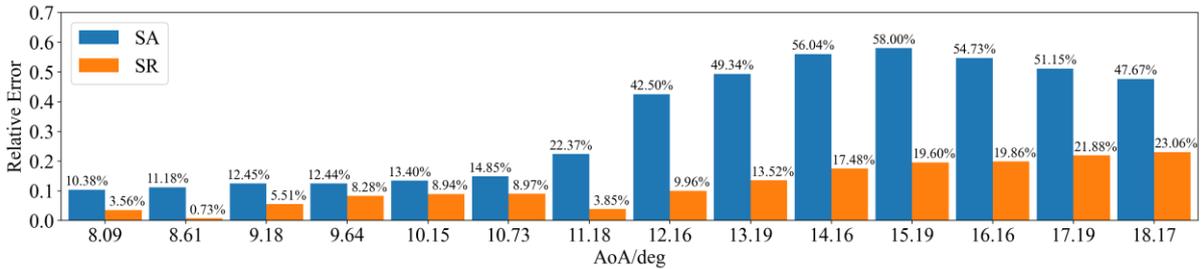

图 42 S814 翼型 Ma=0.15, Re=2.0×10$^6$ 来流状态下失速区域的升力系数相对误差对比

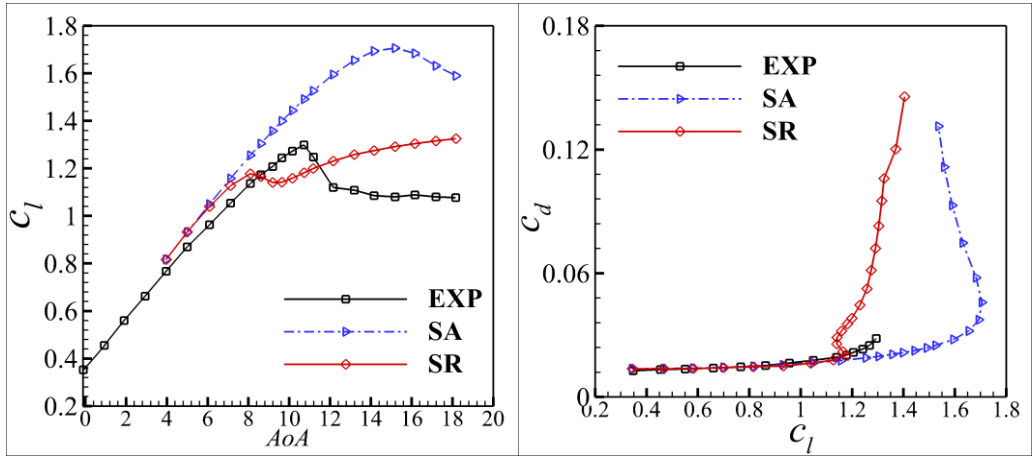

图 43 S814 翼型 Ma=0.15, Re=2.0×10$^6$ 来流状态下升力系数、阻力系数结果对比

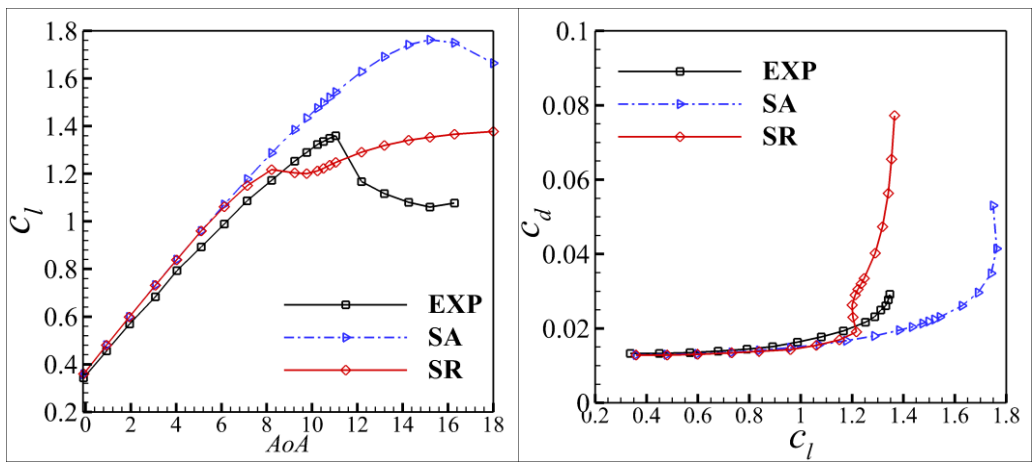

图 44 S814 翼型 Ma=0.15, Re=3.0×10$^6$ 来流状态下升力系数、阻力系数结果对比

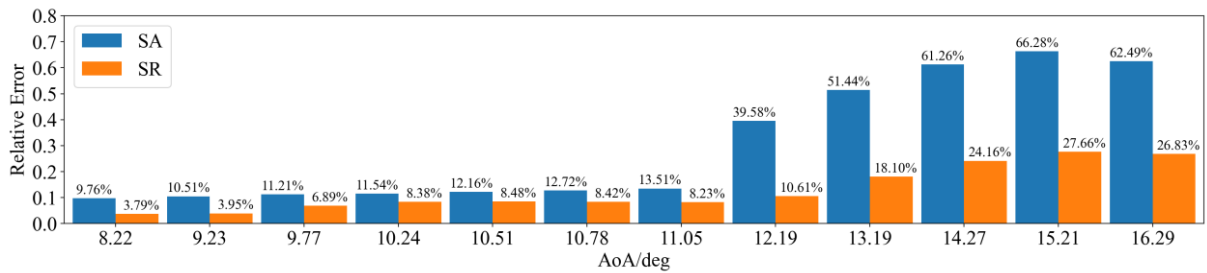

图 45 S814 翼型 Ma=0.15, Re=3.0×10$^6$ 来流状态下失速区域的升力系数相对误差对比

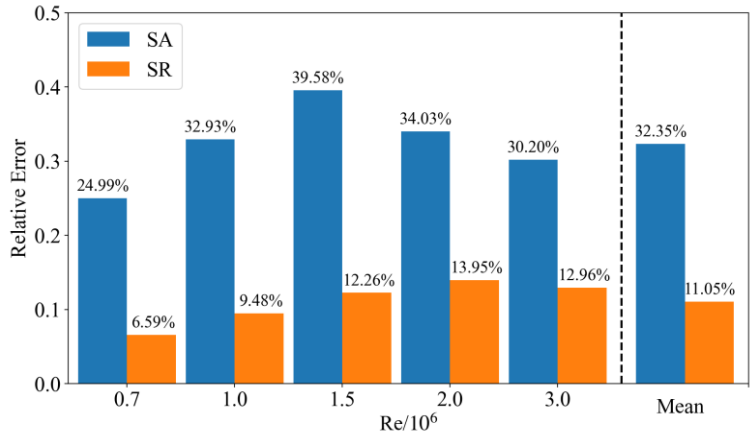

图 46 S814 翼型不同雷诺数下失速区域平均升力系数相对误差对比

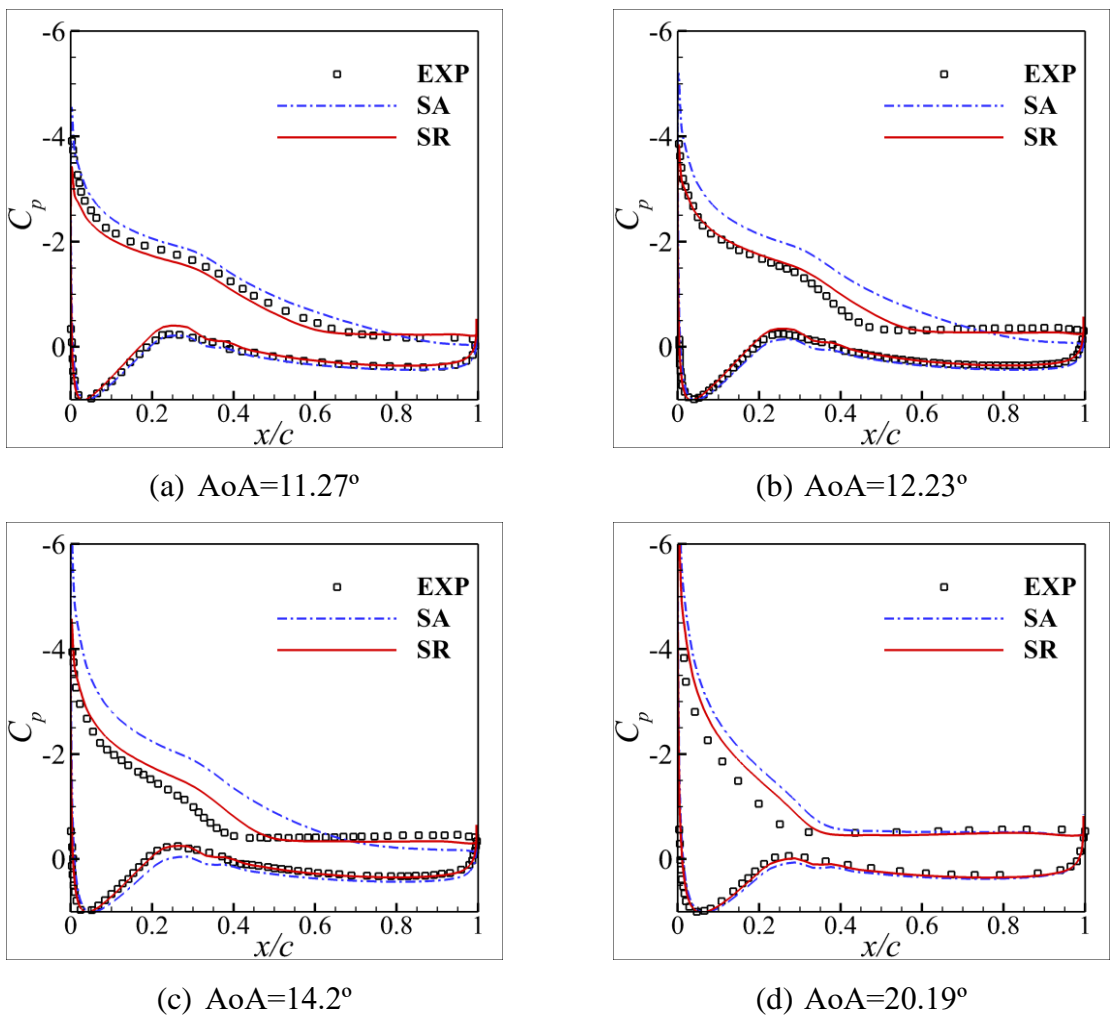

(a) AoA=11.27º  (b) AoA=12.23º

(c) AoA=14.2º  (d) AoA=20.19º

图 47 S814 翼型 Ma=0.15, Re=1.5×10$^6$ 来流状态下不同迎角对应的壁面压力系数结果对比

### 4.6 S805 翼型

本小节展示了基于符号回归的修正模型在 S805 翼型[52]上的计算结果，首先，对三个典型工况下数值模拟与试验测量得到的升力系数曲线、升阻极曲线，以及翼型失速迎角处的升力系数相对误差进行了详细对比，分别如图 48 至图 53 所示。从对比结果可

以看出，修正模型在三个典型工况下的升力系数曲线和升阻极曲线与实验数据的吻合程度明显优于原始 SA 模型。特别是在大迎角分离区域，修正模型能够更准确地捕捉升力系数和升阻极曲线的变化趋势，与实验测量结果表现出更一致的趋势走向，显著减少了预测误差。此外，对于失速迎角附近的升力系数，修正模型能够更好地反映实验数据中的非线性变化特征，进一步验证了其在复杂流动条件下的预测能力。

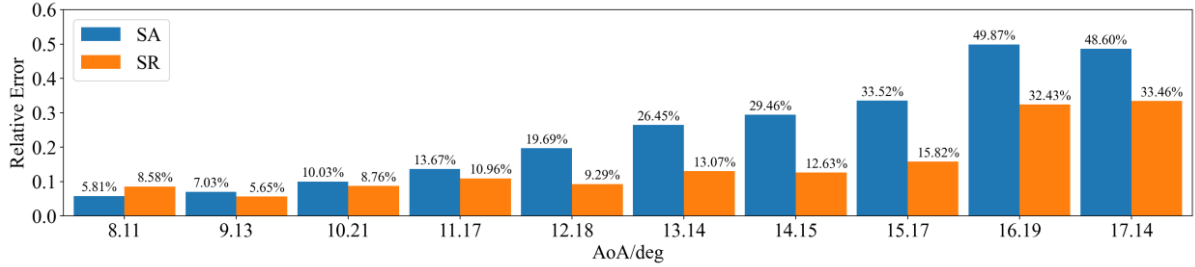

图 48 S805 翼型 Ma=0.15, Re=1.0×10$^6$ 来流状态下失速区域的升力系数相对误差对比

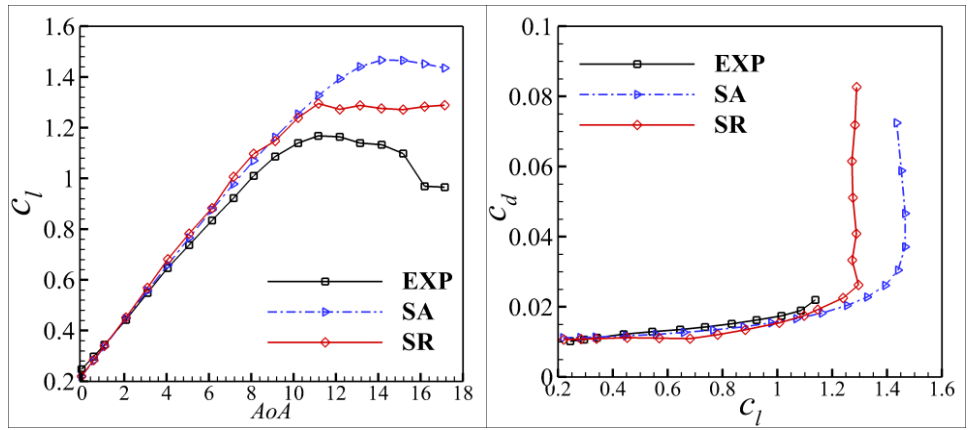

图 49 S805 翼型 Ma=0.15, Re=1.0×10$^6$ 来流状态下升力系数、阻力系数结果对比

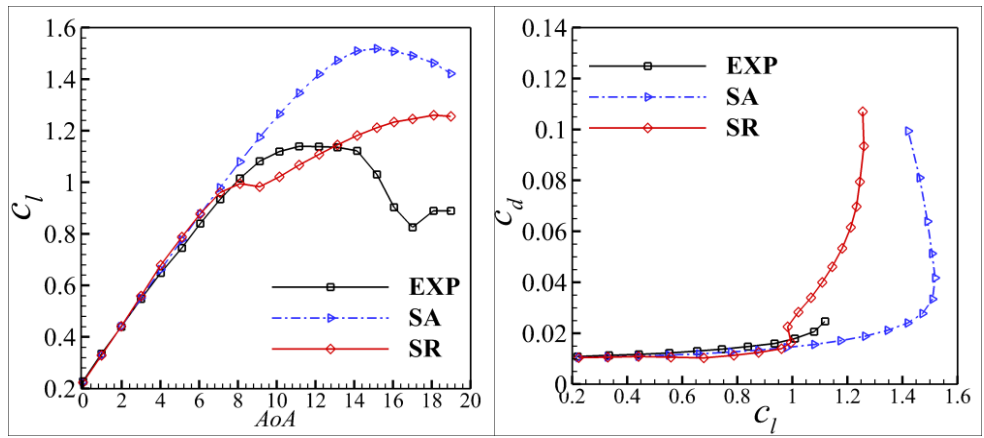

图 50 S805 翼型 Ma=0.15, Re=1.5×10$^6$ 来流状态下升力系数、阻力系数结果对比

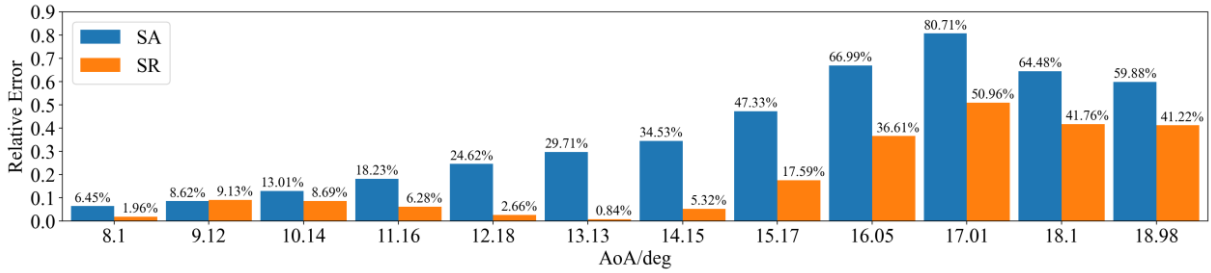

图 51 S805 翼型 Ma=0.15, Re=1.5×10⁶ 来流状态下失速区域的升力系数相对误差对比

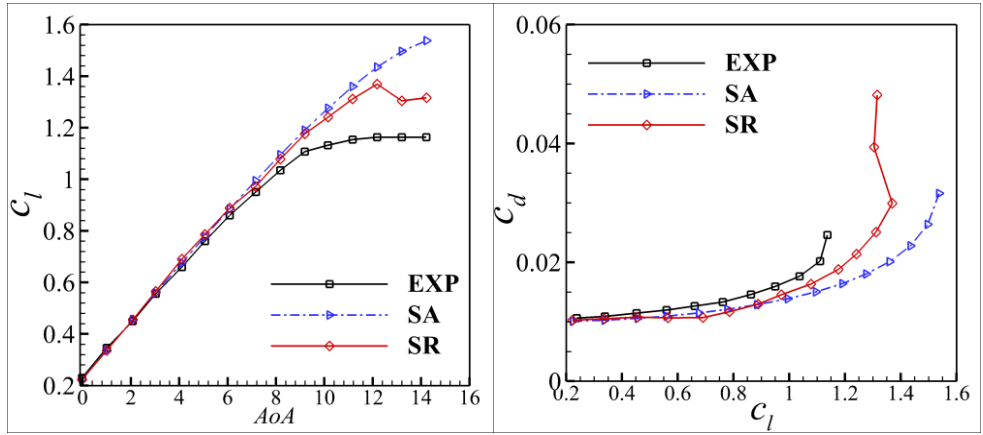

图 52 S805 翼型 Ma=0.15, Re=2.0×10⁶ 来流状态下升力系数、阻力系数结果对比

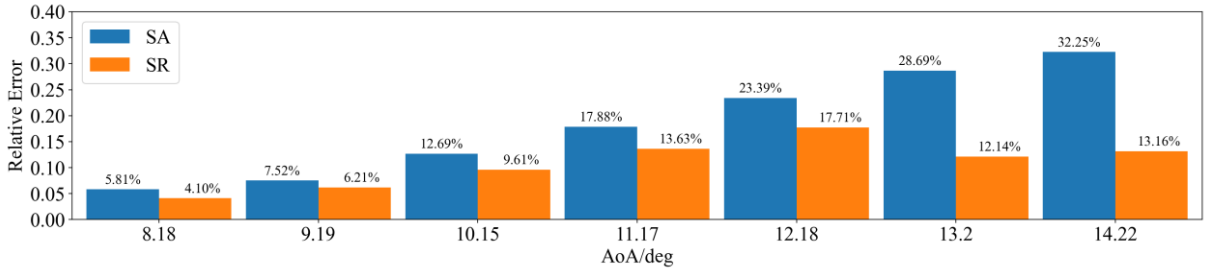

图 53 S805 翼型 Ma=0.15, Re=2.0×10⁶ 来流状态下失速区域的升力系数相对误差对比

进一步分析了 Re=1.0×10⁶ 工况下不同来流迎角对应的壁面压力系数分布，如图 54 所示。从结果来看，修正模型在大迎角条件下的壁面压力系数分布与实验数据更为接近，显著减少了原始 SA 模型预测中出现的偏差。特别是在失速迎角之后，修正模型能够更准确地捕捉到压力梯度的变化趋势，尤其是在翼型吸力面关键区域的压力分布上展现出更优的匹配性。相比之下，原始 SA 模型在这些条件下的预测偏离实验数据较为显著，难以正确反映分离流动对壁面压力的影响。修正模型的这种改进进一步验证了其在大迎角分离流动条件下的预测能力和准确性。

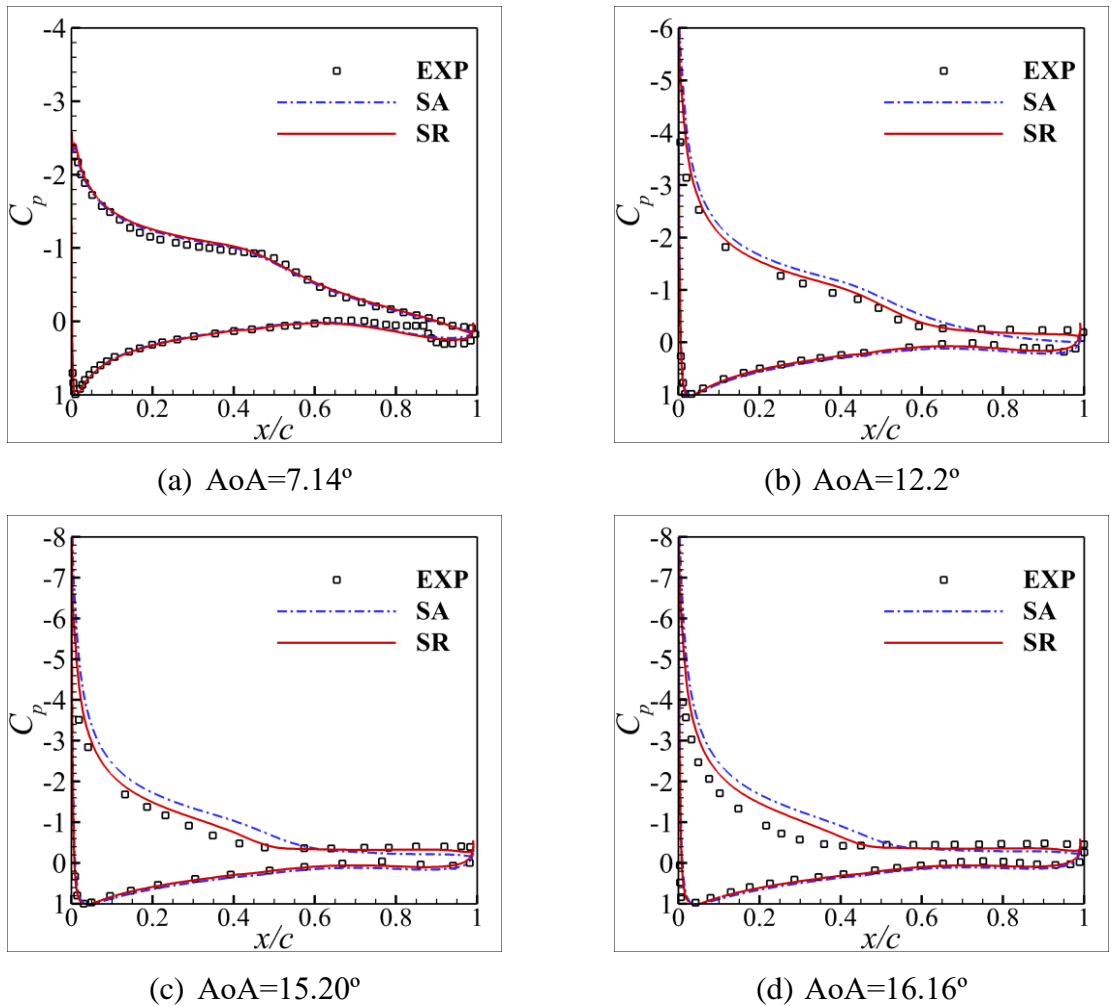

(a) AoA=7.14º  (b) AoA=12.2º

(c) AoA=15.20º  (d) AoA=16.16º

图 54 S805 翼型 Ma=0.15, Re=1.0×10⁶ 来流状态下不同迎角对应的壁面压力系数结果对比

此外，不同雷诺数下失速区域升力系数的平均相对误差对比结果如图 55 所示。可以看到，修正模型在各个雷诺数工况下的平均相对误差相比原始 SA 模型均显著降低。例如，在统计所有雷诺数工况的总体平均相对误差时，修正模型相较于原始 SA 模型的误差减少了 44.66%，计算精度提升至原来的 1.81 倍（图 55 最右两列）。这一结果表明，修正模型不仅能够在单一工况下显著改进预测精度，还具较强的泛化能力。

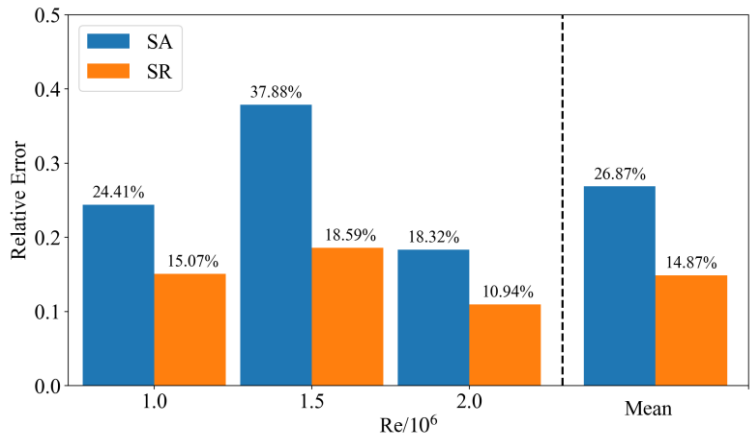

图 55 S805 翼型不同雷诺数下失速区域平均升力系数相对误差对比

综上所述，基于符号回归的修正模型在变外形和变工况条件下表现出了优异的预测能力和泛化性能，尤其在高雷诺数分离流动的模拟中展现了显著的潜力。这为进一步改进 RANS 湍流模型并提升其工程适用性提供了重要依据。

## 4.7 曲线后台阶和驼峰

本节给出所提出的修正模型在二维曲线后台阶（Curved Backward-Facing Step, CBFS）和驼峰（hump）算例上的测试结果。这两个算例均是由于几何外形的突然变化引起了流动分离，其分离点一般是固定的，RANS 方法难以准确预测流动再附点，这一点与前面所测试的翼型、机翼算例有所不同。本小节的目的在于验证所提出的模型在这类分离算例上的泛化能力。图 56 和图 57 分别给出了 CBFS 和 hump 算例的外形及网格示意图，其中 hump 算例的来流马赫数为 0.1，雷诺数为 $9.36\times10^5$；CBFS 算例的来流马赫数为 0.01，雷诺数为 $1.37\times10^4$。

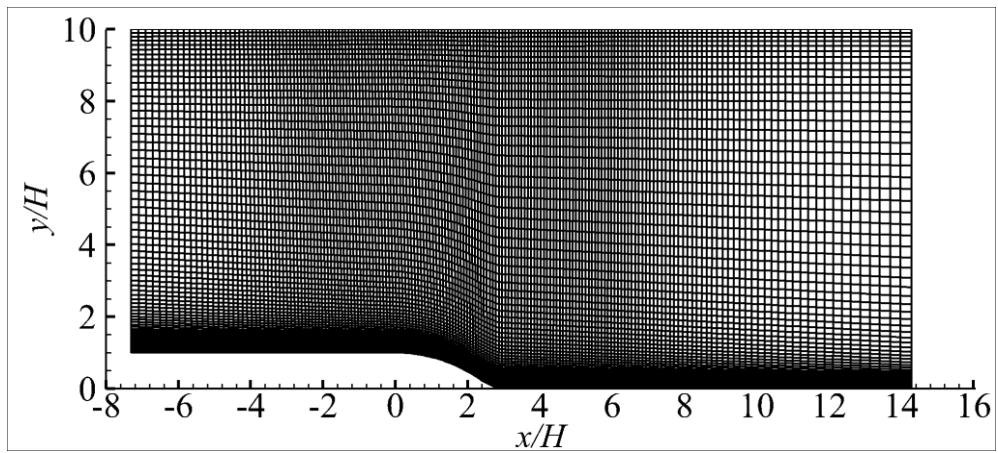

图 56 CBFS 算例外形及计算网格示意图

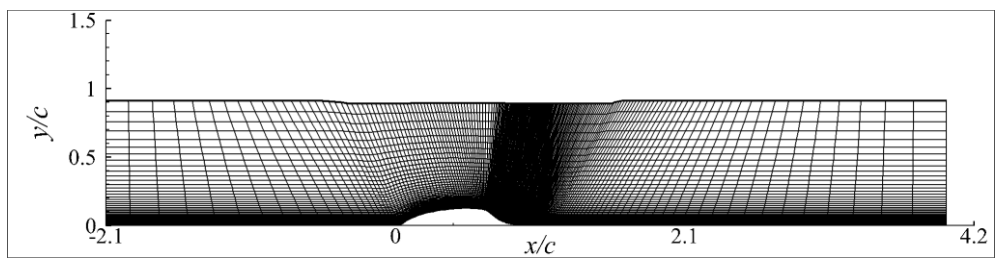

图 57 hump 算例外形及计算网格示意图

图 58 和图 59 分别给出了 CBFS 和 hump 两个算例中不同模型对壁面压力系数和摩擦阻力系数的计算结果对比。从图中可以看出，对于这类流动，修正模型的计算结果与原始 SA 模型基本一致，虽然未有显著的改善，但结果也未出现明显的劣化。这主要归因于这两个算例的流动分离机制与翼型流动存在显著差异，而修正模型的训练数据中未包含此类流动数据，导致其难以针对这类流动进行有效修正。这表明，针对不同分离流动类型的统一建模仍有提升空间。未来的研究可尝试将多种分离流动类型纳入训练数据集，以构建具有更强泛化能力的模型，这也是本文工作的重要拓展方向之一。

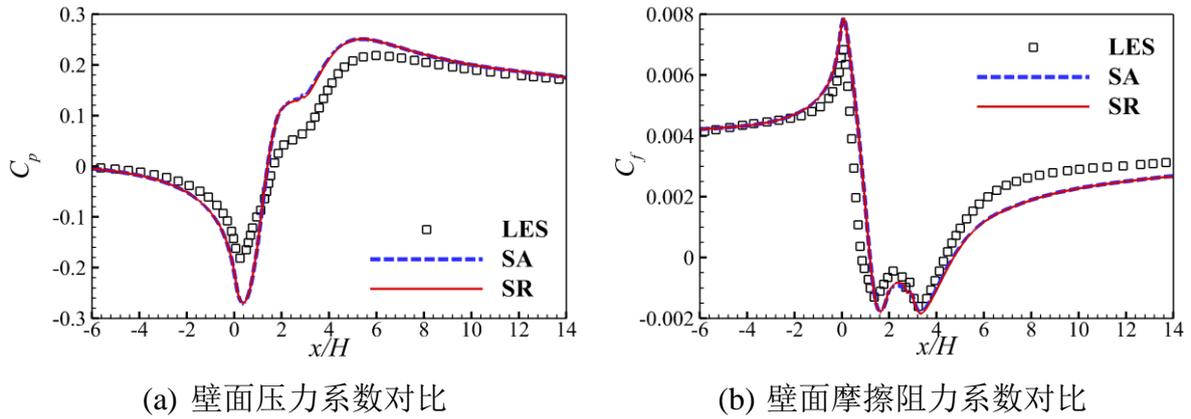

(a) 壁面压力系数对比  (b) 壁面摩擦阻力系数对比

图 58 CBFS 算例壁面压力系数与摩擦阻力系数结果对比

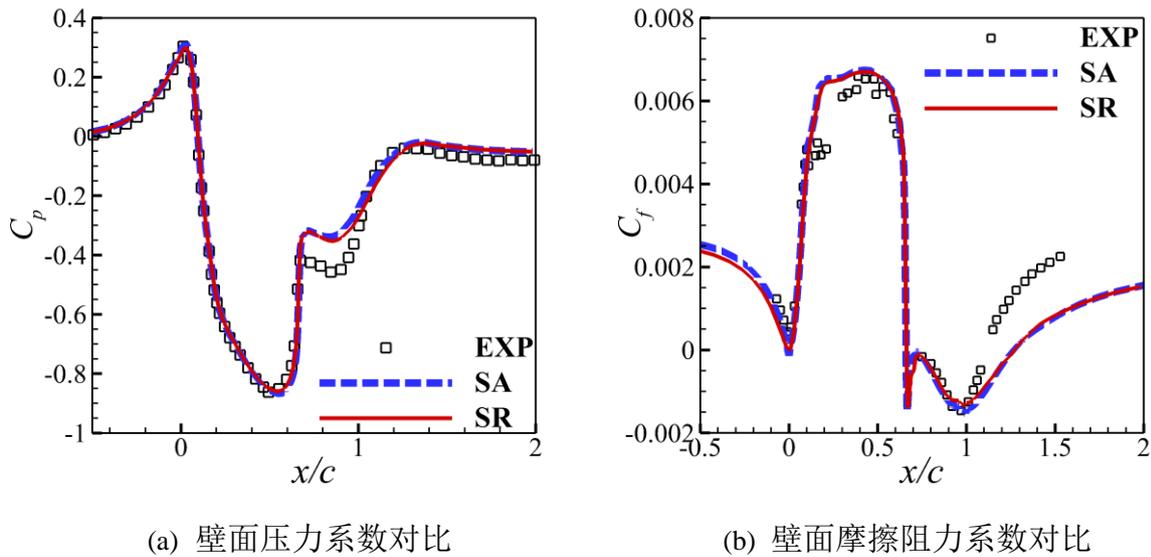

(a) 壁面压力系数对比  (b) 壁面摩擦阻力系数对比

图 59 hump 算例壁面压力系数与摩擦阻力系数结果对比

### 4.8 零压力梯度平板

经典 SA 模型的参数是在简单流动上进行校准的，比如零压力梯度平板，这保证了 SA 模型对于此类附着流动的良好预测能力。而对于湍流模型的修正可能会导致模型在这类问题上的不适定，因此有必要考察修正模型在附着流动上的表现。因为本文提出的对于湍流模型的修正作用于湍流模型生成项，影响了涡粘分布，最终会影响到速度型的分布，因此本小节以零压力梯度平板为例，验证修正模型对于这类简单附着流动的预测能力。

图 60 对比了在零压力梯度平板算例上不同模型的计算结果，包括摩擦阻力系数分布（左图）以及不同站位处的速度型分布（右图）。图中的结果对比表明修正模型保持了预测湍流边界层速度剖面的能力。两个不同流向位置的速度型剖面与基准 SA 模型吻合良好，如右图所示。此外，由左图也可以看出，修正模型对平板摩擦系数的预测也与基准 SA 模型结果非常吻合，说明修正模型保持了基线 SA 模型对附着流的精度。

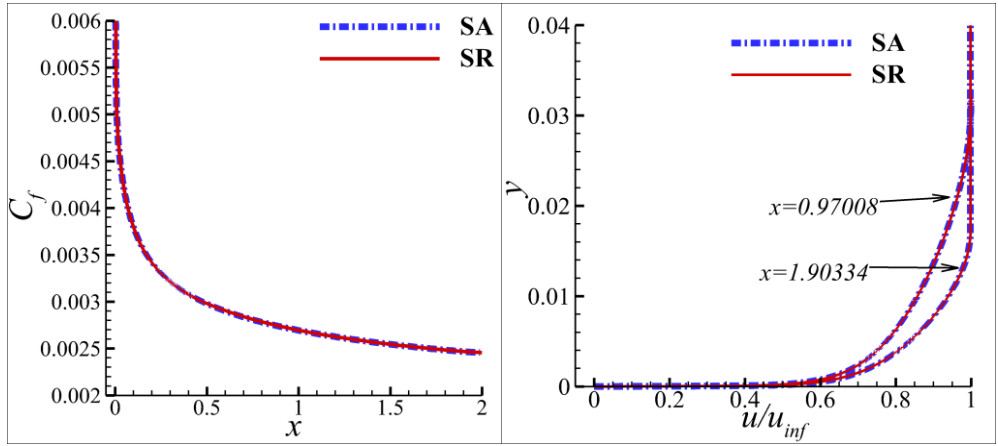

图 60 零压力梯度平板算例不同模型计算结果对比，左图：摩擦阻力系数分布，右图：不同站位处的速度型

## 5 总结与展望

基于数据同化算法开发工作所提供的高质量数据支撑，本文选取了两种典型高雷诺数翼型分离流动作为研究对象：（1）SC1095 薄翼型在跨声速大迎角下的分离流动；（2）DU91-W2-250 厚翼型在大迎角下的分离流动。首先通过数据样本精简与特征构建等预处理手段构建了适用于符号回归算法的样本数据集，其次通过符号回归算法得到了关于 SA 模型生成项修正场的代数表达式。得到的表达式构建了当地时均流场特征到对应修正系数 $\beta$ 之间的关系，其在形式上具有一定的物理合理性：只在近壁面附近才具有修正作用，对远场部分不进行修正，且值域为[0, 2]，确保了生成项系数的非负性。之后为了保证该公式在平板等外形以及翼型附着流上与标准 SA 模型结果一致，本章又对该公式进行了修正，确保其在顺压梯度、零压力梯度下的合理性。最后将修正之后的公式嵌入到 CFD 求解器中在翼型、机翼、曲线后台阶、驼峰、平板等算例上进行后验测试，并与标准 SA 模型的计算结果进行了对比。结果表明，在附着流情况下，修正模型的表现与标准 SA 模型一致；在翼型分离流情况下，修正模型计算结果相比 SA 模型具有明显的改善，其在翼型失速区域升力系数的平均相对误差相比 SA 模型降低 50%以上。同时修正模型也表现出了较好的泛化性，在多个不同外形、不同工况的翼型算例、三维机翼算例上都具有良好的修正效果，表明了本文所提出的面向高雷诺数复杂分离流动数据驱动湍流建模方法的良好前景。

本文基于符号回归方法所构建的修正表达式对于翼型外形因逆压梯度引起的分离流动有较好的作用，未来可在此基础上进一步提升建模数据的丰富性，并考虑实现对于符号回归算法的进一步优化，通过增加物理约束等方式增加模型的表示能力，增强模型泛化性，实现对于其他形式分离流动的有效修正。

# 6 参考文献